\documentclass[fleqn,usenatbib,useAMS]{mnras}

\usepackage{graphicx}	
\usepackage{amsmath}	
\usepackage{amssymb}	
\usepackage{multicol}        
\usepackage{longtable}
\usepackage{lscape} 
\usepackage{bm}		
\usepackage{pdflscape}	
\usepackage[flushleft]{threeparttable}
\usepackage{color}
\usepackage{rotating}

\newcommand{\se}{{\tt SExtractor\ }}
\newcommand{\psf}{{\tt PSFEx\ }}
\newcommand{\dos}{{\tt SExtractor+PSFEx\ }}

\usepackage[T1]{fontenc}
\usepackage{ae,aecompl}

\usepackage{newtxtext,newtxmath}

\title [The VVV near-IR galaxy catalogue]{The VVV Near-IR Galaxy Catalogue beyond the Galactic disk}

\author[L.~D. Baravalle]  
{
\parbox[t]{\textwidth}{Laura D. Baravalle$^{1}$, M. Victoria Alonso$^{1,2}$, Dante Minniti$^{3,4,5}$,  Jos\'e Luis Nilo Castell\'on$^{6,7}$, Mario Soto$^{8}$, Carlos Valotto$^{1,2}$, Carolina Villal\'on$^{1}$, Dar\'io Gra\~na$^{1}$, Eduardo B. Am\^{o}res$^{9}$, F. Milla Castro$^{6}$}
\vspace*{6pt} \
\\
$^{1}$ Instituto de Astronom\'ia Te\'orica y Experimental (IATE, CONICET-UNC), Laprida 854, C\'ordoba, Argentina.\\
$^{2}$ Observatorio Astron\'omico de C\'ordoba, Universidad Nacional de C\'ordoba, C\'ordoba, Argentina.\\
$^{3}$ Departamento de F\'isica, Facultad de Ciencias Exactas, Universidad Andr\'es Bello, Av. Fernandez Concha 700, Las Condes, Santiago, Chile.\\
$^{4}$ Instituto Milenio de Astrof\'isica, Santiago, Chile.\\
$^{5}$ Vatican Observatory, V00120 Vatican City State, Italy.\\
$^{6}$ Departamento de F\'isica y Astronom\'ia,  Facultad de Ciencias, Universidad de La Serena, 
Av. Juan Cisternas 1200 Norte, La Serena, Chile.\\
$^{7}$ Instituto de Investigaci\'on Multidisciplinario en Ciencia y Tecnolog\'ia, Universidad de La Serena. Av. Juan Cisternas 1400, La Serena, Chile.\\
$^{8}$ Instituto de Astronom\'ia y Ciencias Planetarias, Universidad de Atacama, Copayapu 485, Copiap\'o, Chile.\\
$^{9}$ Departamento de F{\'i}sica, Universidade Estadual de Feira de Santana (UEFS), Av. Trans. S/N, CEP 44036-900 Feira de Santana, BA, Brazil.
}

\date{\today}

\begin{document}
\label{firstpage}
\pagerange{\pageref{firstpage}--\pageref{lastpage}}
\maketitle

\begin{abstract}
Knowledge about the large-scale distribution of galaxies is far from complete in the Zone of Avoidance, which is mostly due to high interstellar extinction and to source confusion at lower Galactic latitudes. 
Past near-infrared (NIR) surveys,  such as the Two Micron All Sky Survey (2MASS), have shown the power of probing large-scale structure at these latitudes. 
Our aim is to map the galaxy distribution across the Southern Galactic plane using the VISTA Variables in the V\'ia L\'actea Survey (VVV), 
which reach 2 to 4 magnitudes deeper than 2MASS.
We used \dos to identify extended objects and to measure their sizes, the light concentration index, magnitudes, and colours. Morphological and colour constraints and visual inspection were used to confirm galaxies.  
We present the resulting VVV NIR Galaxy Catalogue of 5563 visually confirmed galaxies, of which only 45 were previously known.  This is the largest catalogue of galaxies towards the Galactic plane, with 99\% of these galaxies being new discoveries. We found that the galaxy density distribution closely resembled the distribution of low interstellar extinction of the existing NIR maps.
We also present a description of the 185 2MASS extended sources 
observed in the region, of which 16\% of these objects had no previous description, which we have now classified. 
We conclude that interstellar extinction  and stellar density are the main limitations for the detection of background galaxies in the Zone of Avoidance. The VVV NIR Galaxy Catalogue is a new data set providing information for extragalactic studies in the Galactic plane. 
\end{abstract}




\begingroup
\let\clearpage\relax
\endgroup
\newpage

\section{Introduction} 
\label{sec:intro}

The identification of galaxies behind the Milky Way (MW) is a difficult task, due to the obscuring effects of both dust and stars, and also the high stellar crowding at low Galactic latitudes (\citealt{2000A&ARv..10..211K}).  
The hidden galaxies can contribute to the identification of large-scale structures beyond our Galaxy 
\citep{Huchra2012, Macri2019}, as these allow us to obtain complete luminosity, mass and density distributions that can be tested against the models of galaxy and structure formation \citep{Peebles1980,Klypin1993}.

At optical wavelengths, systematic searches for galaxies towards the obscured regions in our Galaxy have been performed to reduce the effects of the Zone of Avoidance (ZoA). \cite{Kraan1999}, \cite{Kraan2000} and \cite{Woudt2001} presented the first deep optical search for galaxies using the object´s diameter for the selection of candidates behind the MW, with
these authors identifying overdensities and filaments of galaxies that might be related to possible extragalactic large-scale structures.

The use of NIR surveys minimizes the effects of foreground extinction in the ZoA in comparison with optical passbands.  The Two Micron All Sky Survey (2MASS; \citealt{Skrutskie2006}) produced images in the $J$, $H$, and $K_{s}$ NIR passbands, while \cite{2000AJ....119.2498J} developed an algorithm to detect and characterize extended sources in the 2MASS catalogue.
The procedure to separate point sources from extended 
sources includes tracking the point-spread function, image background removal, photometry and object classification, using a decision tree technique which takes into account the radial shape, surface brightness and symmetry parameters of the sources.  Near the Galactic plane (|$b$| $<$ 3$^{\circ}$), the extended sources are dominated by galaxies,
Galactic HII regions and large low surface brightness nebulae, stellar clusters and multiple stars. The main science products of this survey are the Point Source Catalog (PSC), consisting of over 500 million stars and galaxies, and the Extended Source Catalog (2MASX) with 1.6 million resolved sources complete to $K_{s}$ = 13.5 mag, and covering more than 99\% of the sky.
\cite{2000AJ....120..298J} reported the detection and spectroscopic confirmation of nearby galaxies hidden behind the MW data in the ZoA. Later, \cite{Schroder2007} performed a visual search for galaxies based on the Deep Near Infrared Survey (DENIS, \citealt{Epchtein1997}) and compared these results with objects in common with 2MASX.  These authors presented a catalogue of 122 galaxies and possible galaxy candidates, including morphological types estimated from the galaxy appearance in the $I$, $J$, and $K$ passbands, and total magnitudes and errors derived using the MAG\_AUTO magnitudes from \se (\citealt{Bertin1996}). More recently,  \cite{Schroder2019a} presented a homogeneous bright galaxy catalogue of 3763 objects based on the 2MASX survey at latitudes lower than 10$^{\circ}$, with the main goal of obtaining the Tully-Fisher relation to this
whole-sky sample. 

The 2MASS Redshift Survey (2MRS, \citealt{Huchra2005, Huchra2012, Macri2019}) provides redshifts for the 45,640 brightest 2MASS galaxies (extinction-corrected $K_{s}$ $<$ 11.75 mag) in the regions defined
 as |$b$| $\ge$ 5$^{\circ}$ for 30$^{\circ}$ $<$ l $<$ 330$^{\circ}$ and |$b$| $\ge$ 8$^{\circ}$ for other l values. The radial velocities of this survey might be able to contribute to studies of large-scale structures and cosmic flows in this important region of the hidden sky.  In this sense, the 2MASS Tully-Fisher survey (2MTF, \citealt{Masters2008, Howlett2017}) aims to measure distances of all bright spirals using the TF relation.  Finally, \cite{Lambert2020} applied to this catalogue a modification to the traditional Friends-of-Friends algorithm (\citealt{Huchra1982}), and found new group candidates that might help to define new large-scale structures in the ZoA. 

\cite{Staveley2016} presented the deep HI survey (HIZOA) in the southern ZoA covering the region of 212$^{\circ}$ $<$ $l$ $<$ 36$^{\circ}$ and |$b$| $ < $ 5$^{\circ}$. 
These HI surveys allow the study of sources in this region because they suffer minimal foreground extinction or source confusion when compared with optical and NIR surveys. These surveys are complementary, as NIR selection mainly favors the detection of early-type galaxies, while HI surveys the late-types. 
\cite{Sorce2017} predicted structures towards the ZoA by
performing simulations that yielded a probability distribution of galaxies. They then compared their results with a dozen known galaxy clusters, including the presence of the more distant Vela super-cluster (\citealt{Kraan2017}), and found a remarkable agreement.
Finally, \cite{Schroder2019b} presented
detections of 170 galaxies in the northern ZoA at |$b$| $ < $ 6$^{\circ}$, of which a third of these had no previous HI observations.

Recent NIR surveys have increased the capacity for detecting extragalactic sources in the ZoA, as shown by \cite{Williams2014}, who presented a photometric catalogue of 548 HIZOA galaxies that reach 2 magnitudes deeper than 2MASS in this region.  
 The NIR properties include ellipticities, position angles, isophotal  and extrapolated total magnitudes in the three $J$, $H$ and $K_{s}$ NIR passbands. \cite{Said2016} presented a deep NIR catalogue of 915 galaxies from the HIZOA, that includes
 ellipticities in the $J$ passband and isophotal magnitudes at $K_{s}$ = 20  mag / arcsec$^2$ in the $J$, $H$ and $K_{s}$ passbands.  Their main goal was to obtain accurate NIR photometric parameters for the NIR Tully-Fisher analysis.  This work is an extension of \cite{Williams2014} with the addition of new observations. The photometric and spectroscopic surveys are primordial to understand the large-scale structure in this complicated region due to the presence of the Milky Way.  All the surveys involving NIR data, including 2MASX, 2MRS and 2MTF, and the 'blind' HI data, such as  HIZOA, are complementary. Important structures in the local Universe such as the Great Attractor (for a review of its size and location see  \citealt{Mutabazi2014}) and Norma cluster (\citealt{Kraan1996}) are found in these regions. The combination of all surveys will allow peculiar velocities to be determined, in order to obtain a better understanding of the local dynamics, the cosmic flow fields and the underlying density field (\citealt{Kraan2018}). 

The VISTA Variables in the V\'ia L\'actea (VVV, \citealt{Minniti2010}) is a NIR variability survey of the entire MW Bulge and a large portion of the Southern Galactic Disk, with the main scientific goal being to gain more insight into the inner MW’s origin, structure, and evolution.  This survey has detected objects that are hidden behind Galactic high extinction regions, and these discoveries include variable stars, brown dwarfs (\citealt{Beamin2013}), new stellar open clusters (\citealt{Borissova2011,Borissova2014}; \citealt{Barba2015}), and 
new globular clusters (\citealt{Minniti2011,Minniti2017}). 

In addition to studies of Galactic structure, the VVV survey also offers an excellent opportunity to study extragalactic sources 
behind the MW, such as background galaxies, active galaxies including quasars and blazars, and groups and clusters of galaxies.   In this sense, \cite{Amores2012} identified 214 galaxy candidates in the d003 tile of the VVV survey
($l$ = 298.356$^{\circ}$, $b$ = -1.650$^{\circ}$)  behind  the Galactic disk, by means of visual inspection and a comparison of their sizes and colours 
with field stars.  \cite{Coldwell2014} confirmed the existence of the 
X-ray detected galaxy cluster Suzaku J1759-3450 \citep{Mori2013} at $z$ = 0.13 in the b261 tile ($l$ = 356.597$^{\circ}$, $b$ = -5.321$^{\circ}$), with the photometry of the sources being obtained using 
\se (\citealt{Bertin1996}).

\cite{Baravalle2018} was our first galaxy search study 
in the reddened and crowded fields located at low Galactic latitudes. We presented a photometric method based on \dos (\citealt{Bertin1996, Bertin2011}) to identify and characterize extragalactic sources behind the MW disk. The method was tested in the d010  ($l$ = 308.569$^{\circ}$, $b$ = -1.649$^{\circ}$) and d115 ($l$ = 295.438$^{\circ}$, 
$b$ = 1.627$^{\circ}$) tiles, and revealed 530 new galaxy candidates. 
\cite{Baravalle2019} presented the first confirmed galaxy cluster VVV-J144321.06-611753.9 at $z$ = 0.234 in the VVV d015 tile ($l$ = 315.836$^{\circ}$, $b$ = -1.650$^{\circ}$). 
The photometry was performed with \dos using the \cite{Baravalle2018} procedure, and the selection was based on the Cluster Red Sequence in the colour-magnitude diagrams \citep{Gladders2000}.

More recently, \cite{Saito2019} reported that VVV-WIT-04, a NIR variable source,  has an extragalactic origin and might be the counterpart of the radio source PMN J1515-5559.  \cite{Pichel2020} presented the NIR and mid-IR (MIR) properties of 4 known blazars located in the VVV regions, wich have very different NIR properties in the colour-magnitude and colour-colour diagrams compared with stellar or extragalactic sources, and also exhibit a significant variability in the $K_{s}$ light curves. 

\cite{Baravalle2018} and \cite{Baravalle2019} optimized the photometric procedure in order to identify and characterize extended sources. Here, we used the VVV NIR images across the Southern Galactic disk and we present the photometric catalogue of galaxies in the ZoA. This catalogue will allow the community to pursue a number of different scientific projects such as mapping the total extinction across the Galactic plane, 
and to recognize clear NIR windows of low interstellar extinction (\citealt{Minniti2018, Saito2020}). 
Other related projects involve  associate galaxy hosts for transient sources, including Supernovae, Gamma-ray bursts and gravitational wave events, to search for hidden nearby galaxies at low Galactic latitudes
and to identify candidates for compact groups and clusters of galaxies. The paper is organized as follows: 
Section $\S$2 presents the VVV NIR data, including principally the methodology to identify extragalactic sources and confirmed galaxies, as well as the comparison with other NIR surveys and the correlation with MIR data. Section $\S$3 presents the VVV NIR Galaxy Catalogue in the Southern Galactic disk of the survey and the galaxy distribution map. Section $\S$4 includes the principal conclusions and final remarks.

\section{The VVV survey}
\label{sec:cat}

The NIR observations allow the study of regions at the lower Galactic latitudes, where interstellar extinction  is severe. 
This wavelength regime is less affected by foreground extinction than the optical one. 
It is also sensitive to early-type galaxies, groups and galaxy clusters, which should be less confused with Galactic objects such as young stellar objects and cool cirrus sources (\citealt{Schroder2007}). 

The VVV survey is a NIR variability public survey of the Galactic bulge 
(10$^{\circ}$ $<$ $l$ $<$ 350 $^{\circ}$ 
and -10$^{\circ}$ $<$ $b$ $<$ +5$^{\circ}$)
and an adjacent section of the mid-plane (295$^{\circ}$ $<$ $l$ $<$ 350$^{\circ}$
and  -2$^{\circ}$ $<$ $b$ $<$ +2$^{\circ}$) (\citealt{Minniti2010}). 
The survey  was carried out using the VISTA (Visible and Infrared Survey Telescope for Astronomy) 4m telescope at ESO, which is equipped with a wide-field NIR camera (VIRCAM, \citealt{Dalton2006}) with a pixel scale of 0.34''/ pixel.

The VVV tiles are produced by six single pointing observations with a total field of view of 1.64 square degrees 
in the $Z$ (0.87 $\mu$m), $Y$ (1.02 $\mu$m), $J$ (1.25 $\mu$m), 
$H$ (1.64 $\mu$m), and $K_{s}$ (2.14 $\mu$m) passbands (\citealt{Saito2010}). The VVV images were taken under varying observing conditions, although in general this effect would be minimal because we use images with seeing  $<$ 0.9 arcsec in the $K_{s}$ passbands.  The background sky level varies from image to image, and the source  
density changes with position across the MW disk, depending on the line-of-sight.
The survey area is fully covered by 348 tiles in the MW, of which 196 are in the bulge and 152 in the disk regions. 
In this work, we focused our efforts on the disk parts of 220 deg$^{2}$ covered by the survey.  The images from the Project Programme 179.B-2002 were downloaded from the Cambridge Astronomical Survey Unit\footnote{http://casu.ast.cam.ac.uk/vistasp/imgquery/search} (CASU,
\citealt{Emerson2006}) with the observing status 'Completed', with the same exposure time and the smallest observed seeing.

In the future, the VISTA Variables in the V\'ia L\'actea eXtended Survey (VVVX) will triple the areal coverage of the original VVV survey following the same observational strategy. This extended survey, aimed for completion in late 2021, will provide $J$, $H$ and $K_{s}$ astrometric and photometric catalogues reaching similar magnitudes for the extended area, as well as variability information for the $K_{s}$ passband.  The galaxies found in these regions would connect the VVV galaxy distribution with other surveys at higher galactic latitudes.

\subsection{The methodology}
\label{sec:metodology}

The algorithm to detect and characterize extragalactic sources using \se (\citealt{Bertin1996}) and \psf (\citealt{Bertin2011})  has been extensively described in \citet{Baravalle2018}. Briefly, \se performs the photometry in the $Z$, $Y$, $J$, $H$ and $K_{s}$ VVV images and creates catalogues for each passband.  \psf takes these catalogues and creates the best PSF model by looking for well-defined point sources. Finally, \se applies the PSF model to each
source to obtain the astrometric, photometric, and
morphological properties. 

We applied the custom-built pipeline to the 152 tiles of the Galactic disk
region using only the $J$, $H$ and $K_{s}$ images, as most of the galaxies have no light
contributions in the $Z$ and $Y$ passbands. All detected sources in these
passbands were cross-matched taking the $K_{s}$ passband as a reference, using
the maximum object separation of 1 arcsec. In order to discriminate point
sources from extended objects, we used magnitudes, the stellar index \textit{CLASS\_STAR}, the SPREAD\_MODEL ($\Phi$) parameter, the radius that contains 50\% of the total flux of an object ($R_{1/2}$), and the concentration index ($C$, \citealt{Conselice2000})  from the $K_{s}$ detections.  
The extended sources have to satisfy the morphological criteria:
\textit{CLASS\_STAR} $<$ 0.3; $\Phi$ $>$ 0.002; 
1.0 $<$ $R_{1/2}$ $<$ 5.0 arcsec and 
2.1 $<$ $C$ $<$ 5 (\citealt{Baravalle2018}). 

~\se performs a clean procedure for any source inside the adaptive aperture, in which the fluxes are integrated by applying a mirroring between one side of the ellipse and the other.  Also, the flux 
is defined by taking into account the threshold that separates the source from the background. The stellar density at lower Galactic latitudes makes it difficult to separate close objects degrading this process, and as this makes the adaptive apertures smaller, the magnitudes are underestimated. In order to minimize this effect, the cases with extreme contamination were not included in this study, which in fact represent a very small percentage of our total sample.  The \se photometry allowed us to obtain total magnitude (MAG\_AUTO) estimates using the Kron magnitudes (\citealt{Kron1980}), which represent approximately 94\% of the total flux (\citealt{Bertin1996}). 
\dos provided us with the PSF fitting magnitudes (MAG\_PSF),  using the PSF model mentioned above, and also the circular aperture magnitudes obtained within a fixed aperture of 2 arcsec diameter to avoid strong stellar contamination. The ($J$ - $K_{s}$), ($J$ - $H$) and ($H$ - $K_{s}$) colours were then derived from these aperture magnitudes.

\subsubsection{Interstellar extinction and stellar density}
\label{subsec:extinction}

The extinction at lower Galactic latitudes is high in some parts of the MW. \cite{Schlegel1998} derived  extinction maps from the 100 $\mu$m dust emission by DIRBE/IRAS, which were re-calibrated by  \cite{Schlafly2011}.  At |b| $<$ 5$^{\circ}$, the presence of contaminating sources was not removed and the reddening reported should be taken
with caution (\citealt{Arce1999, Amores2005, Amores2007, Gonzalez2012, Soto2013, Soto2019}). 
\cite{Nagayama2004} studied the distribution of galaxies at low Galactic latitudes (b $\sim$ 1.7$^{\circ}$), and by using a stellar colour-excess technique these authors found a systematic underestimation of A$_{K}$ $\sim$ 0.4 mag  in the map of \cite{Schlegel1998}.  However, despite these concerns, the maps are still widely used to characterize the foreground extinction in studies of the distribution of galaxies in the ZoA. In particular, the work of  \cite{Schroder2007}, \cite{vanDriel_2009} and \cite{Schroder2019a} have used the \cite{Schlegel1998} maps to obtain the extinction-corrected magnitudes, and found them to be consistent with the 2MRS and 2MTF surveys. 
Our magnitudes were corrected by interstellar extinction along the line-of-sight, using the maps of \cite{Schlafly2011} and the VVV NIR relative extinction  coefficients of \cite{Catelan2011}:
A$_{J}$ = 0.280 A$_{V}$, A$_{H}$ = 0.184 A$_{V}$ and 
A$_{Ks}$ = 0.118 A$_{V}$. 
These corrected magnitudes and colours are represented by $J^{\circ}$, $H^{\circ}$ and $K_{s}^{\circ}$ and the colours ($J$ - $K_{s}$)$^{\circ}$ and ($H$ - $K_{s}$)$^{\circ}$ throughout this work.  

\subsubsection{False detections and duplicate objects}

\begin{figure*}
\centering
\includegraphics[width=0.30\textwidth,height=0.30\textwidth]{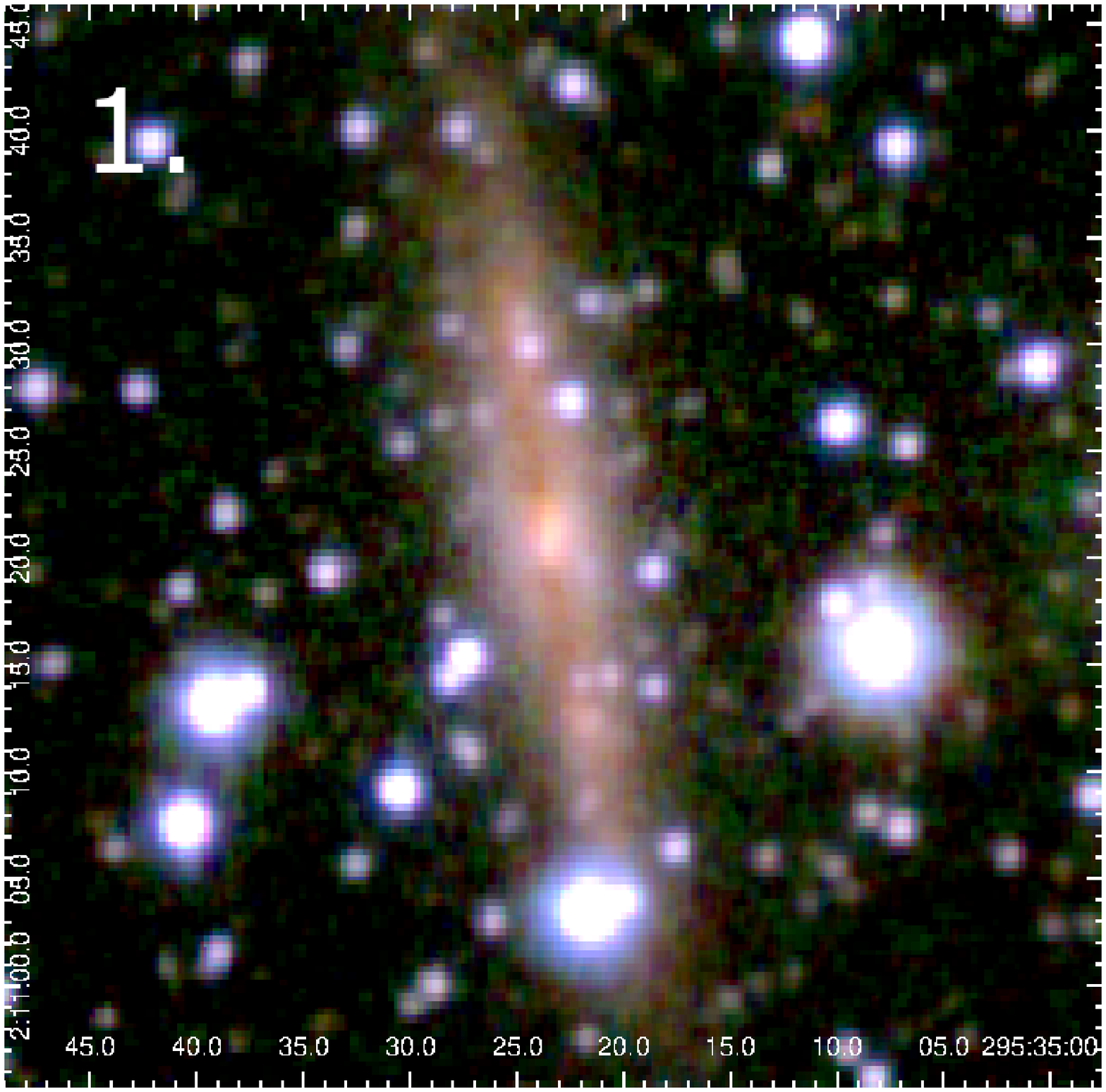}
\includegraphics[width=0.30\textwidth,height=0.30\textwidth]{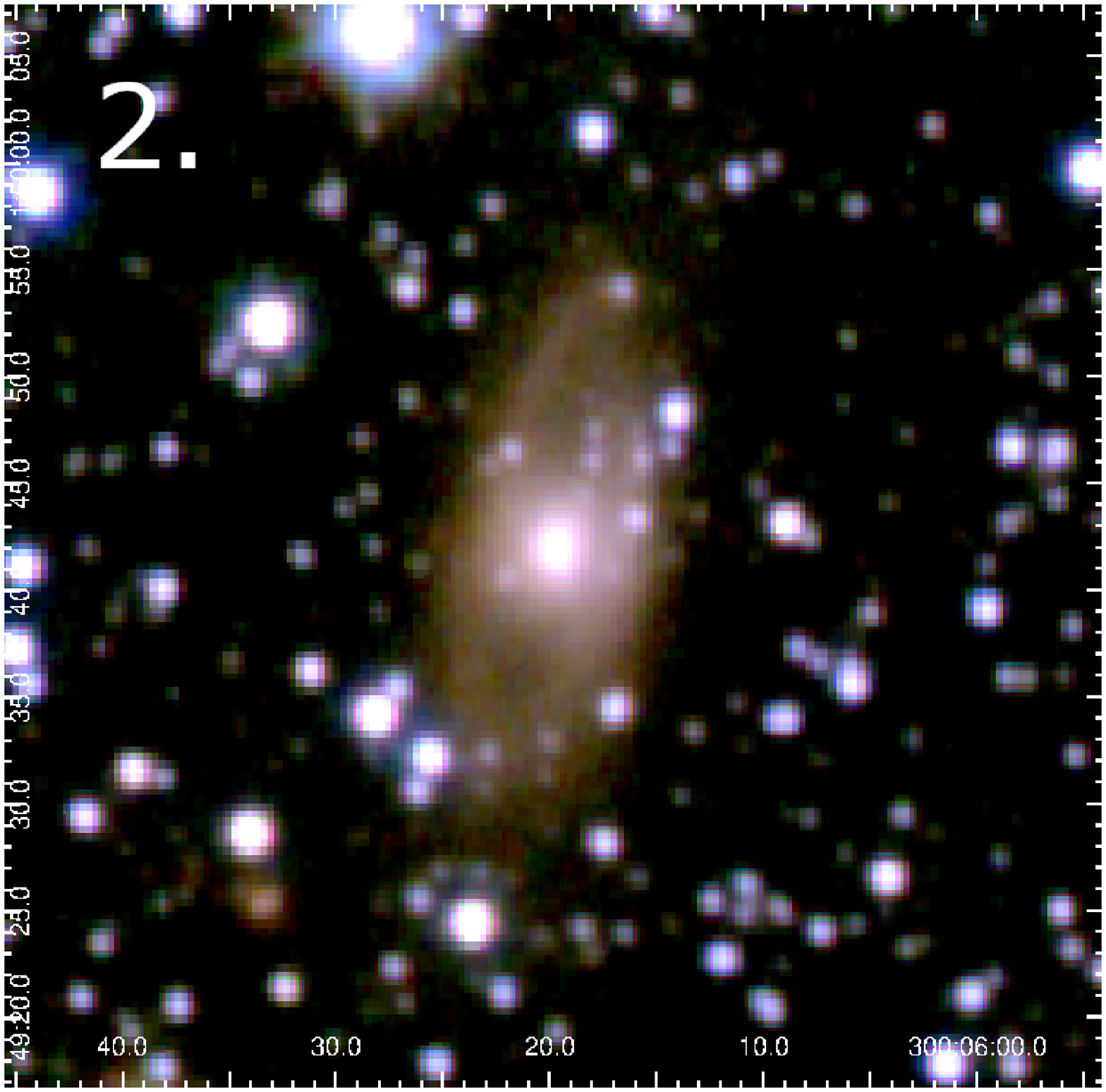}
\includegraphics[width=0.30\textwidth,height=0.30\textwidth]{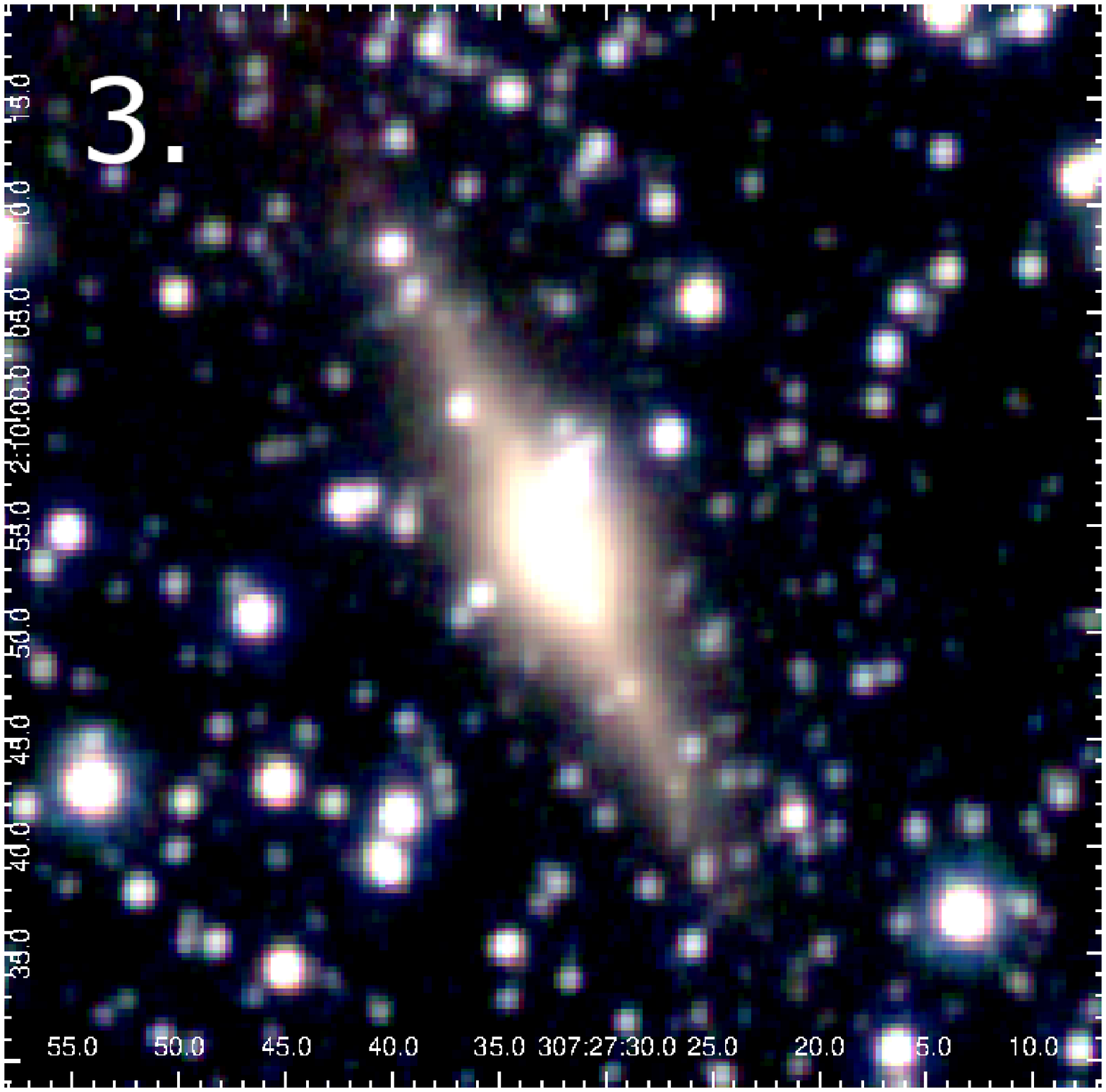}
\includegraphics[width=0.30\textwidth,height=0.30\textwidth]{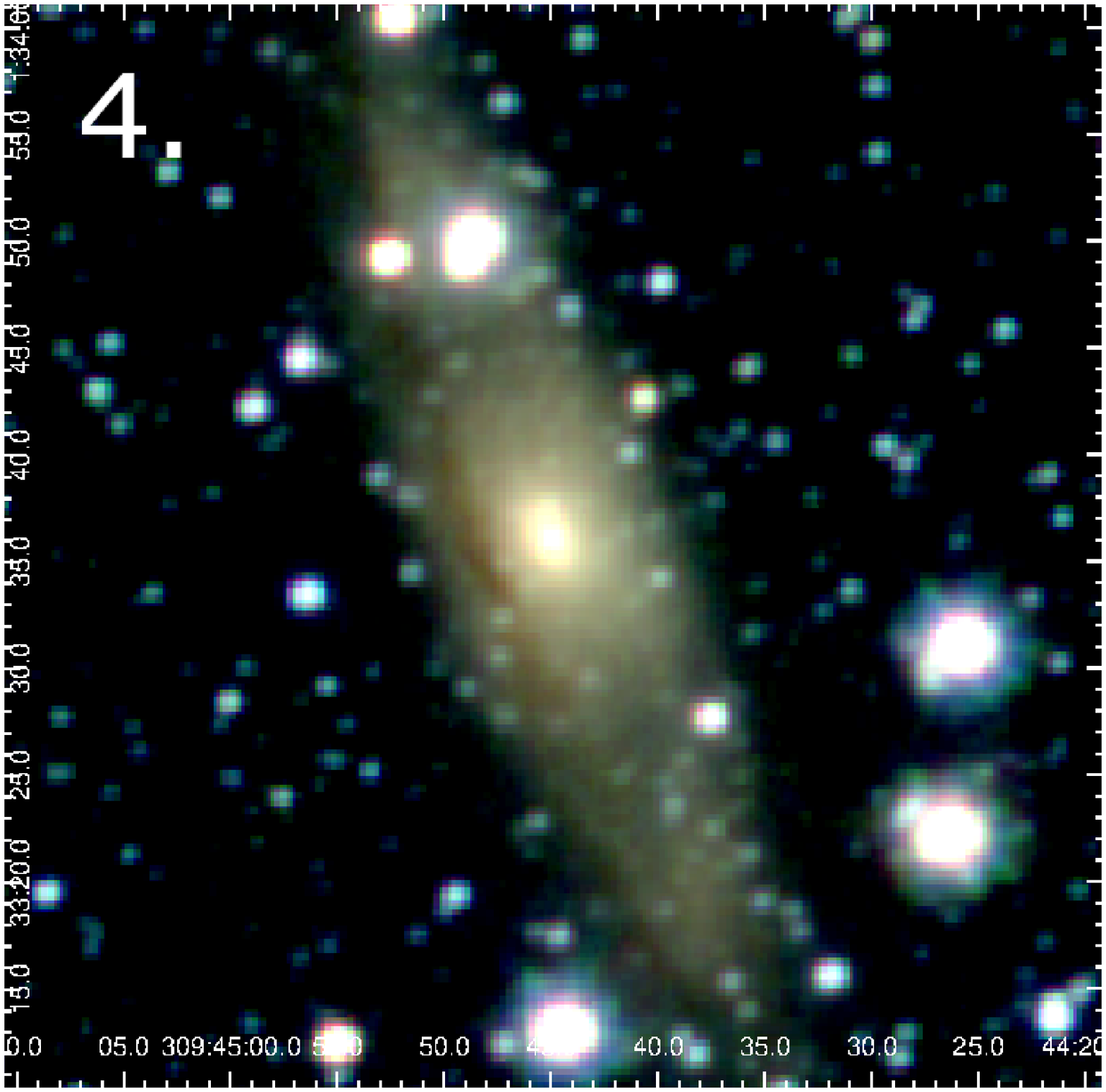}
\includegraphics[width=0.30\textwidth,height=0.30\textwidth]{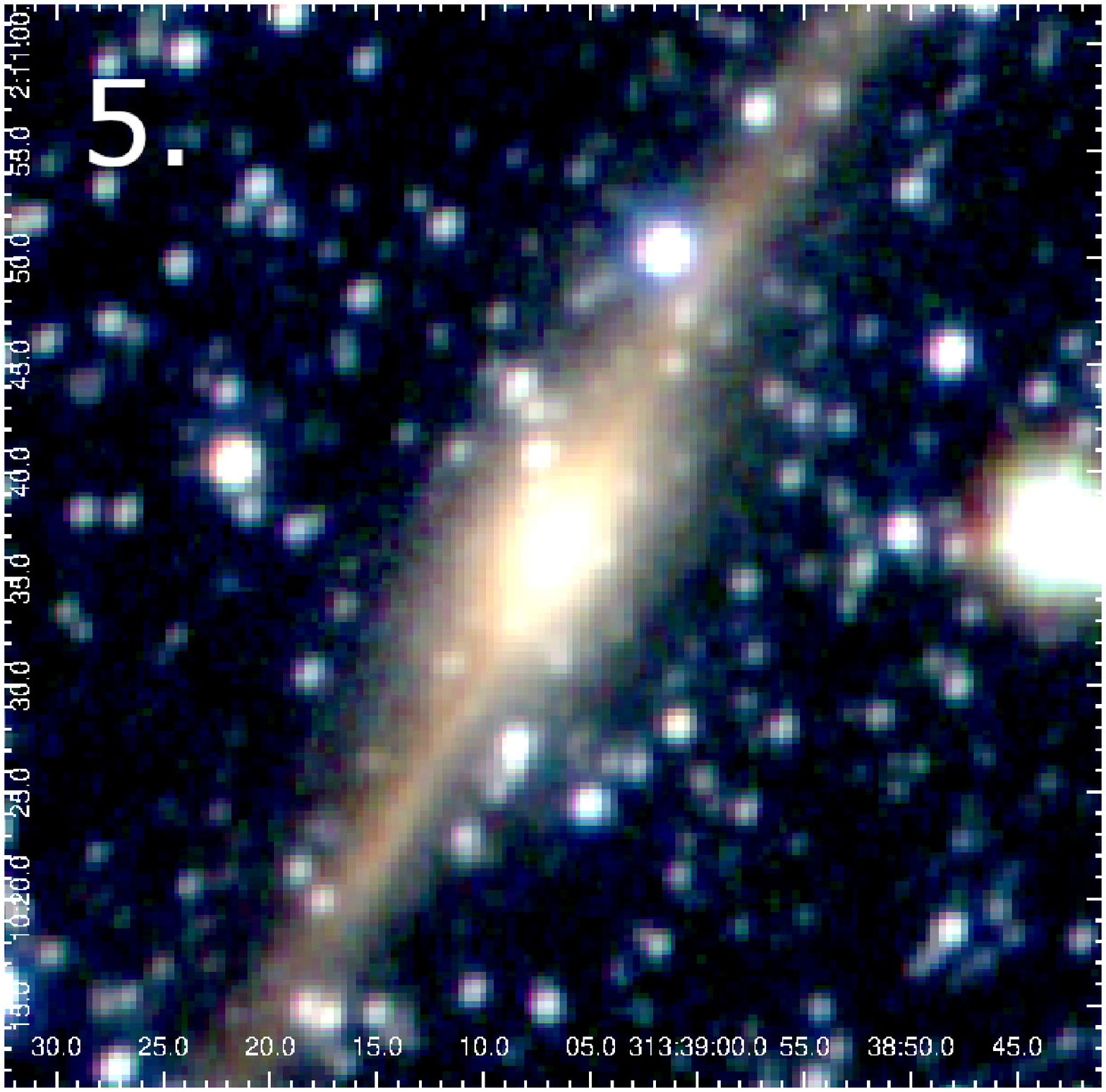}
\includegraphics[width=0.30\textwidth,height=0.30\textwidth]{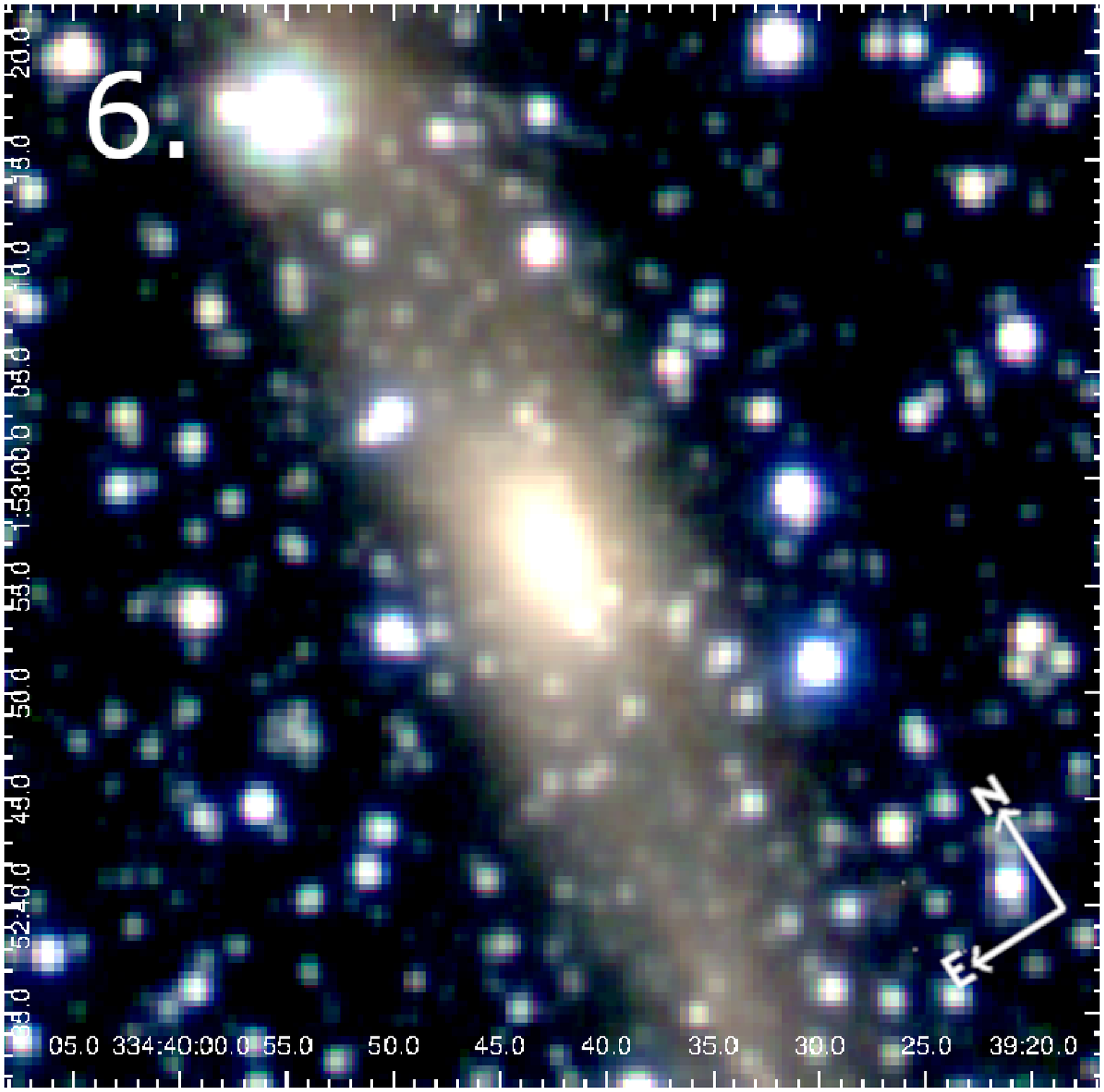}
\caption{The VVV colour composed images of galaxies with strong stellar contamination.
The panels show the sources: 1. J11523245-5950248, 2. J12280968-6054558, 3. J13281021-6022580, 
4. J13471848-6034133, 5. J14155209-5855348, and 6. J16174633-4751584.
}\label{galaxies}
\end{figure*}

In order to minimize false detections, we also considered colour cuts: 
0.5 $<$ ($J$ - $K_{s}$)$^{\circ}$ $<$ 2.0 mag, 
0.0 $<$ ($J$ - $H$)$^{\circ}$ $<$ 1.0 mag,  
0.0 $<$ ($H$ - $K_{s}$)$^{\circ}$ $<$ 2.0 mag, and ($J$ - $H$)$^{\circ}$ + 0.9 ($H$ - $K_{s}$)$^{\circ}$ $>$ 0.44 mag following \cite{Baravalle2018}. These cuts provide a good compromise between avoiding the loss of interesting objects and the inclusion of spurious ones, and these are similar to those used in \cite{2000AJ....119.2498J, 2000AJ....120..298J} and \cite{Amores2012}.
This process classifies these objects as  extragalactic candidates, and the final visual inspection allows us to confirm the galaxies. This inspection was performed by four of the authors of the present  study for all the candidates by looking at the images in the five passbands of the VVV survey using the VISTA Science Archive (VSA\footnote{http://horus.roe.ac.uk/vsa}).  When any doubts, discrepancies and/or comments arose, we searched for false three-colour images generated from the $J$, $H$ and $K_{s}$  passbands.  The combined  colour and the extended nature
allowed us to distinguish better the true galaxies from false ones.  Some objects were also found with important stellar contamination, with Figure~\ref{galaxies} showing examples of the largest galaxies detected with strong stellar contamination.  These objects were also observed by the 2MASX survey. In our first investigation (\citealt{Baravalle2018}), we tried to mask nearby stars and to correct galaxy magnitudes.  However, the strong contamination affected  more than half of the galaxy light contributions and made it difficult to obtain 
accurate magnitude estimates.  Thus, the results  were in general not reliable, and we decided not to correct in the present study for stellar contamination. In addition, the cases with a high number of nearby stars, especially those for small objects, were not considered in this work. 
In contrast, the cases with little contamination were included in the analysis without any correction, resulting in brighter magnitudes.  
Following \citet{Baravalle2018}, we 
obtained the photometric and structural parameters of these galaxies, including the $R_{1/2}$, $C$, ellipticity and the spheroid S\'ersic index ($n$, \citealt{Sersic1968}).

Appendix A includes useful information that characterizes the VVV tiles of the Galactic disk.  For all the tiles, the median interstellar extinctions using the values close to the extended sources, extragalactic sources and galaxies are A$_{Ks}$ = 1.423 mag, 0.656 mag and 0.710 mag, respectively.  In general, the extended objects are distributed across the tiles, and the extragalactic sources and galaxies were found mainly in the regions with lower interstellar extinctions. We also used the stellar density defined as the logarithm of the number of stars with $K_{s} < $ 15 mag per square degrees, using the PSF photometry of \cite{Alonso2018}.

\begin{figure*}
\centering
\includegraphics[width=0.45\textwidth,height=0.40\textwidth]{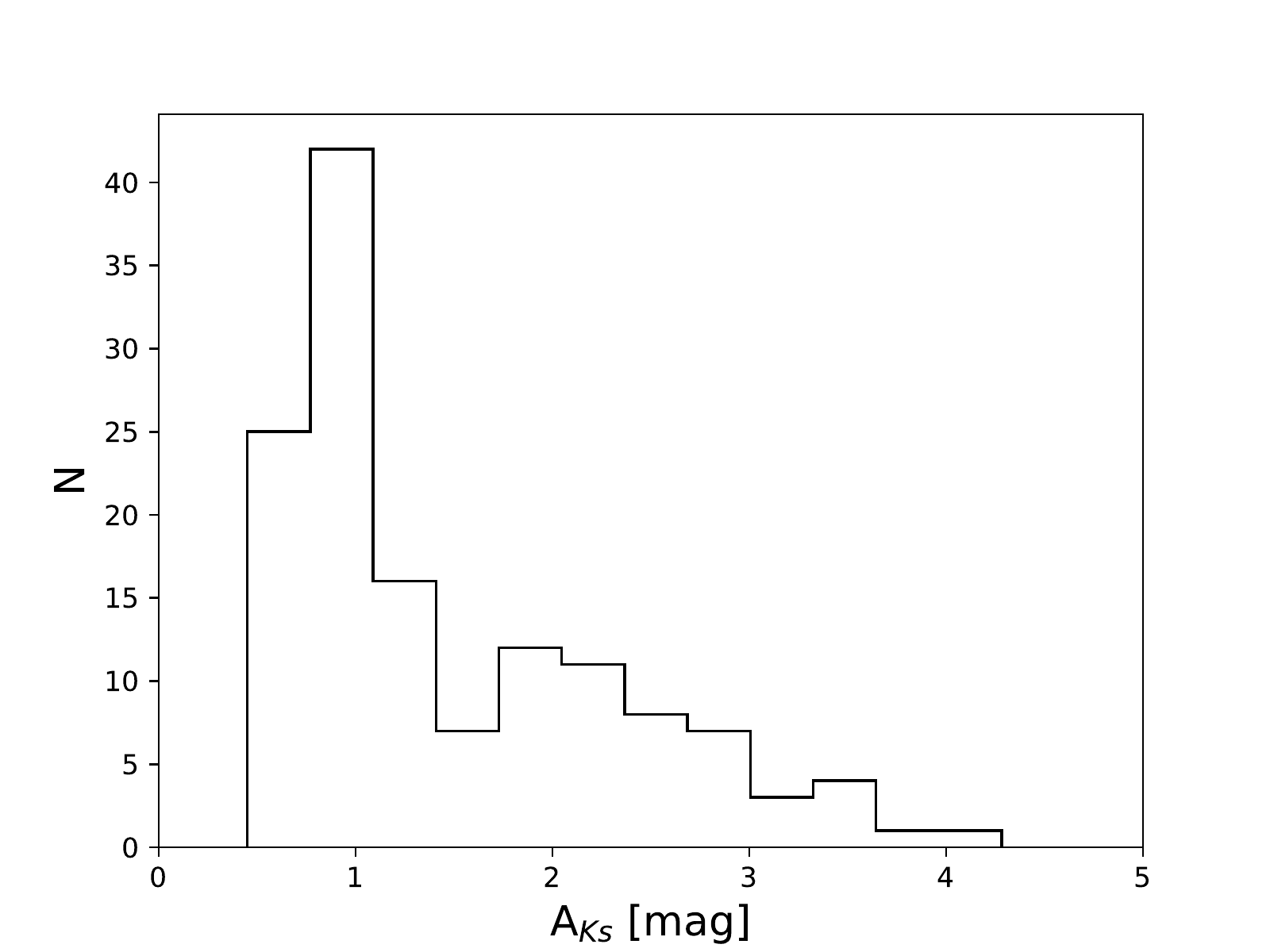}
\includegraphics[width=0.45\textwidth,height=0.40\textwidth]{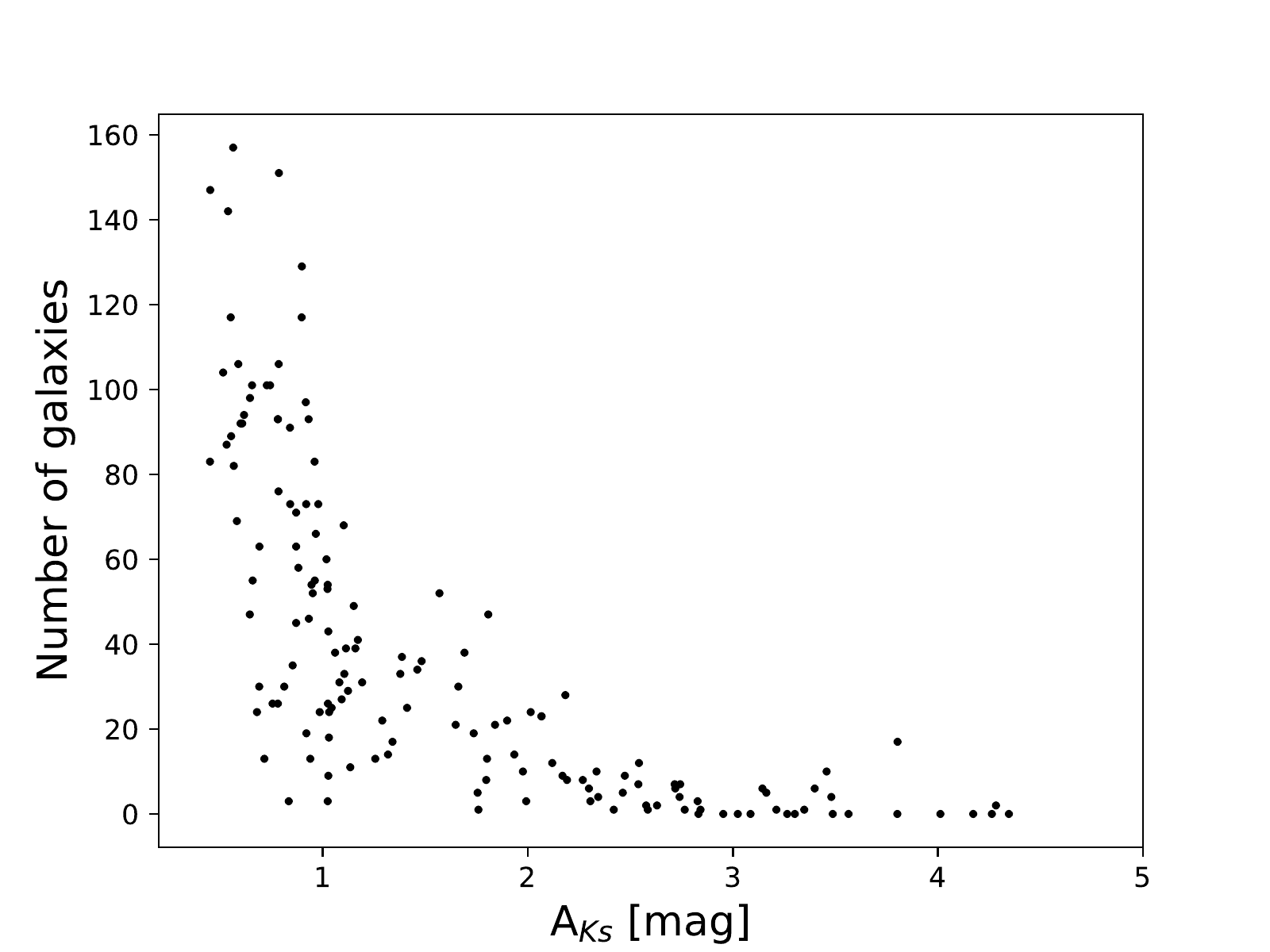}\\
\includegraphics[width=0.45\textwidth,height=0.40\textwidth]{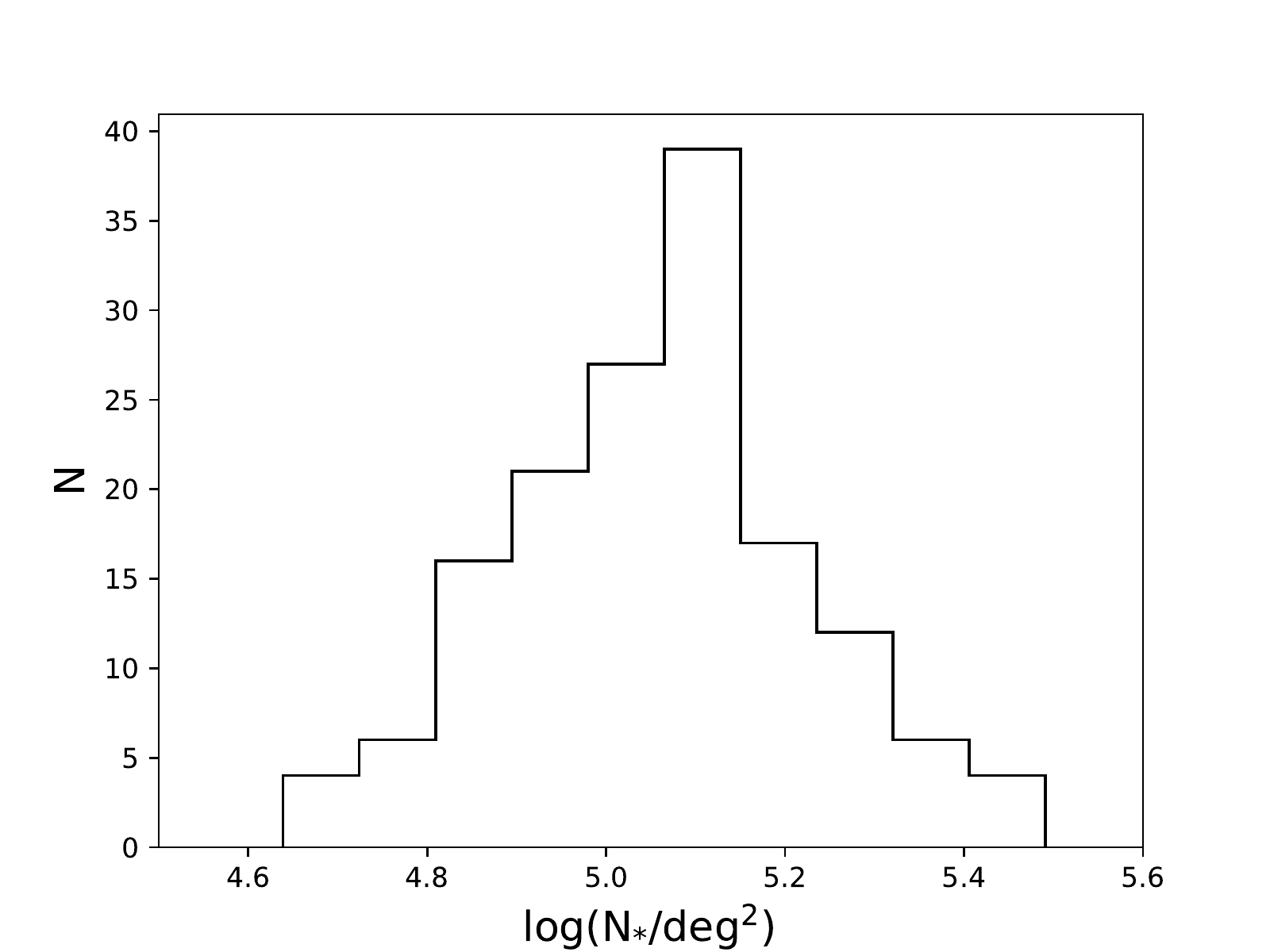}
\includegraphics[width=0.45\textwidth,height=0.40\textwidth]{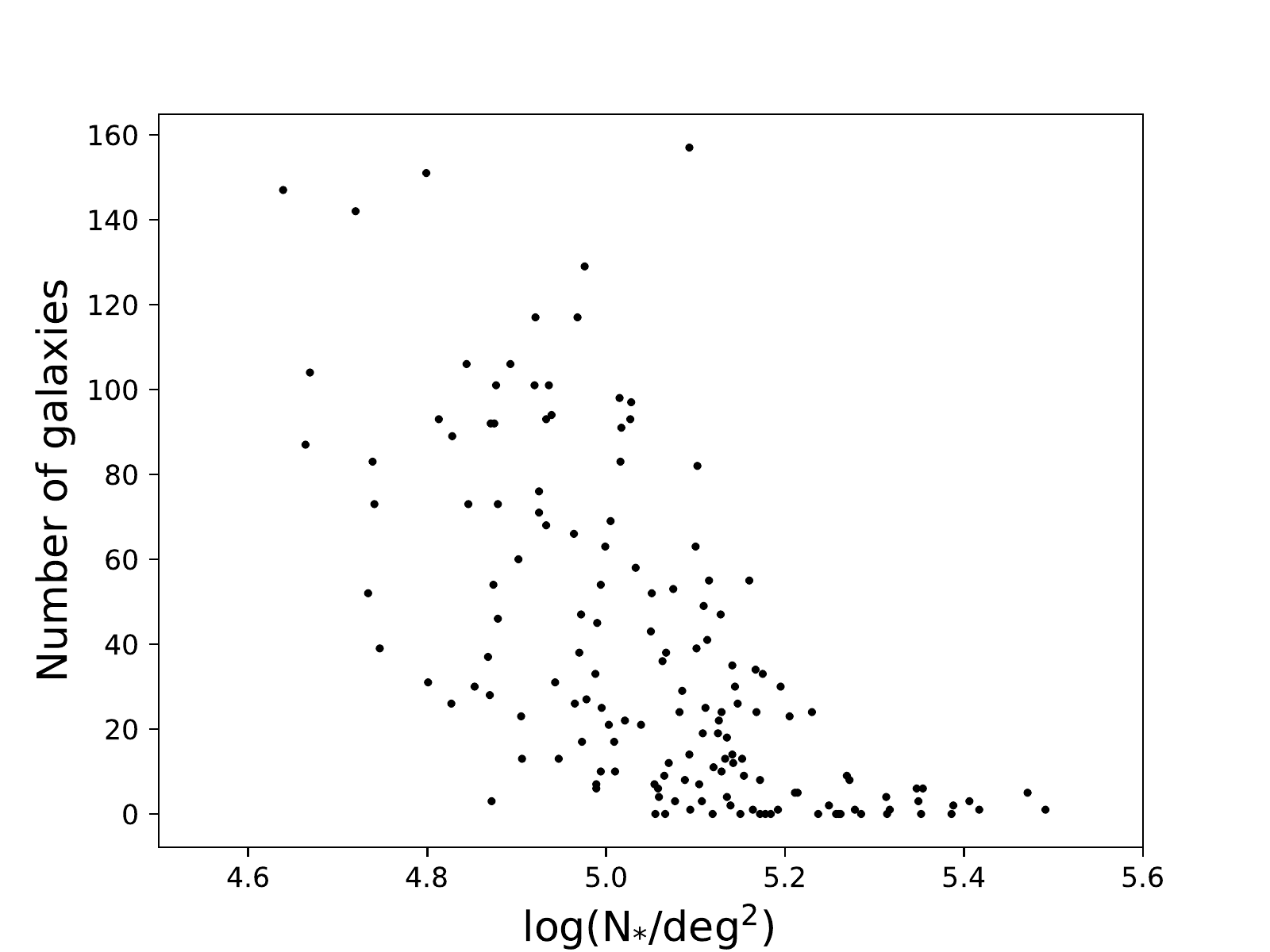}
\caption{VVV tiles in the Galactic disk showing the distributions of the median A$_{Ks}$ interstellar extinction obtained for all the extended objects detected in the tiles (upper panels) and the stellar density estimates (bottom panels).
Left panels display the distributions, and the right panels show the number of galaxy detections for each tile as a function of the extended median A$_{Ks}$ and the stellar density.
}\label{nagal}
\end{figure*}

Figure~\ref{nagal} shows the distribution of median A$_{Ks}$ interstellar extinctions defined by extended sources in the upper panels, and displays the stellar densities  
in the bottom panels.  The distributions are presented
in the left panels, and the number of detected galaxies per tile as a function of these median A$_{Ks}$ interstellar extinctions and stellar densities is shown
in the right panels.   At lower interstellar extinctions (A$_{Ks}$ $<$ 1 mag), the highest number of detected galaxies is 157, while for 1 $\le$ A$_{Ks}$ $\le$ 3 mag, the number drops from 68 to 1.  The median stellar density in the VVV disk, given by log (N$_{*}$/deg$^{2}$), is 5.066 $\pm$ 0.167, which is higher than the peak of the distribution of about 4.5 reported by \cite{Schroder2019a} in the ZoA sample.   These results show  severe contamination in our studied regions.  As expected, as shown in the right panels, the highest number of galaxy detections were obtained in the outermost parts of the disk and the lowest values were found in regions with high interstellar extinction and stellar contamination. No  extragalactic candidates were detected in the d100, d108 or d110 tiles.  After the visual inspections, all the extragalactic sources in the d065, d069, d070, d071, d072, d073, d099, d100, d101, d102, d108, d109, d110, d112 and d114 tiles were classified as false detections. These regions have lower Galactic latitudes and mainly higher interstellar extinctions, with star crowding being extreme with median A$_{K_s}$ values from 3 mag up to 6.19 mag and logarithmic stellar densities from 5.055 up to 5.384. On the other hand, more than 150 galaxies were found in the d022 and d122 tiles, with lower interstellar extinctions of 0.56 and 0.79 mag and logarithmic stellar densities of 5.093 and 4.799, respectively.  

\begin{figure*}
\centering
\includegraphics[width=0.45\textwidth,height=0.40\textwidth]{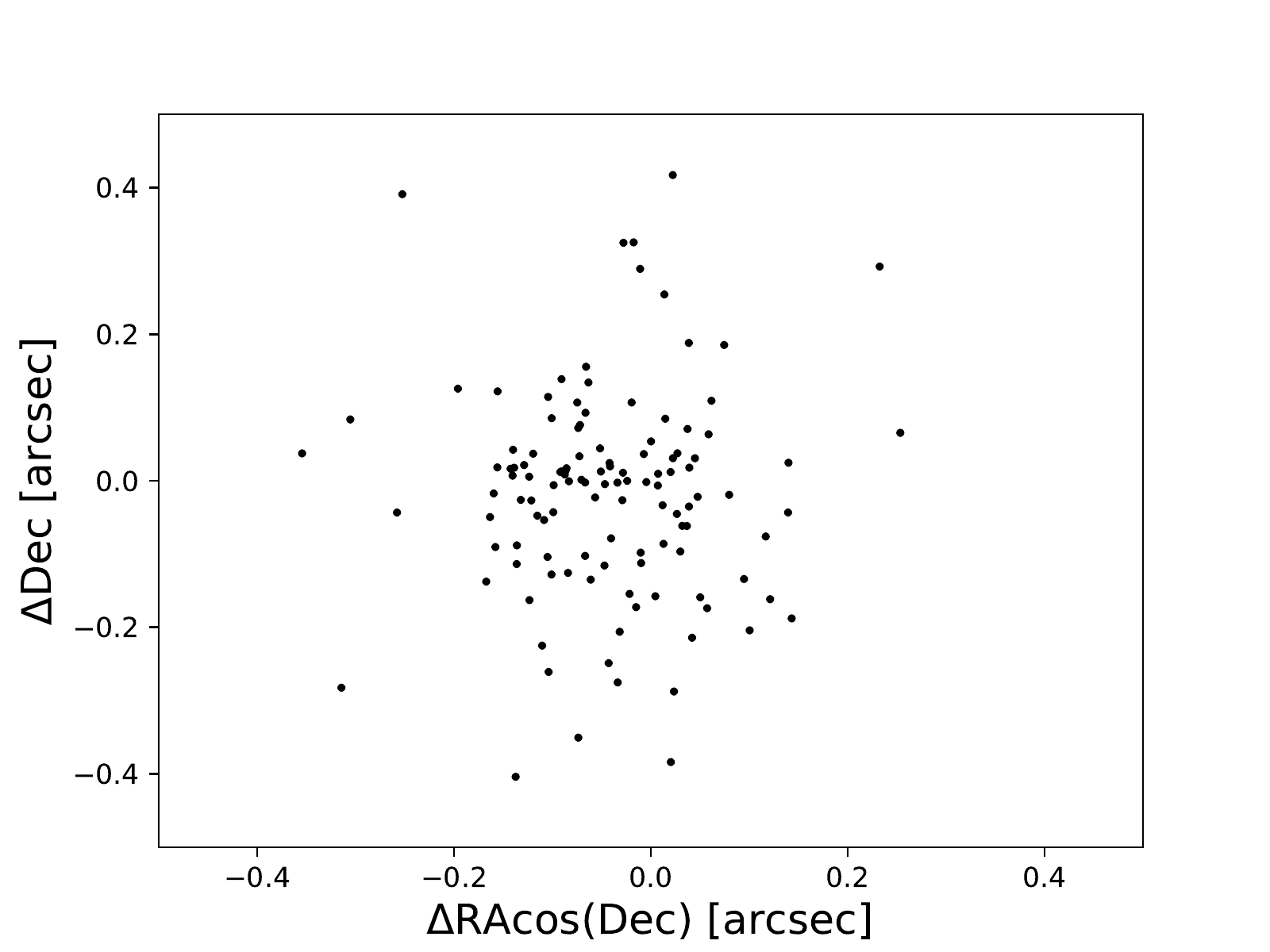}
\includegraphics[width=0.45\textwidth,height=0.40\textwidth]{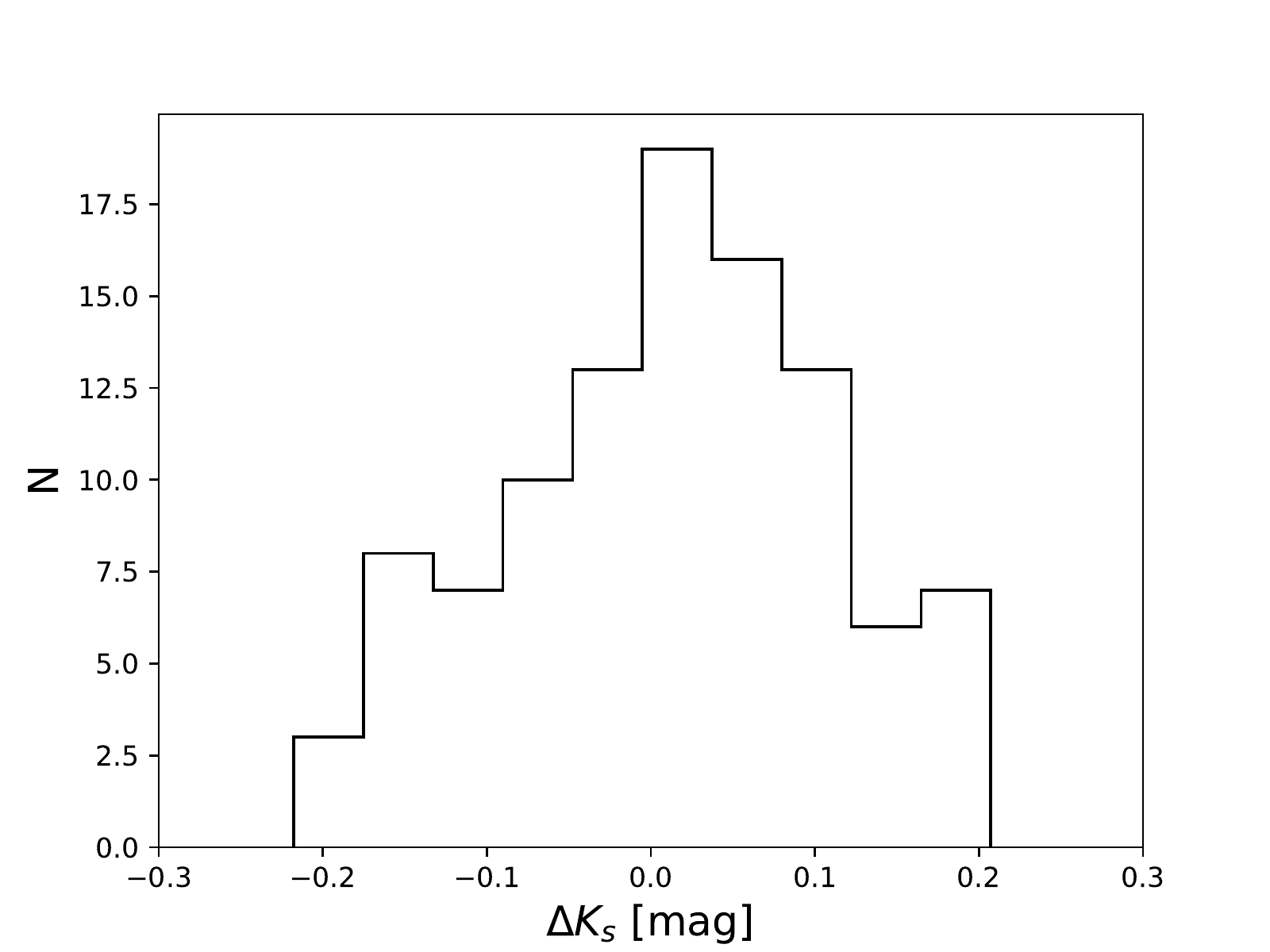}
\caption{Astrometric and magnitude differences for duplicate galaxies.  Left panel shows differences in the coordinates and right panel display the distribution of the $K_{s}$ differences. 
}\label{duplicates}
\end{figure*}

We also have 
repeated objects, mostly found in the overlapping tile boundaries due to the observing procedure and tile construction.  Within the angular separations of 0.75 arcsec, 122 duplicated galaxies were found. Figure~\ref{duplicates} shows the differences in the coordinates and $K_{s}$ magnitudes for these duplicated sources.  The median differences after 1$\sigma$--clipping are 
$\Delta$RA cos(Dec) = (-0.062 $\pm$ 0.060) arcsec, 
$\Delta$Dec = (0.009 $\pm$ 0.066) arcsec and $\Delta K_{s}$ = (0.009 $\pm$ 0.102) mag.  These values might be considered the uncertainty estimates for the astrometry and photometry. 

\subsubsection{Photometric results} 

The different NIR 
photometric systems in general did not match exactly, as was expected, given that the observations were carried out at different sites with different telescopes, IR cameras, detectors and filters. 
We used the photometric transformations detailed in  \cite{Soto2013} to translate our magnitudes to the 2MASS photometric system. The transformations were obtained on a tile by tile basis, and these were based on the selection of stars with a good photometry. Briefly, the procedure can be summarized as follows: First, the $K_{s}$ VVV photometry was used to clean the CASU catalogue of non-stellar sources by selecting stars in a specific intensity range, weighted by second moments. Subsequently, stars in relative isolation were selected by using a radius of 2 arcsec to remove multiple 
sources in close proximity.  This VVV catalogue was then cross-referenced with 2MASS sources, with the signal to noise ratio greater than 7,  
and the result was then matched with the VVV catalogue with a small radius of 0.1 arcsec in the $J$ and $H$ passbands. The resulting clean VVV-2MASS catalogue was then used to derive the photometric transformations using a linear fit with an iterative clipping algorithm and adaptive bins.  The VVV aperture and total magnitudes of the detected galaxies in the Galactic disk were transformed to the 2MASS photometric system using \cite{Soto2013}.

\begin{figure*}
\centering
\includegraphics[width=0.45\textwidth,height=0.40\textwidth]{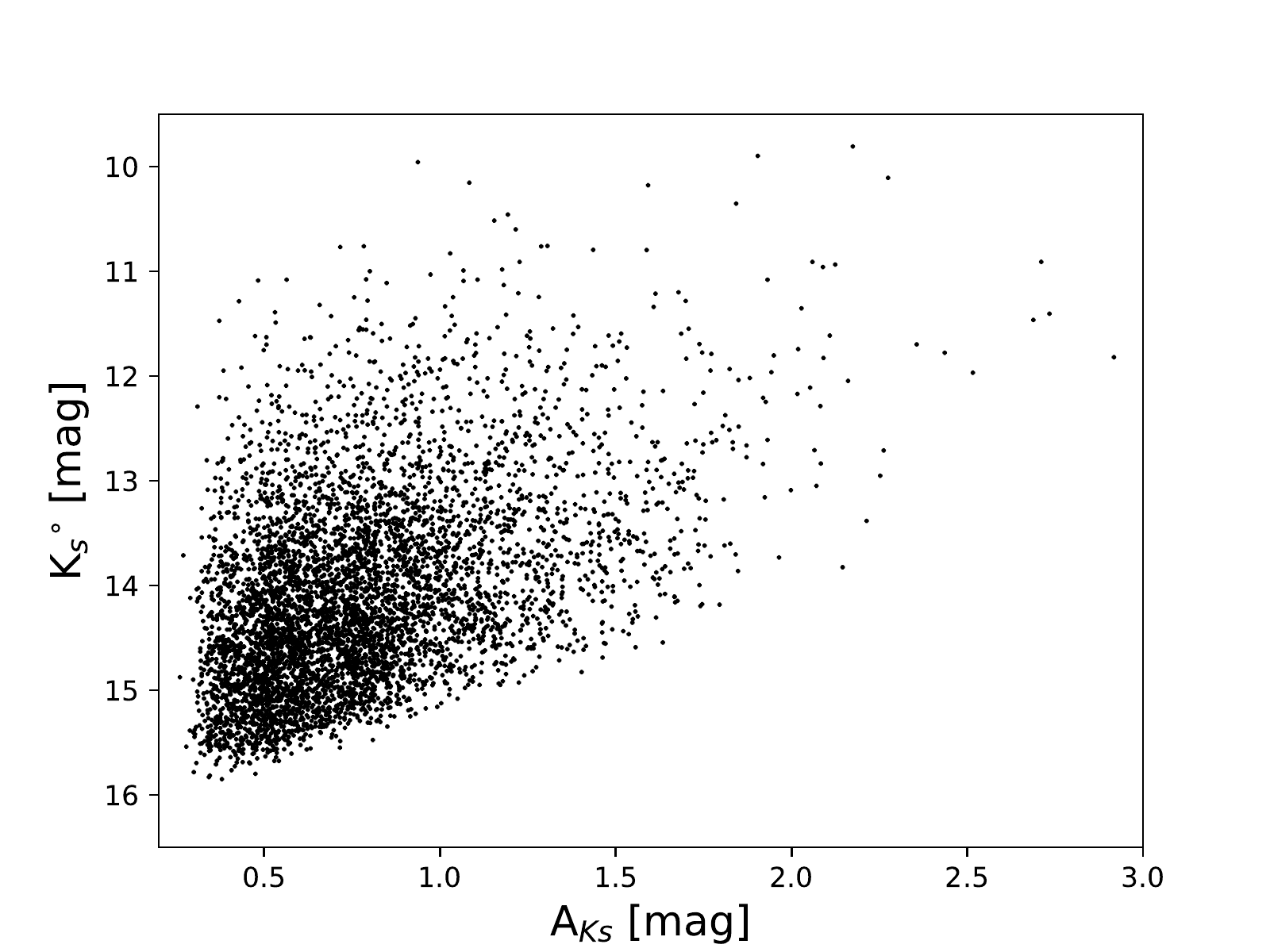}
\includegraphics[width=0.45\textwidth,height=0.40\textwidth]{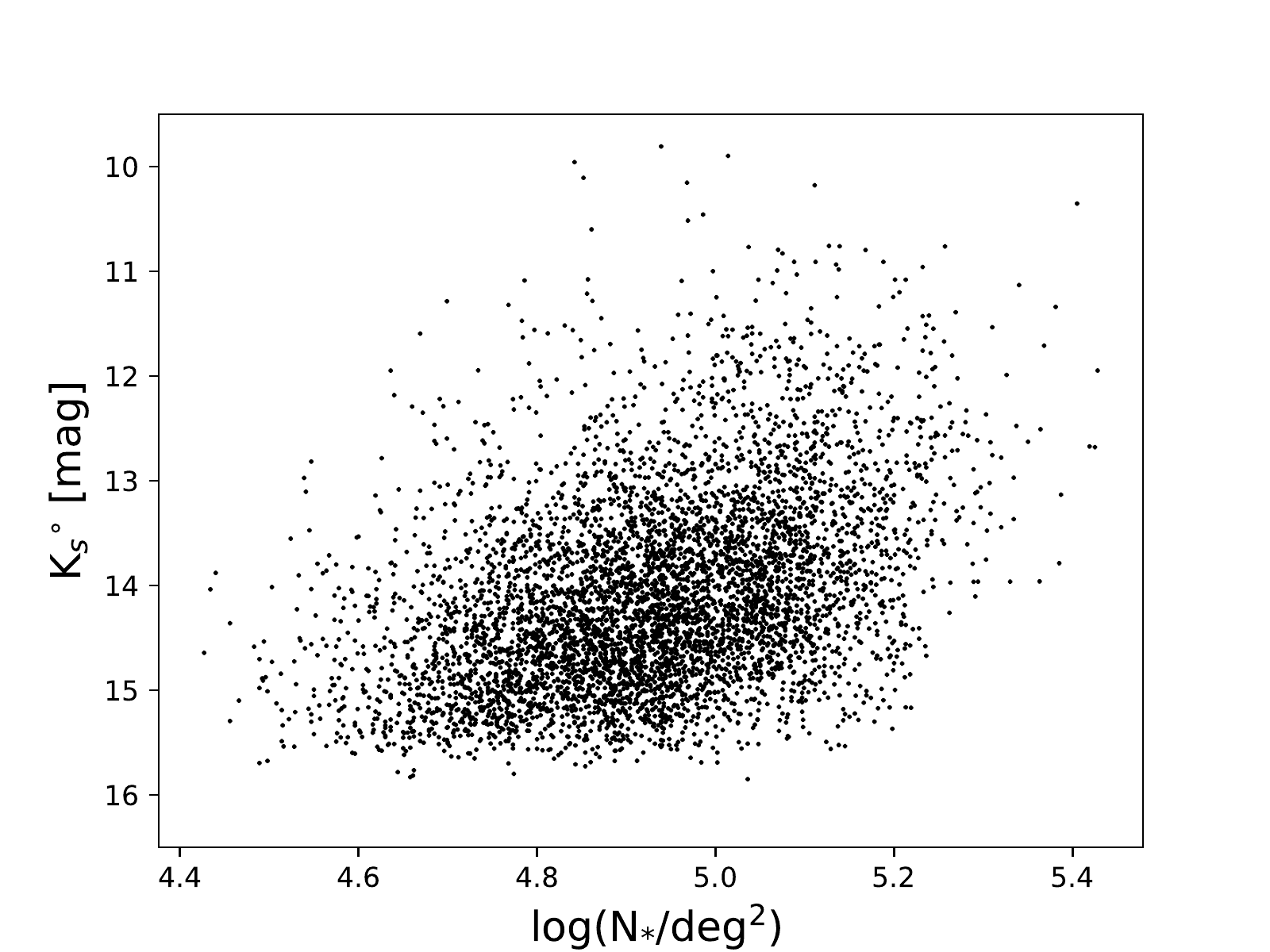}\\
\includegraphics[width=0.45\textwidth,height=0.40\textwidth]{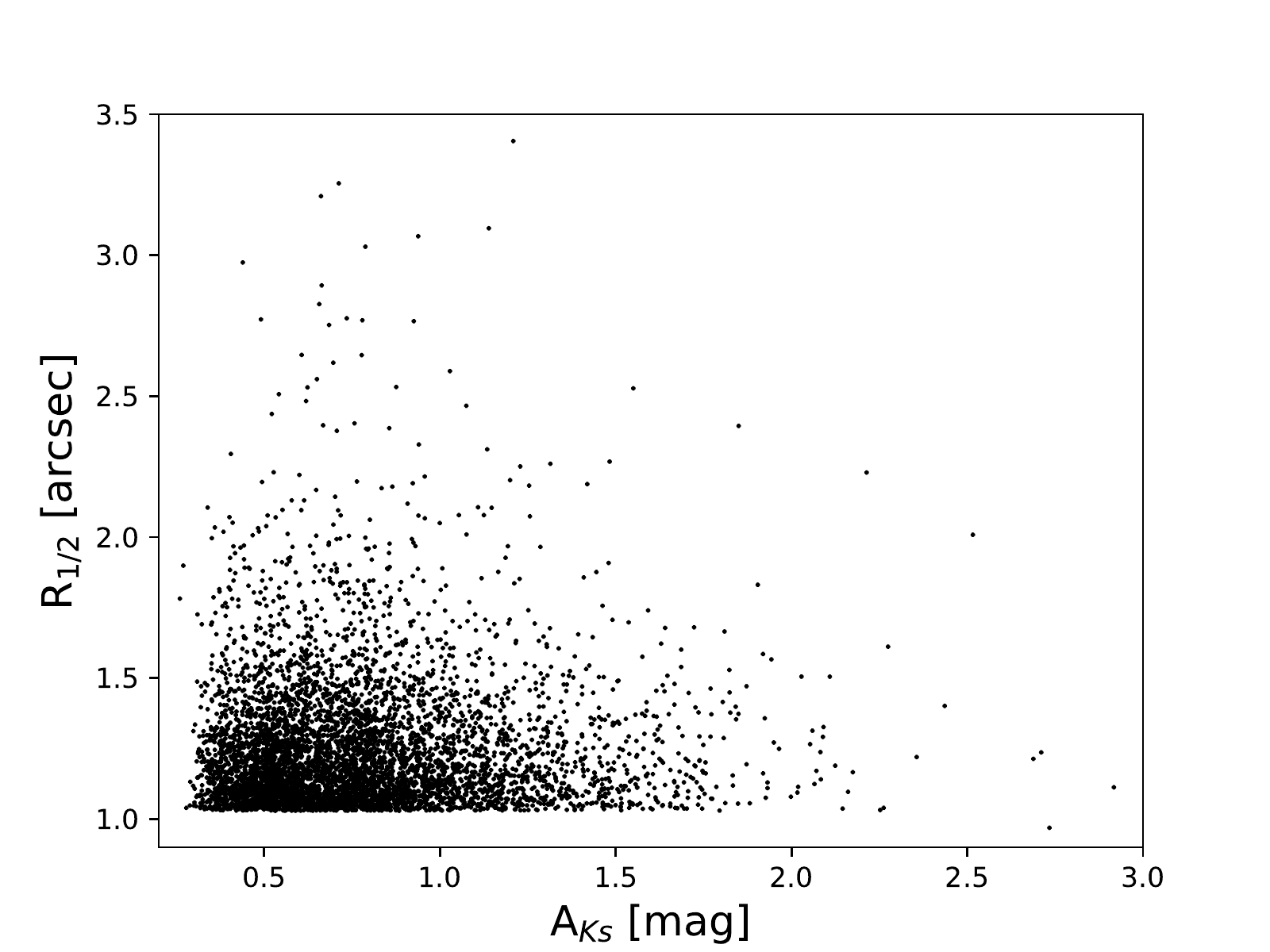}
\includegraphics[width=0.45\textwidth,height=0.40\textwidth]{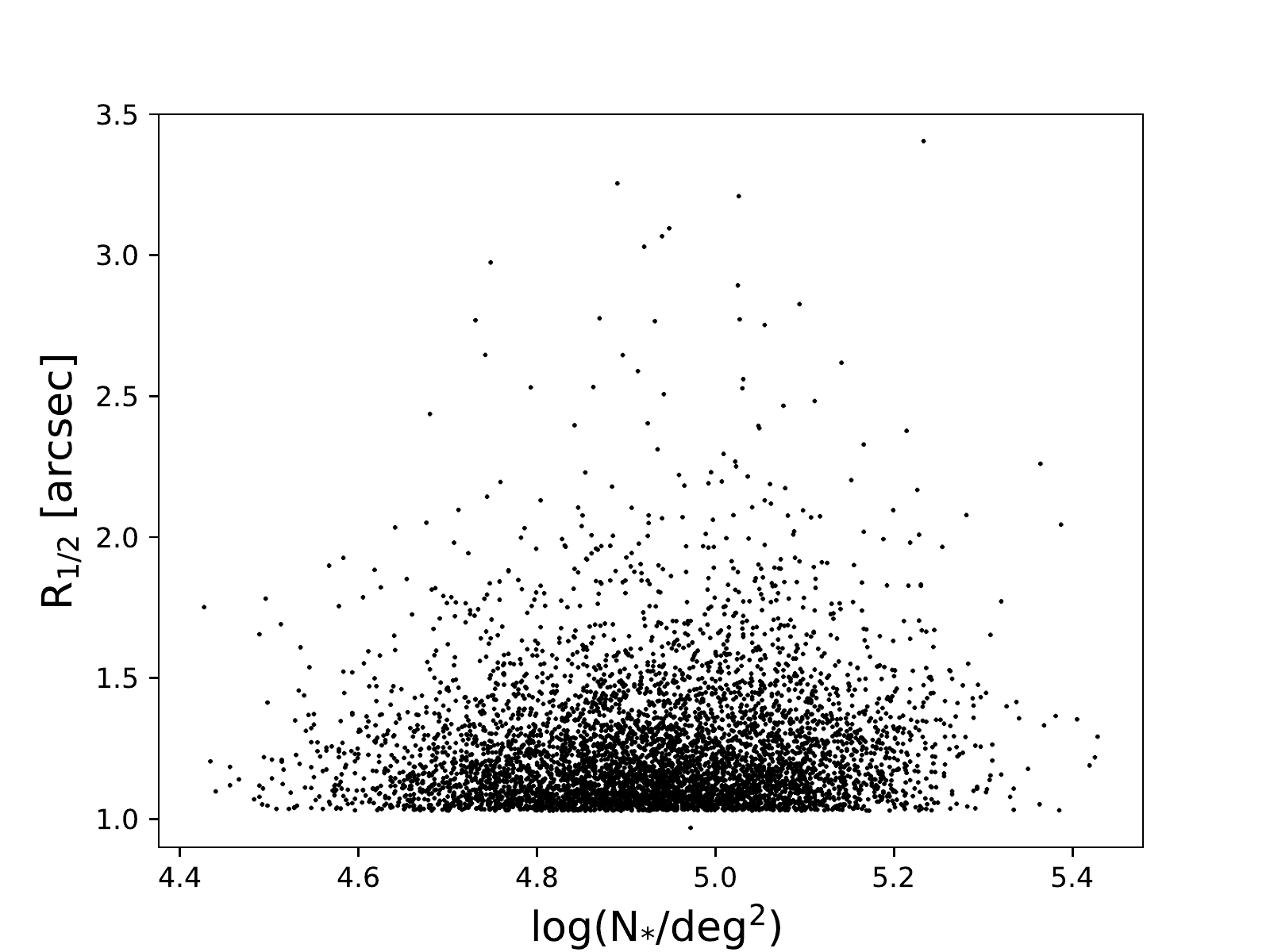}
\caption{Upper panels show the extinction-corrected  $K_{s}^{\circ}$  magnitudes 
as a function of their A$_{Ks}$ interstellar extinctions (left) and stellar density (right).  Bottom panels show the size R$_{1/2}$ as a function of these parameters.  
}\label{relations}
\end{figure*}

For the detected galaxies, Figure~\ref{relations} shows the $K_{s}^{\circ}$ magnitudes and the 
half-light radius as a function of the A$_{Ks}$ interstellar extinctions and stellar densities.  
The estimated stellar density was defined here as the number of stars with $K_{s}^{\circ} < $15 mag in an area of $\sim$ 43 arcmin$^2$ around each galaxy scaled to an area of 1 deg$^2$.  The median value is log(N/deg$^2$) = 4.933 $\pm$ 0.154, which as expected was significantly larger than the values reported in surveys based on the 2MASS data that cover higher Galactic latitudes. 
It is clear that we reached $K_{s}^{\circ}$  magnitudes of
about 15.5 mag in regions with lower interstellar extinctions and stellar densities. In all, 96\% of the detections had an $R_{1/2}$ value of about 1.18 arcsec for A$_{Ks}$ interstellar extinctions smaller than 1.5 mag, which then reached median values of 1.22 arcsec.  
We saw a tendency for brighter galaxies to be observed in regions which displayed higher stellar densities, while at the same time, the size of the detected galaxies did not reveal a clear correlation.   
In \cite{Baravalle2018}, we analysed the completeness of the $K_{s}$ photometry for the two studied tiles: d010 and d115.  At $K_{s}^{\circ}$ = 15.5 mag, we attained a completeness of 80 and 95\% (see their Figure 3) for these tiles, respectively. Comparing the extinctions, the d115 tile had the lowest A$_{Ks}$ values. A conservative completeness of 80\% might be estimated at these faint magnitude levels.  These results show that both interstellar extinction and stellar density are the main limitations and critical concerns for the detection of galaxies at these low Galactic latitudes.   

\subsection{Comparison with NIR surveys}
\label{subsec:2masx}

Over the VVV disk region, 185 extended sources from the 2MASX were recovered and visually checked. These comprise 21 galaxies and 164 Galactic objects that are: bright single and double stars or stellar associations and gas and dust regions.  
In Appendix B, we describe these 2MASX sources by dividing them into these categories. We found that about 16\% of the Galactic objects have no previous description in the literature.  
Our galaxies were cross-matched with the 21 galaxies detected by 2MASX using angular separations of about 2 arcsec. Some examples of the common galaxies between the two surveys are shown in Figure~\ref{galaxies}.
The median astrometric differences after 1$\sigma$--clipping between the positions of the VVV galaxies and 2MASX are very small: 
$\Delta$RA cos(Dec) = (0.087 $\pm$ 0.333) arcsec and
$\Delta$Dec = (0.036 $\pm$ 0.372) arcsec. 
The galaxies with higher differences are those with a strong contamination by nearby stars. 

In this VVV region, there are 22 galaxies reported from \cite{Schroder2007}, 41 from \cite{Williams2014}, 46 from \cite{Said2016} and 19 from \cite{Schroder2019a}. Of these galaxies, we have 12 galaxies in common
with \cite{Schroder2007}, 20 with \cite{Williams2014}, 22 with \cite{Said2016} and 19 with \cite{Schroder2019a}. In total, after the cross-match, we have 45 galaxies in common with other authors (11 with only one, 25 with two, 5 with three and 4 with four authors). 
We visually checked all the objects, including those that were not detected with our procedure (which are diffuse objects, faint galaxies, objects with strong contamination by bright stars or not included in our colour criteria). 

Our magnitudes were only transformed to the 2MASS photometric system (\citealt{Soto2013}) and compared with 2MASS, 2MASX, \cite{Schroder2007} and \cite{Said2016}. 
The galaxies reported by \cite{Schroder2019a} have 2MASX photometry,  and those of \cite{Williams2014} are included in \cite{Said2016}. From these, we used 2MASS aperture magnitudes at a fixed radius aperture of 4 arcsec (2MASS PSC), the isophotal magnitudes within the $K_{s}$ = 20 mag / arcsec$^2$ ($K_{20}$) fiducial elliptical aperture from 2MASX and  \cite{Said2016}, and MAG\_AUTO magnitudes from \cite{Schroder2007}.  

\begin{figure*}
\centering
\includegraphics[width=0.42\textwidth,height=0.34\textwidth]{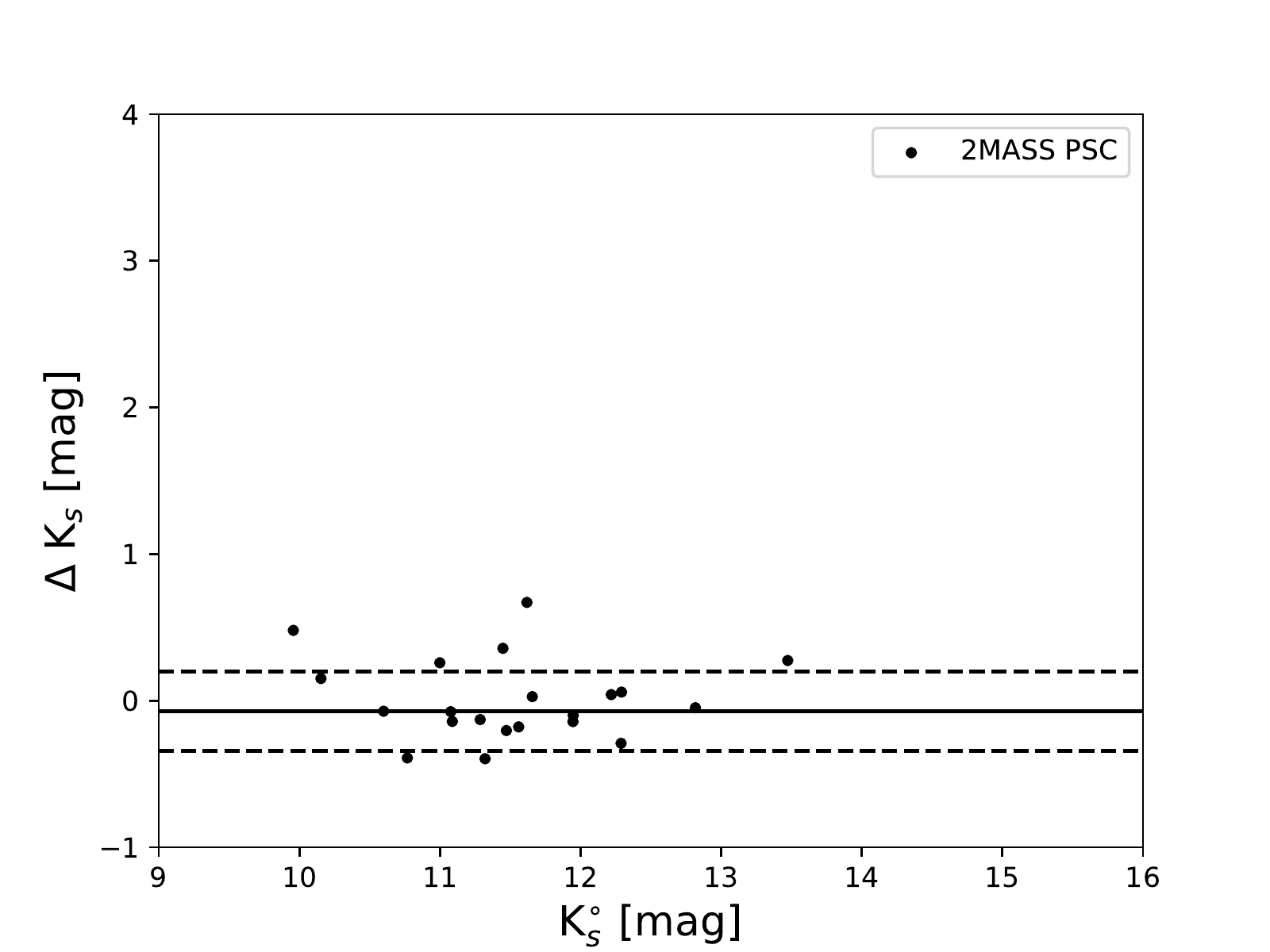}
\includegraphics[width=0.42\textwidth,height=0.34\textwidth]{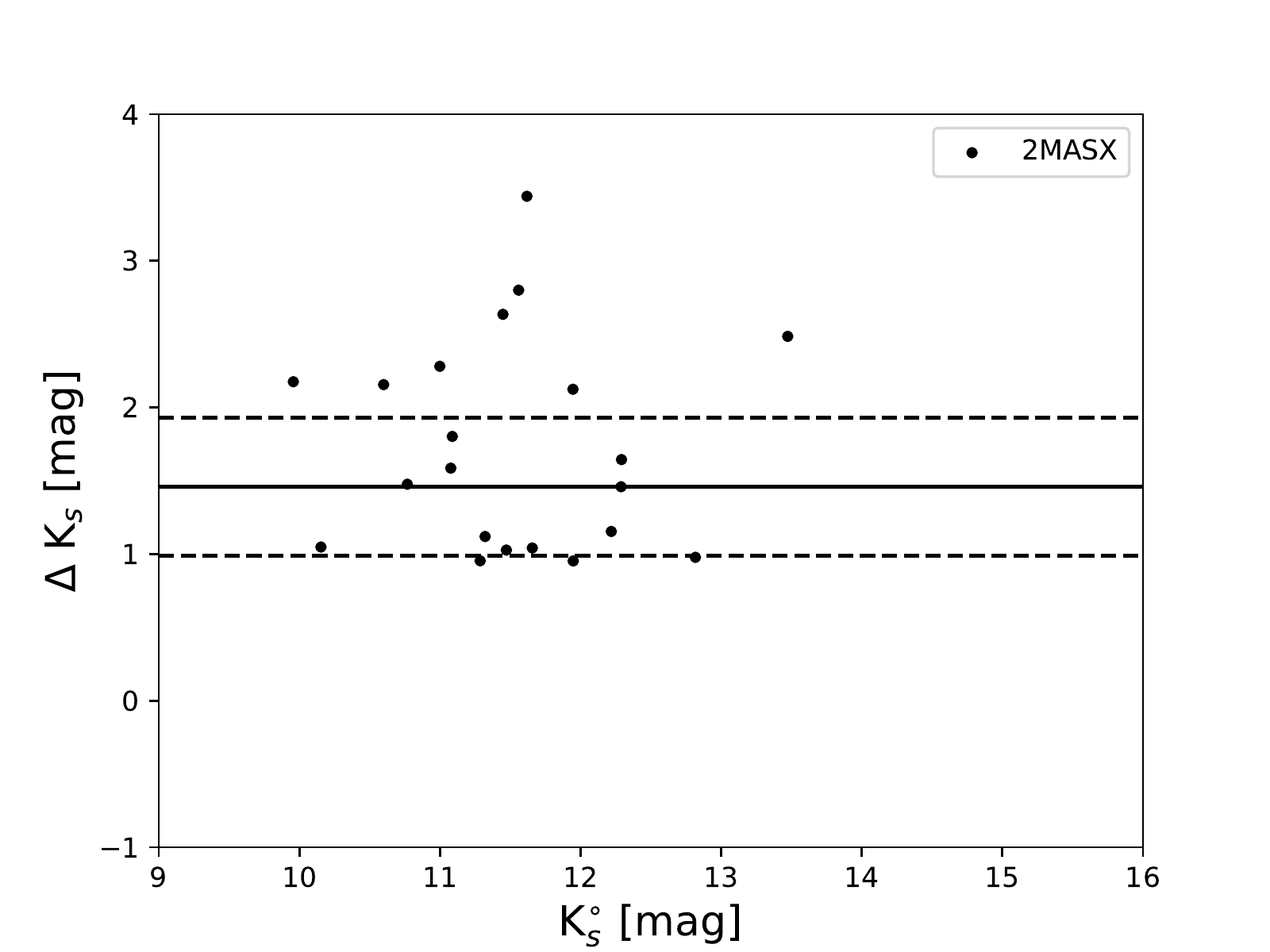}
\includegraphics[width=0.42\textwidth,height=0.34\textwidth]{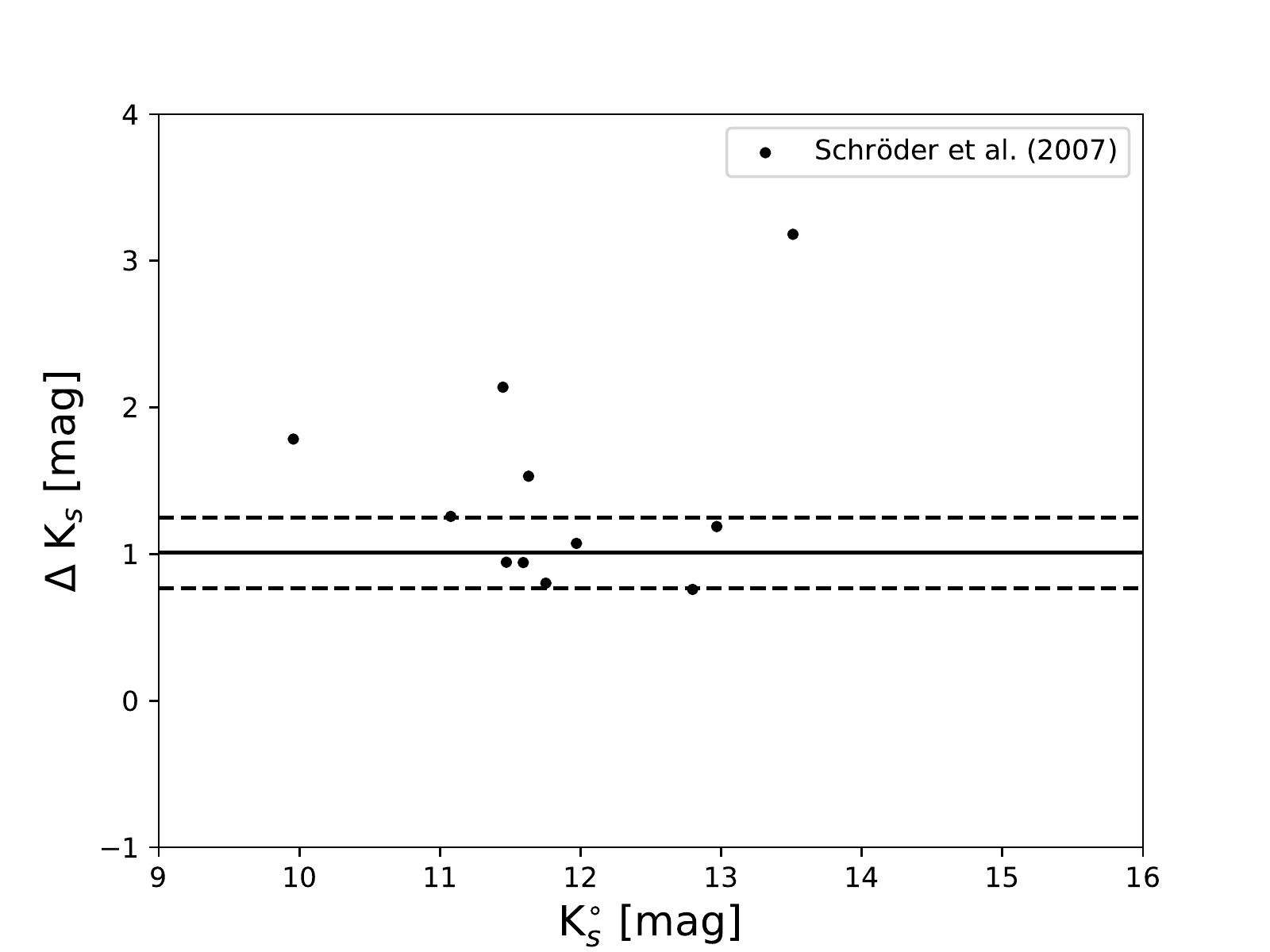}
\includegraphics[width=0.42\textwidth,height=0.34\textwidth]{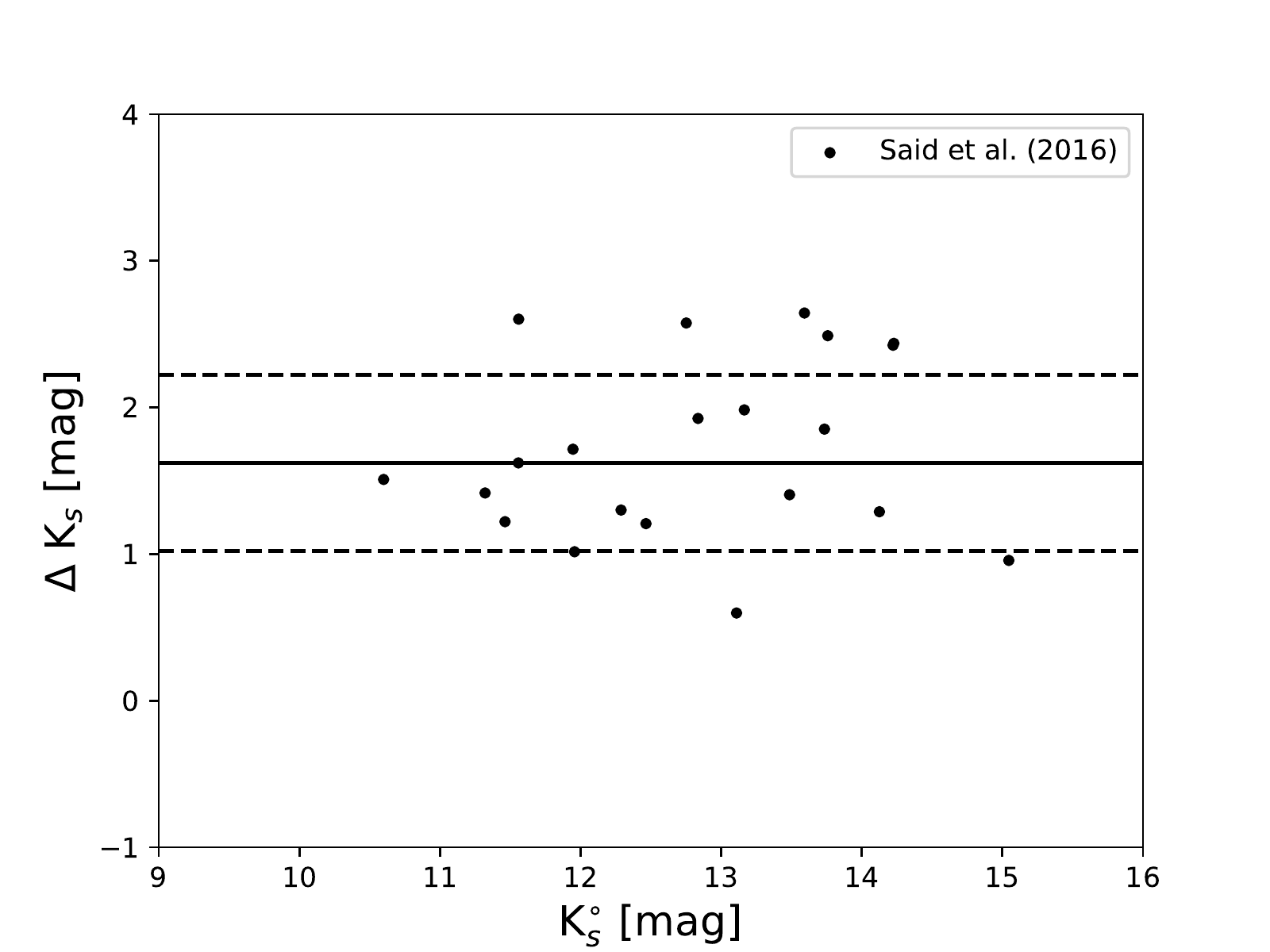}
\caption{Magnitude comparisons in the $K_{s}$ passband for galaxies in common for our  magnitudes and 2MASS (PSC), 2MASX ($K_{20}$), \citet{Schroder2007} (MAG\_AUTO) and \citet{Said2016} ($K_{20}$). Solid lines represent the median differences quoted in Table~\ref{table2}, and dashed lines show the 1$\sigma$ dispersion. 
}
\label{deltaM}
\end{figure*}

The magnitude differences were calculated from our total and literature magnitudes
($\Delta$m = m(VVV) - m(literature)) for the $J$, $H$ and $K_{s}$ passbands, and the statistical differences are the result of 2$\sigma$--clipping. Table~\ref{table2} shows the median magnitude differences in the $J$, $H$ and $K_{s}$ passbands, and includes the final number of objects considered after the $\sigma$--clipping in parenthesis. The number of compared galaxies for $K_{s}$ is higher than those for $J$ and $H$. 
In some cases, the authors did not present estimates for the magnitudes in these passbands or the differences are higher than the 2$\sigma$--clipping.  Figure~\ref{deltaM} shows only the comparisons for the $K_{s}^{\circ}$ magnitudes, as it is the most reliable passband.  In general, the uncertainties are important and might be a result of stellar contamination.  The best comparisons are with 2MASS magnitudes, and we conclude that our total magnitude estimates are comparable with those that are  aperture magnitudes within a fixed aperture of 4 arcsec radius.  The comparison with the isophotal $K_{s}$ magnitudes yielded offsets higher than one magnitude in all cases.  We attribute these observed offsets to the different photometric methods, the different procedures used to define the centre positions and magnitudes and the stellar contamination in these dense regions with high interstellar extinctions. 
 
\begin{table*}
\center
\caption{Median   $J$,  $H$ and $K_{s}$ magnitude differences.}
\begin{tabular}{|lccc|}
\hline 
Literature          & $\Delta$ $J$       &  $\Delta$ $H$ & $\Delta$ $K_{s}$ \\
                    & [mag] & [mag] & [mag] \\
\hline 
2MASS PSC          &  -0.20 $\pm$ 0.52 (18) & -0.19 $\pm$ 0.14 (15) & -0.09 $\pm$ 0.11 (14)\\ 
$K_{20}$ (2MASX)    &  2.13  $\pm$ 0.64 (4) & 1.44  $\pm$ 0.64 (16) &  1.46 $\pm$ 0.47 (17)\\ 
MAG\_AUTO (\citealt{Schroder2007}) &  1.24  $\pm$ 0.18 (5)  &  --                   &  1.01 $\pm$ 0.24 (8)\\
$K_{20}$ (\citealt{Said2016})  &  1.82  $\pm$ 0.40 (15) & 1.53  $\pm$ 0.31 (13) &  1.62 $\pm$ 0.60 (21)\\
\hline 
\end{tabular}
\label{table2}
\end{table*}

\subsection{Correlation with MIR data}
\label{subsec:wise}

The Wide-field Infrared Survey Explorer mission (WISE; \citealt{Wright2010}) mapped the whole sky in the MIR (3.4, 4.6, 12, and 22 $\mu$m) passbands.  
For red sources such as galaxies with older stellar populations, WISE goes one magnitude deeper than the 2MASS $K_{s}$ magnitude for the 3.4 $\mu$m passband (\citealt{Wright2010}).  WISE also detects the most luminous galaxies in the Universe, such as the  Ultra-luminous IR  galaxies which  might be related to mergers  that  lead to  dust-enshrouded star formation (\citealt{Shanshan2013}). In addition,  they might be related to  active galactic nuclei (AGN) activity, such as gas disturbed from stable circular orbits and falling into the central super-massive black hole in merging galaxies (\citealt{Massaro2012}). 

In the regions of the VVV disk, we have found 509 galaxies in common with the WISE survey using a cross-match with angular separations of 2 arcsec, implying that 10\% of our galaxies have also been seen in the MIR regime by WISE.

\section{The VVV NIR Galaxy Catalogue }
 
We created the VVV NIR Galaxy Catalogue (VVV NIRGC), which contains 5563 galaxies across an area of 220 deg$^{2}$ within 295$^{\circ}$ $<$ $l$ $<$ 350$^{\circ}$,
-2.25$^{\circ}$ $< b <$ +2.25$^{\circ}$, using the procedure detailed in the previous section. In this region, we have only a total of 45 galaxies in common with other NIR surveys, such as the 2MASX, \citet{Schroder2007}, \citet{Williams2014} and \citet{Said2016} catalogues. Consequently, in VVV NIRGC, about 99\% of the galaxies are new discoveries. 

Table~\ref{catalogue} shows the first ten galaxies of VVV NIRGC, listing the identification in column (1), the J2000 equatorial coordinates in columns (2) and (3), the Galactic coordinates in columns (4) and (5), the A$_{Ks}$ interstellar extinction in column (6), total extinction-corrected $J^{\circ}$, $H^{\circ}$, and $K_{s}^{\circ}$ magnitudes in columns (7) to (9), the extinction-corrected 
$J^{\circ}$, $H^{\circ}$, and $K_{s}^{\circ}$ aperture magnitudes within a fixed aperture of 2 arcsec diameter in columns (10) to (12), the morphological parameters: $R_{1/2}$, $C$, ellipticity and $n$ in columns (13) to (16) and the WISE flag in column (17). 
If the galaxy has a WISE counterpart, this flag is set to 1, otherwise 
it is set to 0. All magnitudes were extinction-corrected and transformed to the 2MASS photometric system. This table is available in its entirety in a machine-readable form\footnote{https://catalogs.oac.uncor.edu/vvv\_nirgc/}.

 \begin{table*}
  \caption{The VVV NIRGC.}
  \label{catalogue}
  \begin{turn}{90}
  \begin{tabular}{ccccccccccccccccc}
    \hline
     \hline
    ID & RA  & Dec & l & b & A$_{Ks}$ & $J^{\circ}$ & $H^{\circ}$ & $K_{s}^{\circ}$ & $J_{2\arcsec}^{\circ}$ & $H_{2\arcsec}^{\circ}$ & $K_s$ $_{2\arcsec}^{\circ}$ & $R_{1/2}$ & $C$ & $\epsilon$ & $n$  & WISE \\ 
     & (J2000) &  (J2000) & [deg] & [deg] & [mag] & [mag] & [mag] & [mag] & [mag] & [mag] & [mag] & [arcsec] & &  &  & flag\\ 
    \hline
    VVV-J113548.43-635434.3 & 11:35:48.43 & -63:54:34.3 &  294.7325 & -2.2484 & 0.82 & 15.11 & 14.69 & 14.36 & 15.89 & 15.42 &  15.26  &  1.18 & 2.64 & 0.31 & 1.19 & 0\\ 
    VVV-J113550.51-634933.7 & 11:35:50.51 & -63:49:33.7 & 294.7119 & -2.1673 & 0.91 & 15.28 & 14.42 & 13.90 & 16.47 & 15.51 &  15.10  &  1.46 & 2.81 & 0.57 & 2.48 & 0\\
    VVV-J113550.71-635444.0 & 11:35:50.71 & -63:54:44.0 & 294.7371 & -2.2497 & 0.82 & 16.68 & 16.08 & 15.29 & 16.95 & 16.45 &  16.17  &  1.12 & 2.12 & 0.46 & 3.50 & 0\\
    VVV-J113636.55-635556.4 & 11:36:36.55 & -63:55:56.4 & 294.8234 & -2.2449 & 0.88 & 16.19 & 15.43 & 14.98 & 16.63 & 16.02 &  15.83  &  1.12 & 2.36 & 0.47 & 2.28 & 0\\ 
    VVV-J113722.99-635650.0 & 11:37:22.99 & -63:56:50.0 & 294.9092 & -2.2350 & 0.90 & 15.59 & 15.20 & 15.10 & 16.52 & 16.08 &  15.93  &  1.14 & 3.15 & 0.60 & 7.62 & 0\\
    VVV-J113851.89-635321.8 & 11:38:51.89 & -63:53:21.8 & 295.0494 & -2.1339 & 1.47 & 14.69 & 14.03 & 13.86 & 15.76 & 15.18 &  14.94  &  1.25 & 2.14 & 0.20 & 0.42 & 0 \\
    VVV-J113905.29-635803.6 & 11:39:05.29 & -63:58:03.6 & 295.0947 & -2.2023 & 0.82 & 15.77 & 15.00 & 14.71 & 16.22 & 15.70 &  15.56  &  1.12 & 2.55 & 0.38 & 1.64 & 0\\
    VVV-J113926.76-635711.6 & 11:39:26.76 & -63:57:11.6 & 295.1285 & -2.1776 & 0.82 & 15.43 & 14.90 & 14.59 & 16.07 & 15.58 &  15.36  &  1.05 & 2.32 & 0.35 & 1.52 & 0 \\
    VVV-J113946.21-640120.3 & 11:39:46.21 & -64:01:20.3 & 295.1817 & -2.2342 & 0.59 & 16.38 & 15.78 & 15.33 & 16.78 & 16.29 &  16.14  &  1.11 & 2.10 & 0.10 & 3.82 & 0 \\
    VVV-J114000.97-625333.3 & 11:40:00.97 & -62:53:33.3 & 294.8991 & -1.1400 & 0.91 & 14.65 & 14.10 & 14.08 & 15.53 & 14.85 &  14.85  &  1.04 & 2.30 & 0.41 & 3.03 & 0\\
     \hline
    \hline
  \end{tabular}
  \end{turn}
 \end{table*}

\begin{figure*}
\centering
\includegraphics[width=0.33\textwidth,height=0.32\textwidth]{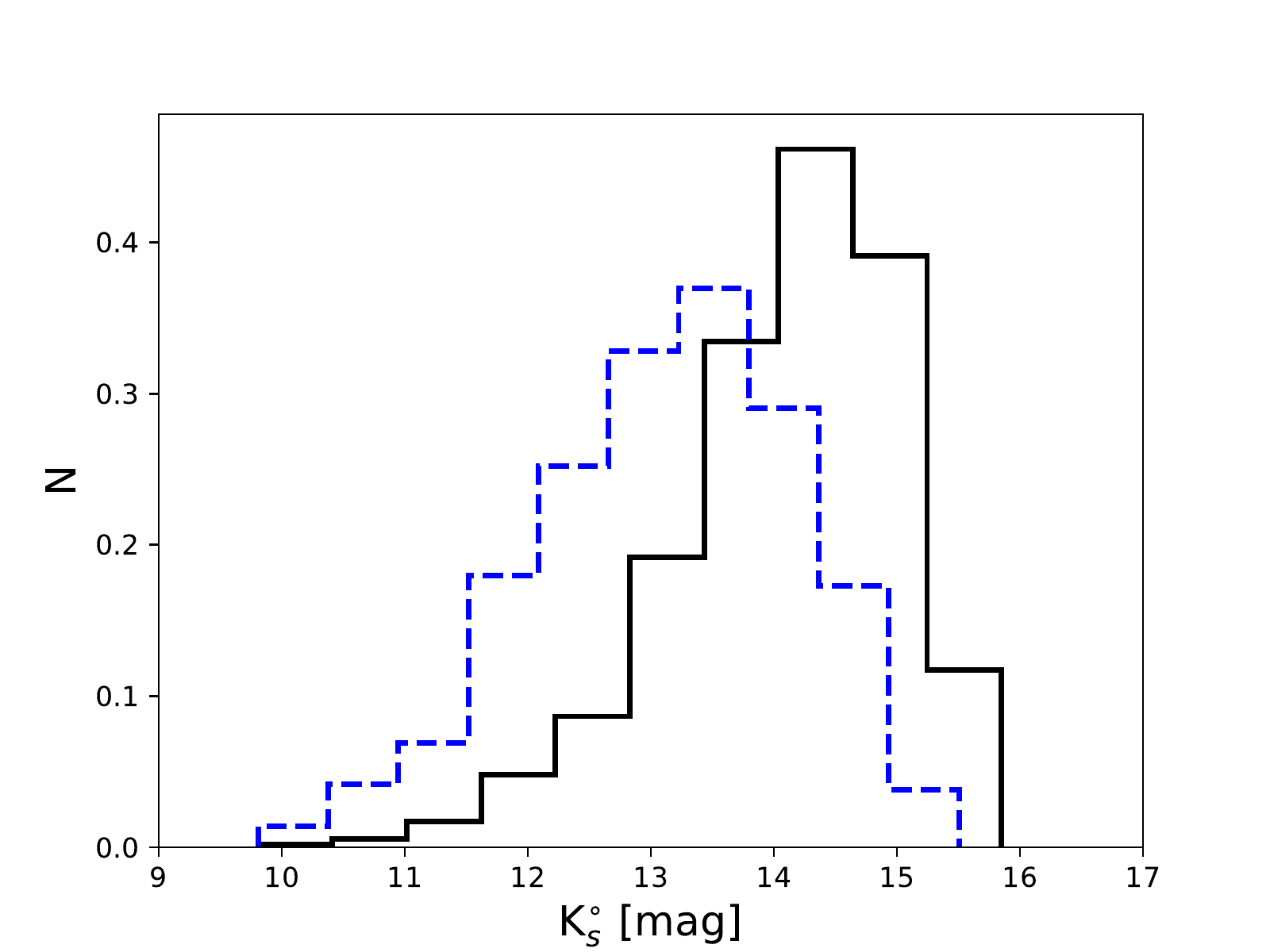}
\includegraphics[width=0.33\textwidth,height=0.32\textwidth]{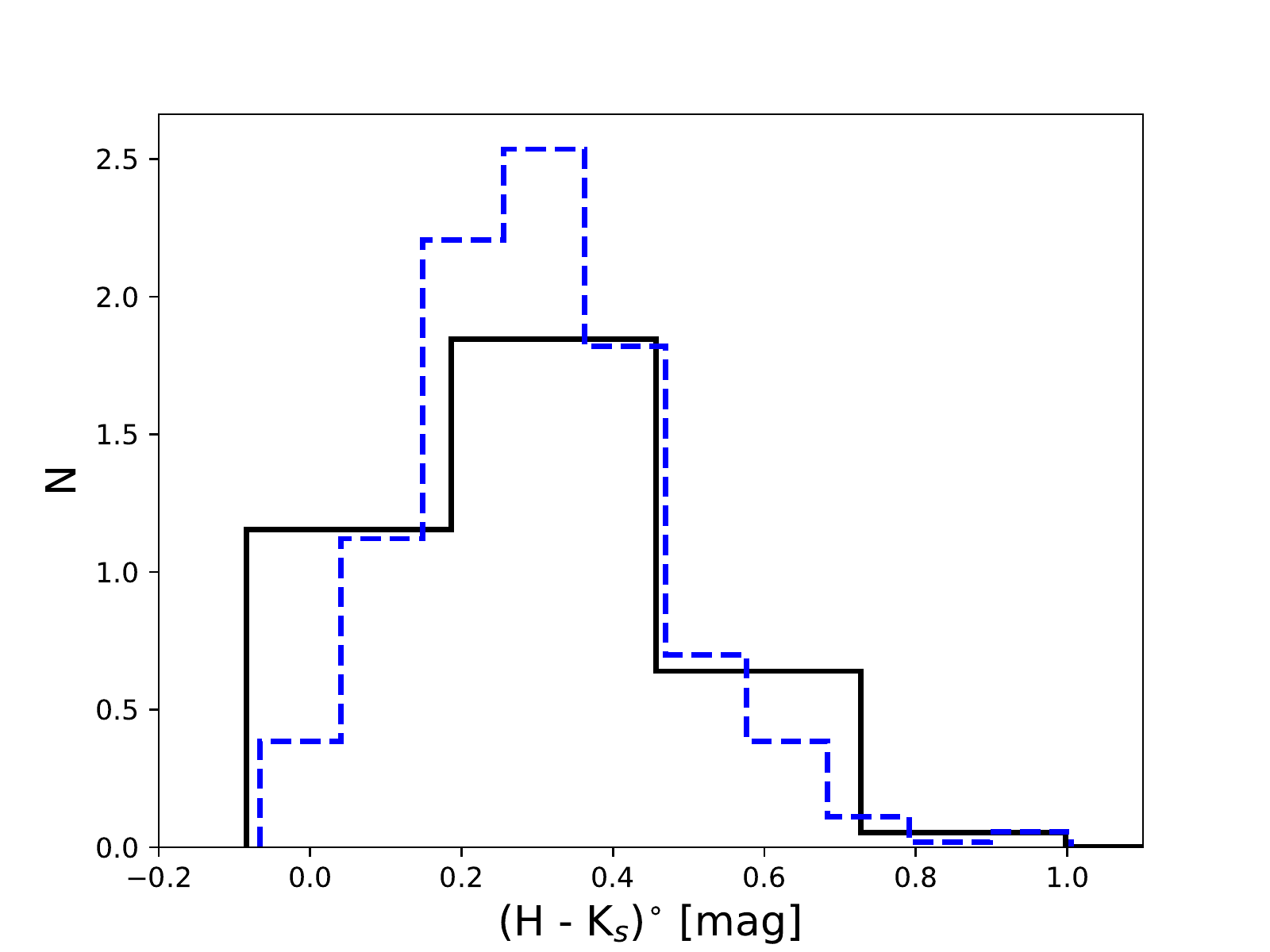}
\includegraphics[width=0.33\textwidth,height=0.32\textwidth]{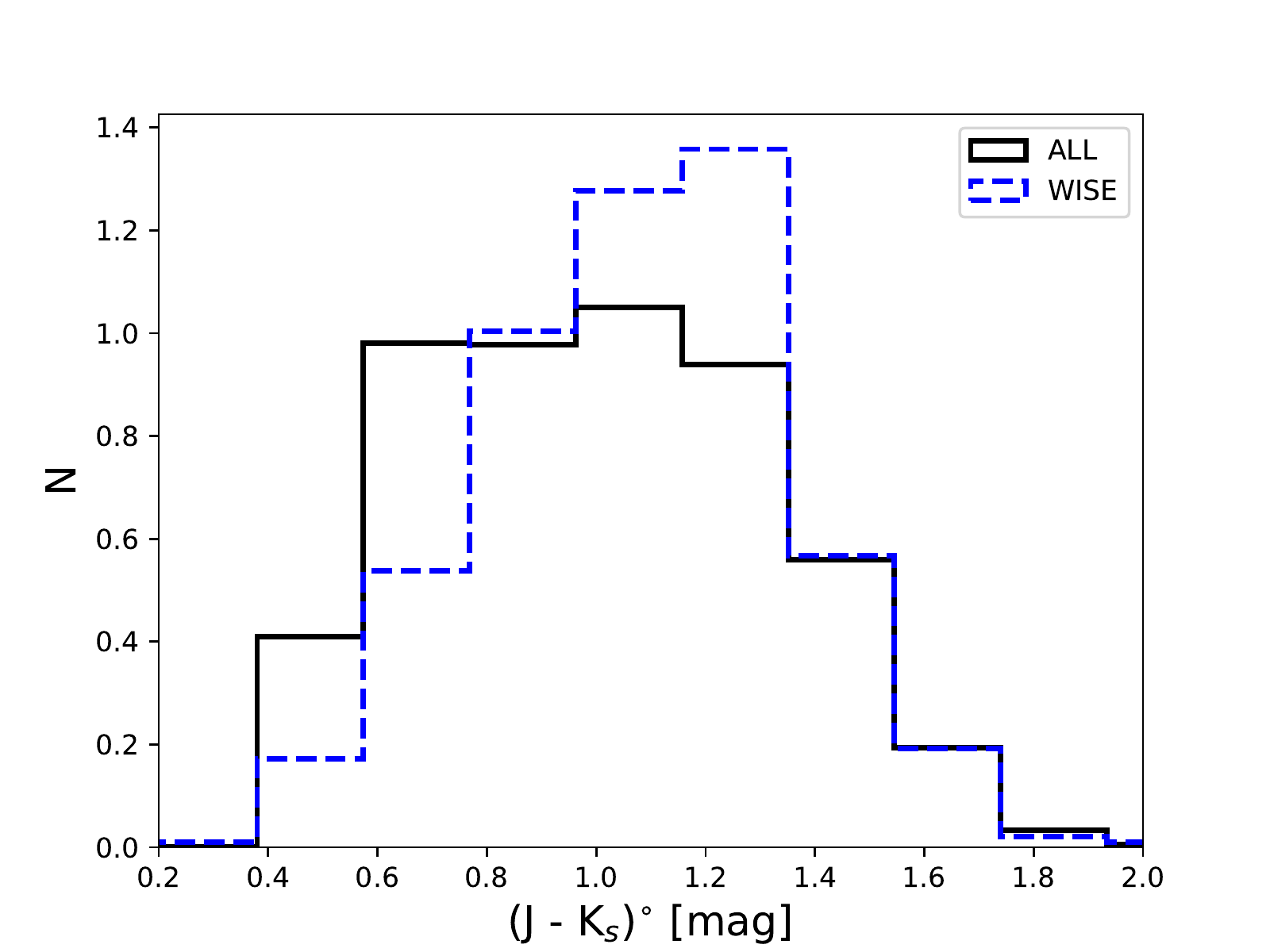}
\caption{Normalized distributions of the photometric parameters  of the galaxies in the VVV NIRGC. The panels show, from left to right, the histograms for $K_{s}^{\circ}$ magnitudes and the ($H$ - $K_{s}$)$^{\circ}$ and ($J$ - $K_{s}$)$^{\circ}$ NIR colours. All the galaxies in the catalogue and those also with WISE data are shown in solid and dashed histograms, respectively. 
}\label{hist2}
\end{figure*}

\begin{figure*}
\centering
\includegraphics[width=0.33\textwidth,height=0.32\textwidth]{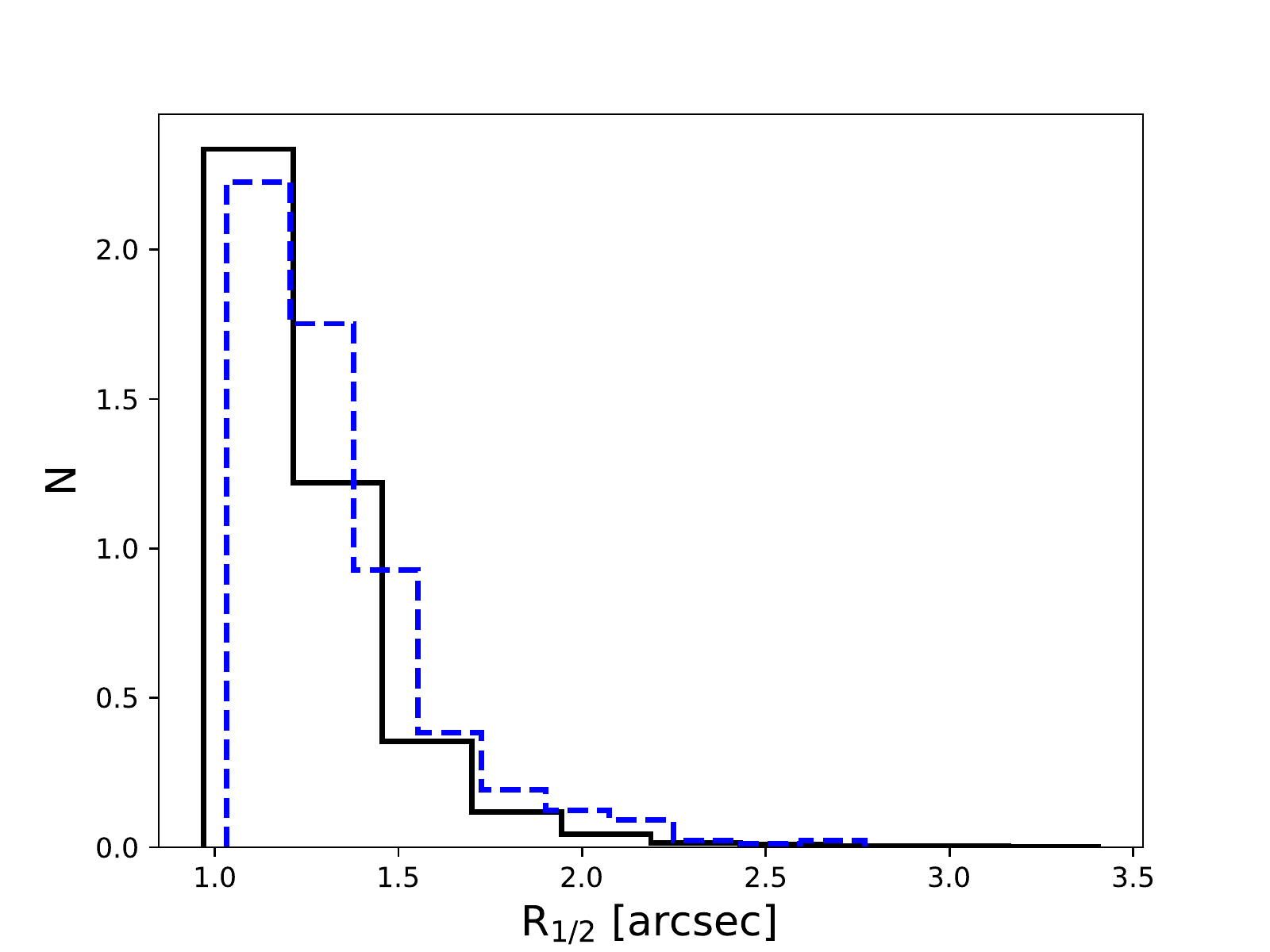}
\includegraphics[width=0.33\textwidth,height=0.32\textwidth]{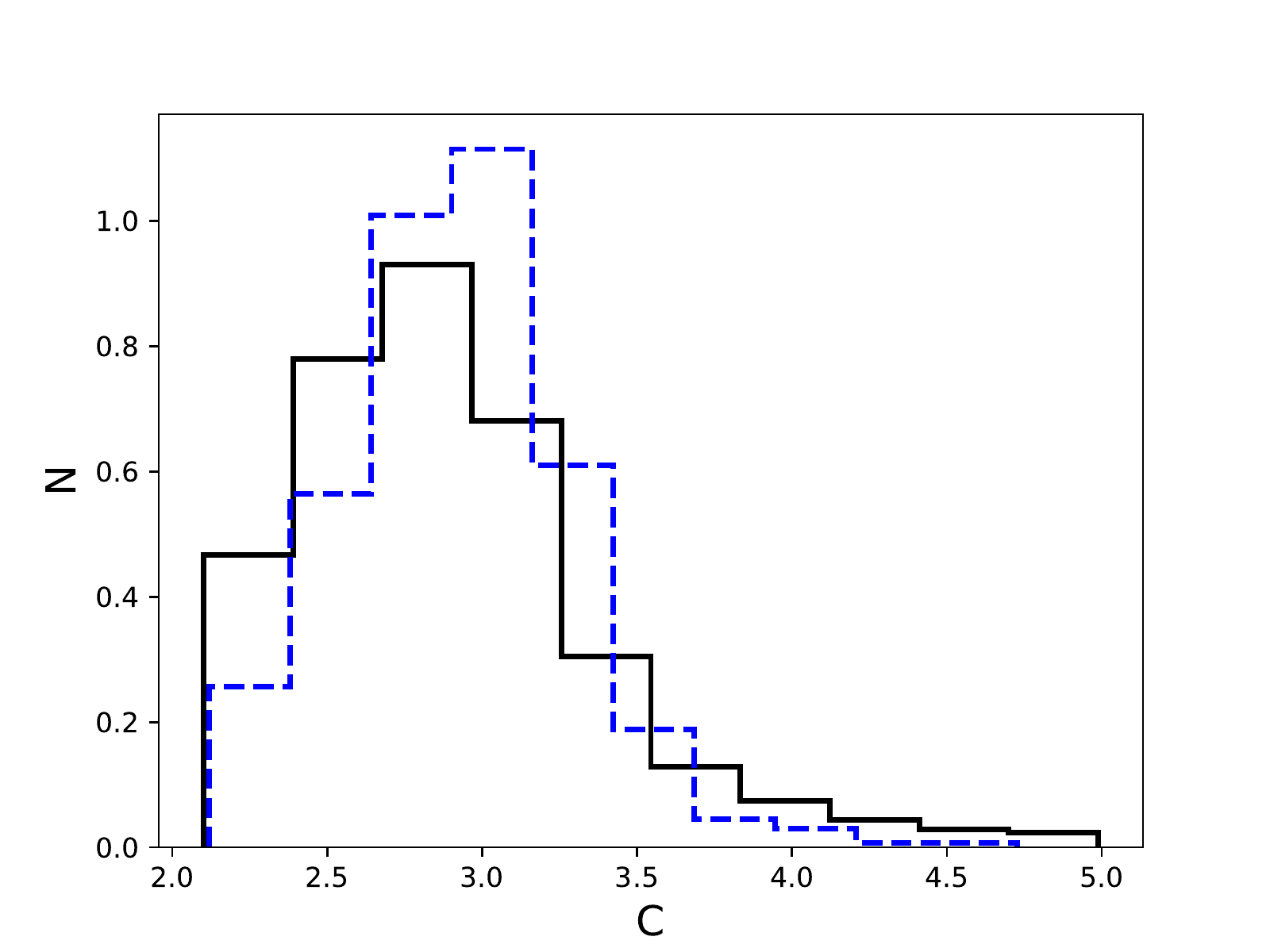}
\includegraphics[width=0.33\textwidth,height=0.32\textwidth]{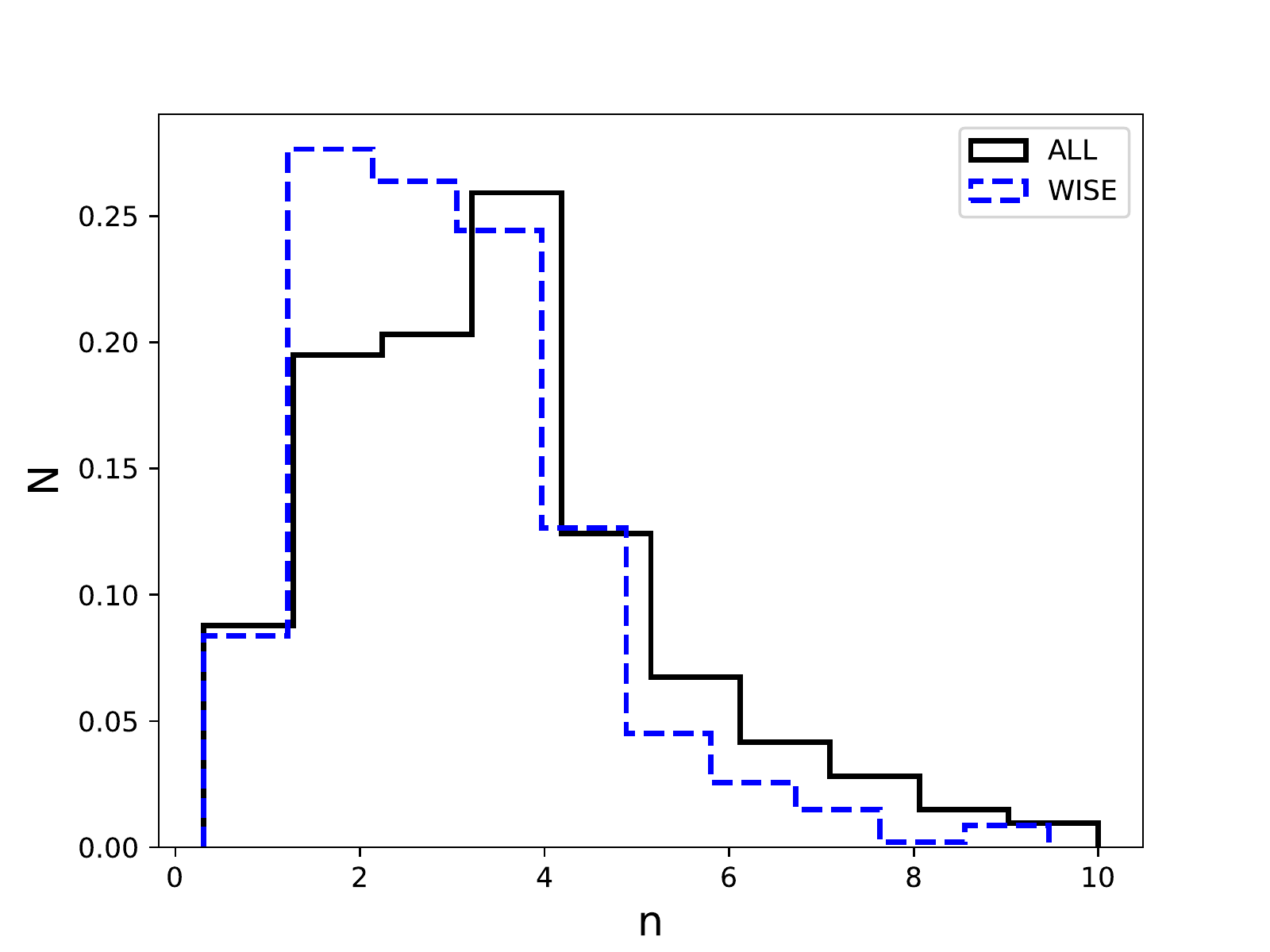}
\caption{Normalized distributions of morphological parameters of the galaxies in the VVV NIRGC.  The panels show, from left to right, the histograms of half-light radius $R_{1/2}$, the concentration index C and the S\'ersic index n.  All the galaxies in the catalogue and those also with WISE data are shown in solid and dashed histograms, respectively. 
}\label{hist1}
\end{figure*}

Figure~\ref{hist2} shows the normalized distributions of the reddening corrected $K_{s}^{\circ}$ magnitudes and the ($H$ - $K_{s}$)$^{\circ}$ and 
($J$ - $K_{s}$)$^{\circ}$ colours of the galaxies in the VVV NIRGC.  Figure~\ref{hist1} also shows the normalized distributions of some of the structural properties for the galaxies, such as  $R_{1/2}$, $C$, and S\'ersic index. We made a distinction between galaxies from the VVV NIRGC (represented with solid lines) and those that also have WISE data (represented with dashed lines) see section~\ref{subsec:wise}.  Table~\ref{tablewise} summarizes the median values of the photometric and structural parameters for the two normalized distributions. For the galaxies in VVV NIRGC, the limiting $K_{s}^{\circ}$ magnitude reaches about 16 mag and the distribution of 
($J$ - $K_{s}$)$^{\circ}$ is nearly constant between 0.5 and 1.2 mag.  These are small
objects with distributions that have peaks with $R_{1/2} \sim $ 1.3 arcsec, $C$ with a value of approximately 3, and $n$ around 4. These results are similar to those found in the tiles analysed in \cite{Baravalle2018, Baravalle2019}, which are characteristic of early-type galaxies. The galaxies with WISE data are brighter and with slightly 
larger sizes with a smaller S\'ersic index than the galaxies, in general, in the VVV NIRGC. These results might be a consequence of the predominance of late-type galaxies.

\begin{table*}
\center
\caption{Median photometric and structural parameters of the
galaxies in the VVV NIRGC. }
\begin{tabular}{|lcc|}
\hline 
\hline 
  Parameter & All galaxies  &  Galaxies with WISE data\\
\hline 
  $K_{s}^{\circ}$ [mag]          & 14.24 $\pm$ 0.92 &  13.22 $\pm$ 1.06\\
  ($J$ - $K_{s}$)$^{\circ}$ [mag]  & 0.99  $\pm$ 0.31 &  1.09  $\pm$ 0.28\\
  ($H$ - $K_{s}$)$^{\circ}$ [mag]  & 0.28  $\pm$ 0.19 &  0.30  $\pm$ 0.16\\
  $R_{1/2}$ [arcsec]     & 1.18  $\pm$ 0.23 &  1.27  $\pm$ 0.27\\
  $C$                    & 2.83  $\pm$ 0.49 &  2.92  $\pm$ 0.36\\
  $n$                    & 3.34  $\pm$ 1.81 &  2.79  $\pm$ 1.47\\
  \hline 
\hline 
\end{tabular}
\label{tablewise}
\end{table*}

\subsection {The galaxy distribution map}

\begin{figure*}
\centering
\includegraphics[width=1\textwidth]{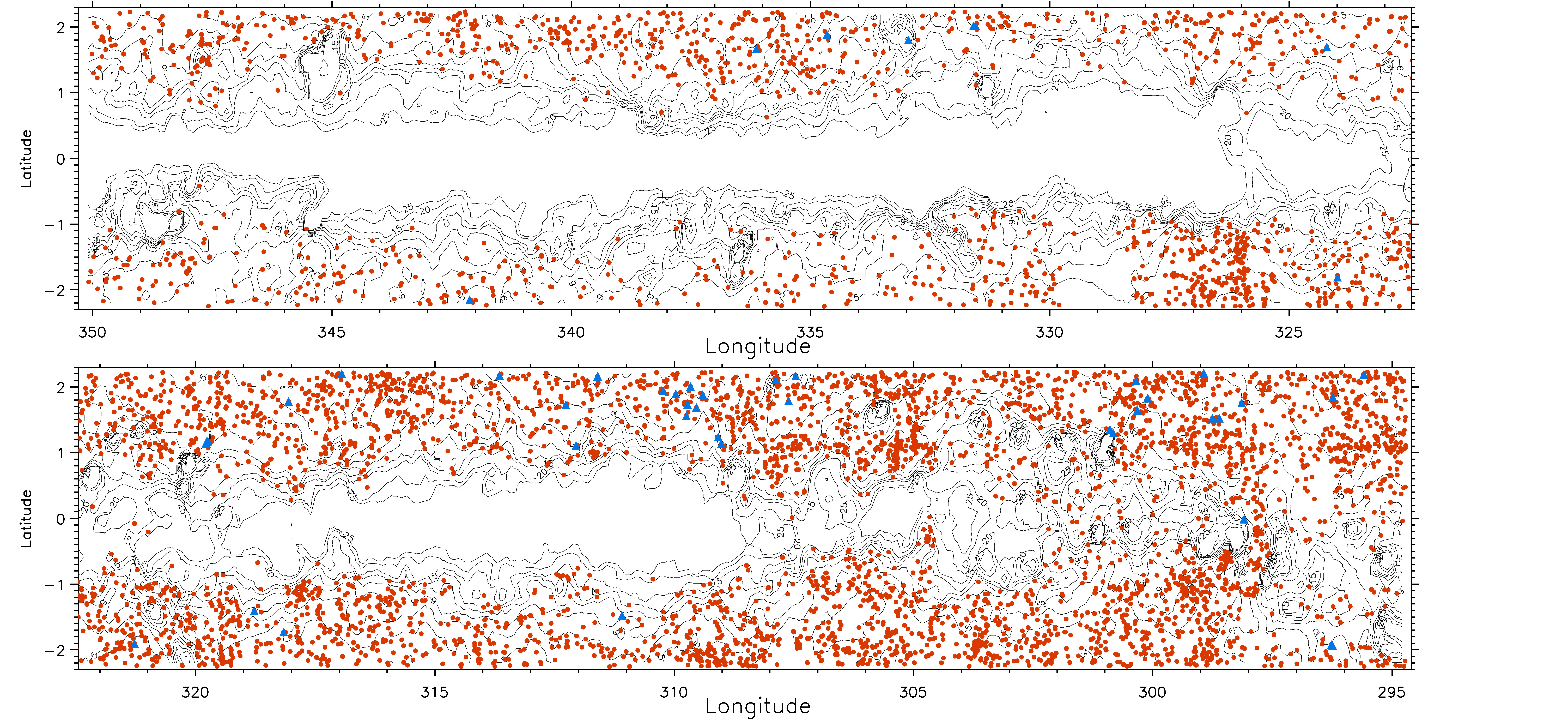}
\caption{Distribution of galaxies in the VVV disk region superimposed with the A$_V$ contours (1, 3, 5, 7, 9, 11, 13, 15, 20 and 25 mag) derived from the extinction map by \citet{Schlafly2011}.  The
galaxies from the VVV NIRGC catalogue are represented with 
red circles, and those in common with other surveys, such as 2MASX, \citet{Schroder2007}, \citet{Williams2014}, \citet{Said2016} and \citet{Schroder2019a}, are shown with blue triangles. For a better visualization, the stellar density gray-scale map was constructed by including all VVV stellar detections.  
}\label{distribution1}
\end{figure*}

\begin{figure*}
\centering
\includegraphics[width=1\textwidth]{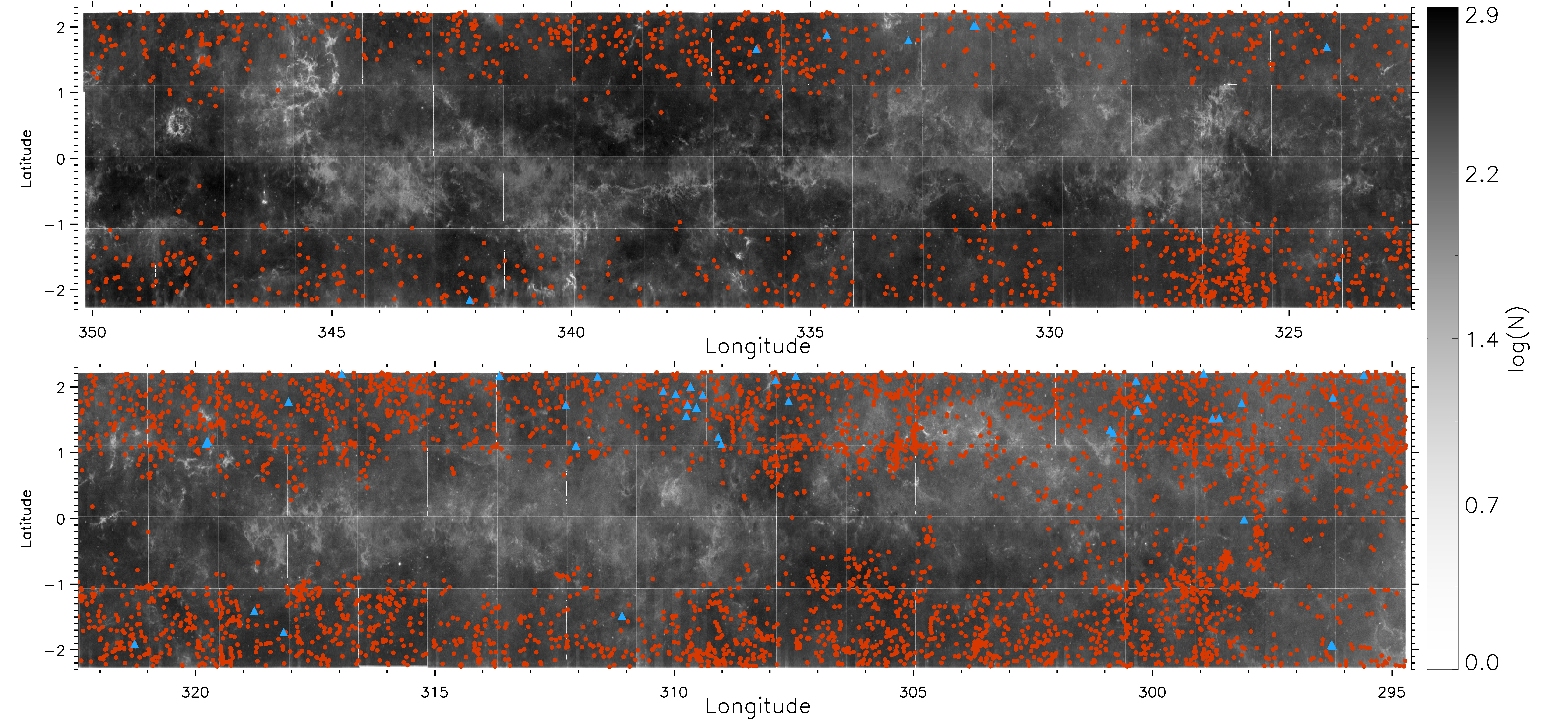}
\caption{Distribution of galaxies in the VVV disk region superimposed with the stellar density gray-scale map from the catalogues by \citet{Alonso2018}. As in previous figure, the 
galaxies from the VVV NIRGC catalogue are represented with 
red circles and those from other surveys with blue triangles.   
 }\label{distribution2}
\end{figure*}

Figures~\ref{distribution1} and \ref{distribution2} show the distribution of the 5563 galaxies in the VVV NIRGC catalogue with the superposition of the A$_V$ contours (1, 3, 5, 7, 9, 11, 13, 15, 20 and 25 mag) derived from the extinction map by \cite{Schlafly2011} and the stellar density gray-scale map from the VVV catalogues by \citet{Alonso2018}, respectively. The star count map has a pixel scale of 0.9 arcmin$^2$ and includes all stellar detections,  regardless of their magnitudes, for a better visualization. The galaxies in the VVV region across the Southern Galactic plane are represented by red circles. We have also included the galaxies in common with previous works: 2MASX, \citet{Schroder2007}, \citet{Williams2014} and \citet{Said2016} with blue triangles. In general, the regions with strong interstellar extinction or a higher stellar density have no galaxy detections. 

\begin{figure*}
\centering
\includegraphics[width=1\textwidth]{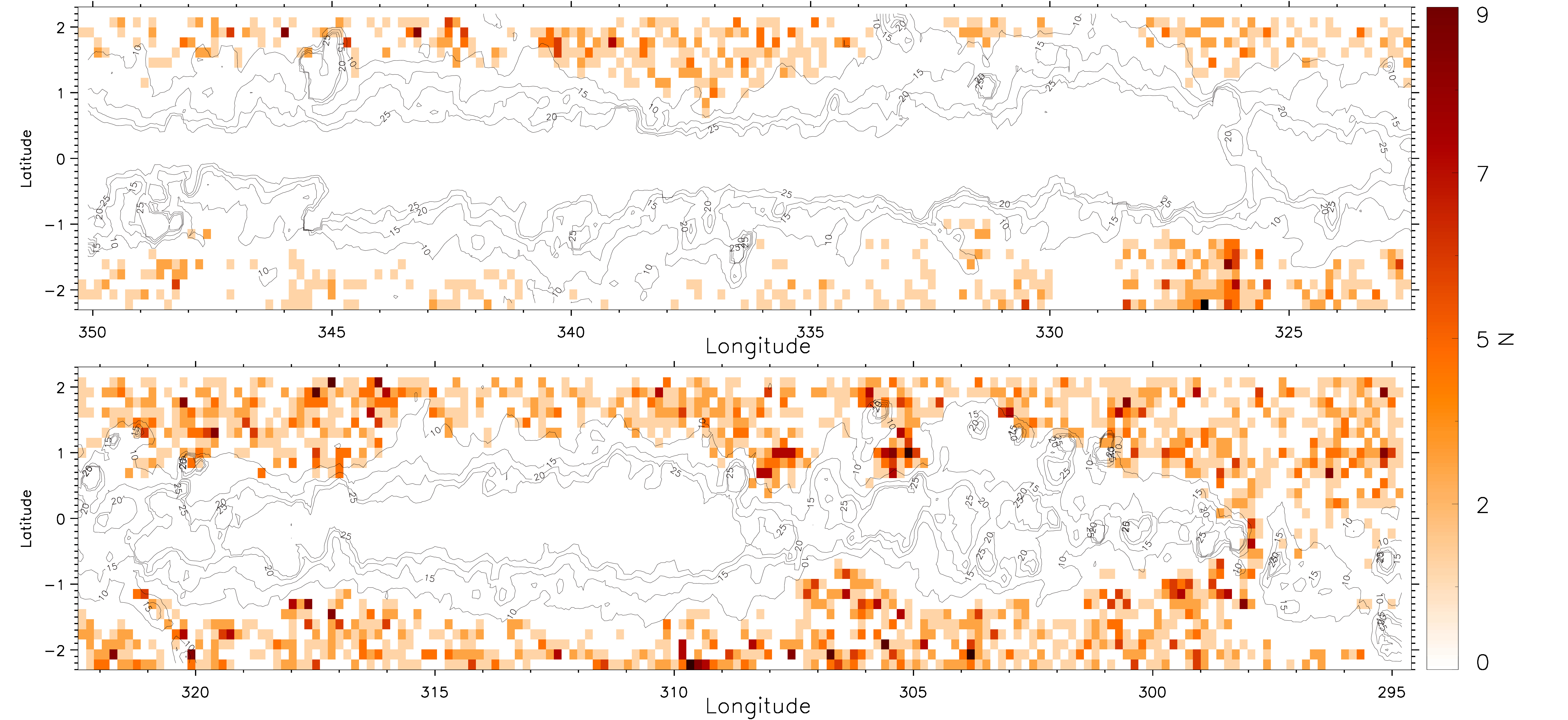}
\caption{Distribution of the brightest galaxies ($K_{s}^{\circ}$ $<$ 15 mag) in the VVV disk region. The density plot has been calculated at a resolution of 90 arcmin$^2$/pixel and shows the superposition of A$_V$ contours (10, 15, 20 and 25 mag) from  \citet{Schlafly2011}.} 
 \label{distribution3}
\end{figure*}

There are 3584 galaxies brighter than $K_{s}^{\circ}$ = 15 mag  
in regions of low interstellar extinction, A$_{Ks}$ $<$ 1 mag. This number represents about 64\% of the galaxies in the VVV NIRGC catalogue.  
Figure~\ref{distribution3} shows the distribution of these brightest galaxies as a density plot with a resolution of 90 arcmin$^2$ per pixel, together with the superposition of some A$_V$ contours (10, 15, 20 and 25 mag) from  \cite{Schlafly2011}. In the Figures~\ref{distribution1}-\ref{distribution3}, possible clustering can be appreciated at different areas of the map.  These regions in Galactic coordinates, in order of density, are the following: (326.0$^{\circ}$, -1.4$^{\circ}$),  (326.7$^{\circ}$, -2.1$^{\circ}$), (299.2$^{\circ}$, -1.0$^{\circ}$), (305.0$^{\circ}$, 1.0$^{\circ}$),
(308.0$^{\circ}$, 1.2$^{\circ}$),
(309.3$^{\circ}$, -2.0$^{\circ}$), 
(317.0$^{\circ}$, 2.0$^{\circ}$)
and (300.8$^{\circ}$, 1.8$^{\circ}$). 
The galaxy cluster VVV-J144321-611754, identified in \citet{Baravalle2019},  is located in a less dense region (the d015 tile, $l$ = 315.8$^{\circ}$ and $b$ = -1.6$^{\circ}$).  As it is 
very important to know which of these observed over-densities of galaxies are real structures, additional observations, i.e. NIR spectroscopy, are needed in order to select and confirm the galaxy cluster candidates.

The VVV NIRGC is the largest collection available to date of galaxies in the Galactic plane within 295$^{\circ}$ $<$ $l$ $<$ 350$^{\circ}$ and -2.25$^{\circ}$ $< b <$ +2.25$^{\circ}$.  In general, dense regions of galaxies are related to lower interstellar extinctions and stellar densities in the Galactic plane.  Thus, Figures~\ref{distribution1}-\ref{distribution3} show that the galaxy distribution closely resembles the distribution of low extinction in the NIR interstellar extinction maps made by the VVV survey \citep{Minniti2018,Soto2019}.

Indeed, the 
main factors driving galaxy detection in the VVV images are the real distribution of galaxies and the total interstellar extinction and stellar density throughout the MW plane. 
The windows of low extinction, in particular, contain a higher number of galaxies per unit area. In fact, this is a useful method for identifying low extinction windows across the MW disk.
Some other secondary factors that favor the detection (or non-detection) of galaxies in the VVV survey are the seeing and the sky brightness variations in the NIR, as discussed earlier.  
For instance, it is more difficult to detect faint galaxies in the crowded regions near the Galactic bulge or along the MW spiral arm tangents.

\section{Main conclusions and Future work}

In this work, we have analysed the VVV NIR images across the Southern Galactic plane in order to 
search for galaxies using {\tt SExtractor+PSFEx}, as described in \cite{Baravalle2018}.
We found that the main limitations for galaxy detections are both the high interstellar extinction and the strong star crowding, which are extremely severe in some cases.  

The VVV NIRGC comprises the photometric and morphological properties of 5563 visually confirmed galaxies, with 99\% of these having no previous identification. This work compiles the largest collection to date of galaxies in the Galactic disk (within 295$^{\circ}$ $<$ $l$ $<$ 350$^{\circ}$, -2.25$^{\circ} <$ $b$ $<$ +2.25$^{\circ}$). 

We have compared galaxy magnitudes with other NIR surveys in the ZoA, namely 2MASS, 2MASX, \cite{Schroder2007} and \cite{Said2016}, and our total magnitude estimates are comparable with the 2MASS magnitudes. The comparison with other isophotal $K_{s}$ magnitudes have offsets higher than one magnitude in all 
cases (Table~\ref{table2}). These differences might be due to the different photometric methods and procedures used to define the magnitudes in these dense regions with high interstellar extinctions. The total number of 185 2MASX extended objects were visually inspected.  The 21 galaxies in common were used for the comparisons, and the remaining 164 extended objects were re-classified as detailed in Appendix B. 

The confirmed galaxies in the VVV NIRGC have colours and morphological parameters that are consistent with early-type galaxies (Table~\ref{tablewise}).
The galaxies with MIR data from WISE are about 10\% of the sample and showing some differences with the total distributions. These are, in general, brighter and 
larger objects with a smaller S\'ersic index than the galaxies in the VVV NIRGC. Thus, they might be late-type galaxies.  

We have also presented the galaxy distribution map across the Southern Galactic plane using the VVV NIRGC. In general, the dense regions of galaxies are related to lower interstellar extinctions in the Galactic plane.
These results confirm that both the interstellar extinction and stellar density are the main limitations for these studies at lower Galactic latitudes. 
This work represents the first step in exploring the distribution of galaxies in the ZoA using the VVV survey. 
The VVV NIRGC may also be used to identify candidate compact groups and clusters of galaxies, which require later spectroscopic confirmation. 

In the future, our procedure might be improved by using co-added images to reach fainter objects and, 
most importantly, be extended to the VVVX area. It is very important to cross-correlate with surveys using other frequencies,  such as 
the Galactic Legacy Infrared Mid-Plane Survey Extraordinaire (GLIMPSE, \citealt{Benjamin2005}) and WISE MIR data,
the Widefield Australian Square Kilometer Array Pathfinder L-band legacy all-sky blind survey (WALLABY, \citealt{Koribalski2020}) of neutral hydrogen, and X-ray and radio source catalogues.

\section*{Acknowledgements}
\addcontentsline{toc}{section}{Acknowledgements}

We would like to thank the anonymous referee for the useful comments and suggestions which has helped to improve this paper. 
This work was partially supported by Consejo de Investigaciones Cient\'ificas y T\'ecnicas (CONICET) and Secretar\'ia de Ciencia y T\'ecnica de la Universidad Nacional de C\'ordoba (SecyT).  We gratefully acknowledge data from the ESO Public Survey program ID 179.B-2002 taken with the VISTA telescope, and products from the CASU. 
DM is supported by the BASAL Center for Astrophysics and Associated Technologies (CATA) through grant AFB 170002 and by Proyecto FONDECYT N$^{\circ}$ 1170121.
This research has made use of the NASA/IPAC Infrared Science Archive, which is funded by the National Aeronautics and Space Administration and operated by the California Institute of Technology.
This publication makes use of data products from the Two Micron All Sky Survey, which is a joint project of the University of Massachusetts and the Infrared Processing and Analysis Center California Institute of Technology, funded by the National Aeronautics and Space Administration and the National Science Foundation. This research has made use of the VizieR catalogue access tool, CDS, Strasbourg, France (DOI: 10.26093/cds/vizier). The original description of the VizieR service was published in 2000, A\&AS 143, 23.


\section*{Data Availability}


The data underlying this article are available in the IATE home page at https://catalogs.oac.uncor.edu/vvv\_nirgc/.





\bibliographystyle{aa.bst} 
\bibliography{Bibliography.bib} 




\appendix

\section{The studied VVV tiles in the Galactic disk}

This is important information for the reader interested in the details of the observed regions.  
Table~\ref{table1} shows some characteristics of the studied tiles,
with the identification in column (1), the Galactic coordinates and the  interstellar extinctions at $V$ passband of the centre in columns (2) to (4), the median $K_{s}$ interstellar extinctions of the extended objects in column (5), and the number of confirmed galaxies in column (6). 

In previous works,  
\citet{Amores2012} reported 214 objects in the d003 tile, of which 72 were considered to be possible galaxies.  These detections are, in general, in agreement with our observations, with any differences 
possible arising from the adopted methodology and the removal of objects with strong stellar contamination.  Also, \citet{Baravalle2018} reported 530 galaxies in the d010 and d115 tiles.  
Here, we were more discriminative in the image selection and more exhaustive in the visual inspection. 

\begin{table}
\center
\caption{VVV tiles in the Galactic disk}
\begin{tabular}{|lccrrr|}
\hline 
 Tile ID &   $l$   & $b$   & A$_{V}$ & A$_{K_s}$ & Number of \\
         &  [deg]  & [deg] & [mag]   & [mag]     & galaxies \\
\hline 
  d001 & 295.4376 & -1.6497 &  8.703  &  1.159 & 39\\
  d002 & 296.8967 & -1.6498 &  7.858  &  1.081 & 31\\
  d003 & 298.3557 & -1.6497 &  4.725  &  0.553 & 89\\
  d004 & 299.8147 & -1.6497 &  4.988  &  0.588 & 106\\
  d005 & 301.2737 & -1.6497 &  4.226  &  0.599 & 92\\
  d006 & 302.7327 & -1.6497 &  5.747  &  0.785 & 106\\
  d007 & 304.1917 & -1.6498 &  4.745  &  0.655 & 101\\
  d008 & 305.6510 & -1.6497 &  4.261  &  0.551 & 117\\
  d009 & 307.1097 & -1.6497 &  4.298  &  0.581 & 69\\
  d010 & 308.5690 & -1.6497 &  9.103  &  0.898 & 129\\
  d011 & 309.9739 & -1.6360 &  7.218  &  0.966 & 66\\
  d012 & 311.4867 & -1.6497 &  8.587  &  1.060 & 38\\
  d013 & 312.9457 & -1.6497 &  5.891  &  0.945 & 54\\
  d014 & 314.4047 & -1.6497 &  6.705  &  0.870 & 63\\
  d015 & 315.8371 & -1.6502 &  6.113  &  0.840 & 91\\
  d016 & 317.2949 & -1.6497 &  5.116  &  0.645 & 98\\
  d017 & 318.7539 & -1.6497 &  5.909  &  0.781 & 93\\
  d018 & 320.2129 & -1.6497 &  4.319  &  0.960 & 83\\
  d019 & 321.6719 & -1.6497 &  7.393  &  0.917 & 97\\
  d020 & 323.1406 & -1.6362 &  7.693  &  0.881 & 58\\
  d021 & 324.5899 & -1.6497 &  10.229 &  1.027 & 43\\
  d022 & 326.0489 & -1.6497 &  3.653  &  0.563 & 157\\
  d023 & 327.5079 & -1.6497 &  4.355  &  0.566 & 82\\
  d024 & 328.9669 & -1.6497 &  6.620  &  0.834 & 3\\
  d025 & 330.4259 & -1.6497 &  5.124  &  0.853  & 35\\
  d026 & 331.8849 & -1.6497 &  9.314  &  0.781  & 26\\
  d027 & 333.3439 & -1.6497 &  5.381  &  0.715  & 13\\
  d028 & 334.8029 & -1.6497 &  9.009  &  0.985  & 24\\
  d029 & 336.1531 & -1.5276 &  8.087  &  0.920  & 19\\
  d030 & 337.7209 & -1.6497 &  8.056  &  1.027  & 9\\
  d031 & 339.1799 & -1.6497 &  9.367  &  1.134  & 11\\
  d032 & 340.6389 & -1.6497 &  10.498 &  1.318  & 14\\
  d033 & 342.0979 & -1.6497 &  8.322  &  1.031  & 24\\
  d034 & 343.5569 & -1.6497 &  6.974  &  1.030  & 18\\
  d035 & 345.0159 & -1.6498 &  6.417  &  0.812  & 30\\
  d036 & 346.4749 & -1.6498 &  8.109  &  0.939  & 13 \\
  d037 & 347.9339 & -1.6497 &  5.365  &  0.690  & 30 \\
  d038 & 349.3929 & -1.6497 &  5.398  &  0.679  & 24 \\
  d039 & 295.4374 & -0.5576 &  7.001  &  1.024 &  3\\
  d040 & 296.8962 & -0.5576 &  10.775 &  1.256 &  13\\
  d041 & 298.3548 & -0.5575 &  4.748  &  0.931 &  93\\
  d042 & 299.8135 & -0.5570 &  9.725  &  1.102 &  68\\
  d043 & 301.2720 & -0.5575 &  10.005 &  1.192 &  31\\
  d044 & 302.7308 & -0.5575 &  20.385 &  1.976 &  10\\
  d045 & 304.1896 & -0.5576 &  19.626 &  1.840 &  21\\
  d046 & 305.6481 & -0.5575 &  10.511 &  1.113 &  39\\
  d047 & 307.1068 & -0.5575 &  10.236 &  1.461 &  34\\
  d048 & 308.5653 & -0.5575 &  19.512 &  2.473 &  9\\
  d049 & 310.0249 & -0.5544 &  28.317 &  3.480 &  4\\
  d050 & 311.4828 & -0.5575 &  39.246 &  2.743 &  7\\
  d051 & 312.9415 & -0.5576 &  31.868 &  2.298 &  6\\
  d052 & 314.4001 & -0.5576 &  28.625 &  2.305 &  3\\
  d053 & 315.8588 & -0.5575 &  28.934 &  2.268 &  8\\
  d054 & 317.3175 & -0.5575 &  21.962 &  2.014 &  24\\
  d055 & 318.7761 & -0.5576 &  18.618 &  2.539 &  7\\
  d056 & 320.2348 & -0.5576 &  24.226 &  2.542 &  12\\
  d057 & 321.6935 & -0.5575 &  17.682 &  2.066 &  23\\
  d058 & 323.1522 & -0.5575 &  17.893 &  2.191 &  8\\
  d059 & 324.6109 & -0.5575 &  22.517 &  3.163 &  5\\
    \hline 
\end{tabular}
\label{table1}
\end{table}

\begin{table}
\contcaption{}
\begin{tabular}{|lccrrr|}
\hline 
 Tile ID &   $l$   & $b$   & A$_{V}$ & A$_{K_s}$ & Number of \\
         &  [deg]  & [deg] & [mag]   & [mag]     & galaxies \\
         \hline 
  d060 & 326.0695 & -0.5575 & 42.993  & 2.463  & 5 \\
  d061 & 327.5282 & -0.5576 & 29.493  & 3.457  & 10\\
  d062 & 328.9868 & -0.5576 & 24.148  & 2.577  & 2\\
  d063 & 330.4455 & -0.5575 & 25.052  & 2.169  & 9\\
  d064 & 331.9041 & -0.5575 & 15.324  & 1.797  & 8\\
  d065 & 333.3629 & -0.5576 & 39.527  & 2.832  & 0\\
  d066 & 334.8215 & -0.5575 & 25.793  & 2.419  & 1\\
  d067 & 336.2801 & -0.5576 & 21.919  & 2.828  & 3\\
  d068 & 337.7388 & -0.5576 & 18.504  & 2.765  & 1\\
  d069 & 339.1975 & -0.5576 & 35.709  & 4.263  & 0\\
  d070 & 340.6561 & -0.5576 & 34.913  & 5.366  & 0\\
  d071 & 342.1148 & -0.5576 & 12.754  & 3.302  & 0\\
  d072 & 343.5735 & -0.5575 & 27.715  & 3.802  & 0\\
  d073 & 345.0321 & -0.5576 & 30.089  & 2.953  & 0\\
  d074 & 346.4908 & -0.5576 & 16.443  & 1.759  & 1 \\
  d075 & 347.9495 & -0.5576 & 17.169  & 1.755  & 5 \\
  d076 & 349.4082 & -0.5575 & 14.235  & 2.585  & 1 \\
  d077 & 295.4375 &  0.5346 &  7.330  & 0.781  & 93\\
  d078 & 296.8963 &  0.5346 &  6.419  & 0.755  & 26\\
  d079 & 298.3552 &  0.5345 &  4.322  & 0.841  & 73\\
  d080 & 299.8141 &  0.5346 &  11.560 & 1.386  & 37\\
  d081 & 301.2729 &  0.5346 &  10.349 & 1.661  & 30\\
  d082 & 302.7318 &  0.5346 &  14.507 & 2.183  & 28\\
  d083 & 304.1908 &  0.5346 &  17.355 & 2.067  & 23\\
  d084 & 305.6495 &  0.5346 &  17.463 & 1.807  & 47\\
  d085 & 307.1084 &  0.5345 &  18.849 & 1.569  & 52\\
  d086 & 308.5672 &  0.5346 &  33.014 & 1.691  & 38\\
  d087 & 310.0261 &  0.5346 &  42.790 & 3.144  & 6\\
  d088 & 311.4849 &  0.5346 &  42.121 & 3.803  & 17\\
  d089 & 312.9438 &  0.5346 &  26.151 & 2.716  & 7\\
  d090 & 314.4027 &  0.5346 &  22.500 & 2.335  & 10\\
  d091 & 315.8616 &  0.5346 &  22.336 & 2.119  & 12\\
  d092 & 317.3204 &  0.5346 &  13.721 & 1.482  & 36 \\
  d093 & 318.7793 &  0.5346 &  16.766 & 1.736  & 19 \\
  d094 & 320.2382 &  0.5346 &  28.764 & 1.899  & 22 \\
  d095 & 321.6970 &  0.5346 &  12.878 & 1.934  & 14 \\
  d096 & 323.1559 &  0.5346 &  18.679 & 2.343  & 4 \\
  d097 & 324.6148 &  0.5346 &  20.111 & 2.630  & 2 \\
  d098 & 326.0736 &  0.5346 &  25.883 & 2.842  & 1 \\
  d099 & 327.5325 &  0.5346 &  33.170 & 4.346  & 0\\
  d100 & 328.9913 &  0.5346 &  79.078 & 6.185  & 0 \\
  d101 & 330.4502 &  0.5346 &  47.601 & 4.172  & 0 \\ 
  d102 & 331.9091 &  0.5346 &  27.476 & 3.564  & 0 \\
  d103 & 333.3679 &  0.5346 &  20.665 & 3.212  & 1 \\
  d104 & 334.8269 &  0.5346 &  16.362 & 2.740  & 4 \\
  d105 & 336.2857 &  0.5346 &  19.643 & 2.719  & 6 \\
  d106 & 337.7445 &  0.5346 &  14.083 & 1.992  & 3\\
  d107 & 339.2034 &  0.5346 &  24.339 & 4.283  & 2\\
  d108 & 340.6622 &  0.5346 &  19.564 & 3.086  & 0 \\ 
  d109 & 342.1212 &  0.5346 &  37.578 & 3.024  & 0\\
  d110 & 343.5801 &  0.5346 &  28.236 & 3.265  & 0 \\
  d111 & 345.0388 &  0.5346 &  19.256 & 3.348  & 1 \\
  d112 & 346.4977 &  0.5346 &  34.060 & 4.012  & 0 \\
  d113 & 347.9566 &  0.5346 &  22.919 & 3.399  & 6 \\
  d114 & 349.4155 &  0.5346 &  23.303 & 3.487  & 0 \\
  d115 & 295.4380 &  1.6270 &   3.520 & 0.451  & 147\\
  d116 & 296.8973 &  1.6268 &   4.355 & 0.531  & 87\\
  d117 & 298.3569 &  1.6268 &   3.756 & 0.514  & 104\\
  d118 & 299.8156 &  1.6266 &   4.546 & 0.538  & 142\\
  d119 & 301.2761 &  1.6268 &   3.258 & 0.450  &  83\\
  \hline 
\end{tabular}
\end{table}
  
  \begin{table}
\contcaption{}
\begin{tabular}{|lccrrr|}
\hline 
 Tile ID &   $l$    & $b$   & A$_{V}$ & A$_{K_s}$ & Number of \\
         &   [deg]  & [deg] & [mag] & [mag]   & galaxies \\
\hline 
  d120 & 302.7357 & 1.6268 &  7.564   & 0.978 & 73 \\
  d121 & 304.1952 & 1.6268 &  12.774  & 0.951 & 52 \\
  d122 & 305.6592 & 1.6765 &  41.703  & 0.786 & 151 \\
  d123 & 307.1145 & 1.6268 &  5.067   & 0.607 & 92 \\
  d124 & 308.5740 & 1.6268 &  4.553   & 0.727 & 101 \\
  d125 & 310.0336 & 1.6268 &  7.753   & 0.919 & 74 \\
  d126 & 311.4932 & 1.6268 &  8.216   & 0.932 & 46\\
  d127 & 312.9528 & 1.6268 &  7.704   & 1.024 & 54\\
  d128 & 314.4124 & 1.6268 &  7.331   & 1.018 & 60\\
  d129 & 315.8719 & 1.6268 &  7.606   & 0.897 & 117 \\
  d130 & 317.3316 & 1.6268 &  4.509   & 0.616 & 94 \\
  d131 & 318.7912 & 1.6268 &  6.303   & 0.784 & 76 \\
  d132 & 320.2508 & 1.6268 &  6.011   & 0.743 & 101 \\
  d133 & 321.7104 & 1.6268 &  8.537   & 0.870 & 71 \\
  d134 & 323.1699 & 1.6268 &  6.516   & 1.025 & 26\\
  d135 & 324.6295 & 1.6268 &  8.388   & 1.092 & 27\\
  d136 & 326.0891 & 1.6268 &  6.649   & 0.870 & 45\\
  d137 & 327.5487 & 1.6268 &  11.189  & 1.378 & 33\\
  d138 & 328.3753 & 1.6268 &  38.469  & 1.801 & 13\\
  d139 & 330.4679 & 1.6268 &  11.014  & 1.648 & 21 \\ 
  d140 & 331.9275 & 1.6268 &  9.788   & 1.411 & 25 \\
  d141 & 333.3871 & 1.6268 &  12.664  & 1.340 & 17 \\
  d142 & 334.8467 & 1.6268 &  7.171   & 1.023 & 53 \\
  d143 & 336.3062 & 1.6268 &  5.663   & 0.691 & 63 \\
  d144 & 337.7658 & 1.6268 &  5.742   & 0.658 & 55 \\
  d145 & 339.2255 & 1.6268 &  5.136   & 0.644 & 47 \\
  d146 & 340.6850 & 1.6268 &  11.347  & 1.171 & 41 \\
  d147 & 342.1446 & 1.6268 &  9.055   & 1.151 & 49 \\
  d148 & 343.6042 & 1.6268 &  8.695   & 1.043 & 25 \\
  d149 & 345.0639 & 1.6268 &  47.313  & 1.290 & 22 \\
  d150 & 346.5234 & 1.6268 &  10.183  & 1.123 & 29 \\
  d151 & 347.9830 & 1.6267 &  8.418   & 0.961 & 55 \\
  d152 & 349.4426 & 1.6267 &  9.097   & 1.105 & 33 \\
\hline 
\end{tabular}
\end{table}

\section{The 2MASS extended sources in the VVV disk region}

In Section~\ref{subsec:2masx}, we reported 185 extended sources from the 2MASX catalogue in the regions of the VVV disk. We visually checked all of these, with this list including 21 galaxies and 164 objects that are mainly stars and 
stellar associations.

Table~\ref{table2massgal} shows the 
2MASX galaxies that were used to compare our photometry, listing the 2MASX identification in column (1), the Galactic coordinates in columns (2) and (3) from the 2MASX  catalogue, the 2MASX and WISE classification types in columns (4) and (5), the heliocentric radial velocity in column (6), and other literature references taken from the NASA Extragalactic Database (NED)\footnote{The NASA/IPAC Extragalactic Database (NED) is operated by the Jet Propulsion Laboratory, California Institute of Technology, under contract with the National Aeronautics and Space Administration.} in column (7). For the classification, we used G for a galaxy and IrS for an IR source, taken from 2MASX and WISE surveys.  
The radial velocities were mainly obtained from the HI survey (\citealt{Staveley2016}). Following \cite{Schroder2019a}, all these sources have flag=1 or 'obvious galaxy' with the exception of 2MASX J12164837-6103579 with flag=2 or 'galaxy'.

\begin{figure*}
\centering
\includegraphics[width=0.33\textwidth,height=0.31\textwidth]{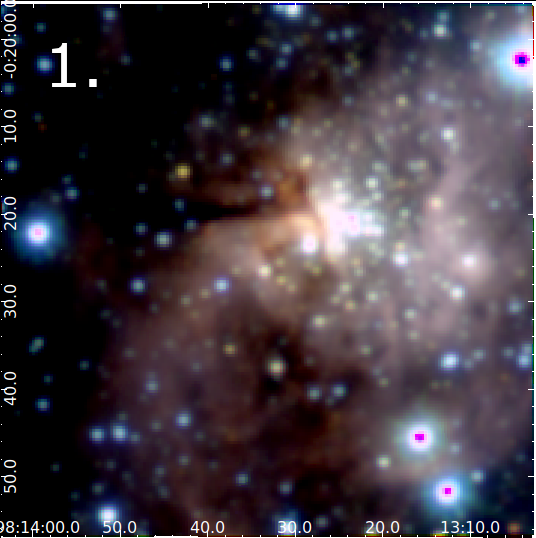}
\includegraphics[width=0.33\textwidth,height=0.31\textwidth]{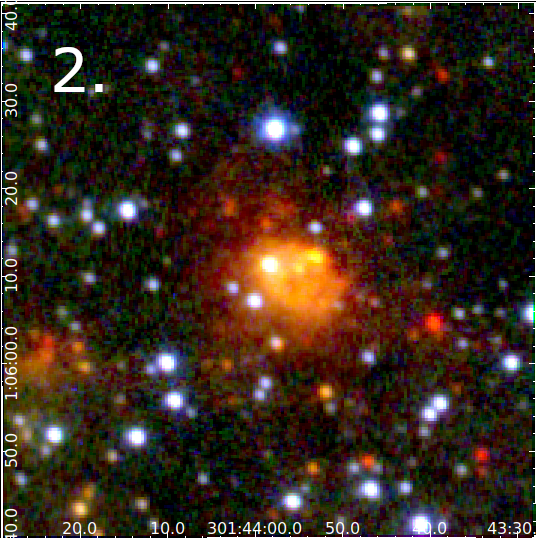}\\
\includegraphics[width=0.33\textwidth,height=0.31\textwidth]{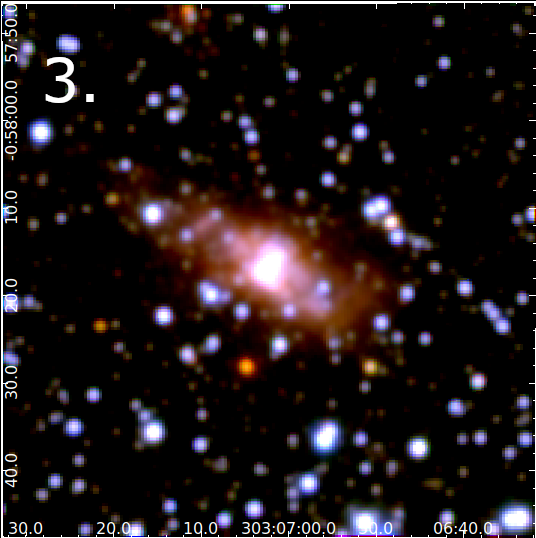}
\includegraphics[width=0.33\textwidth,height=0.31\textwidth]{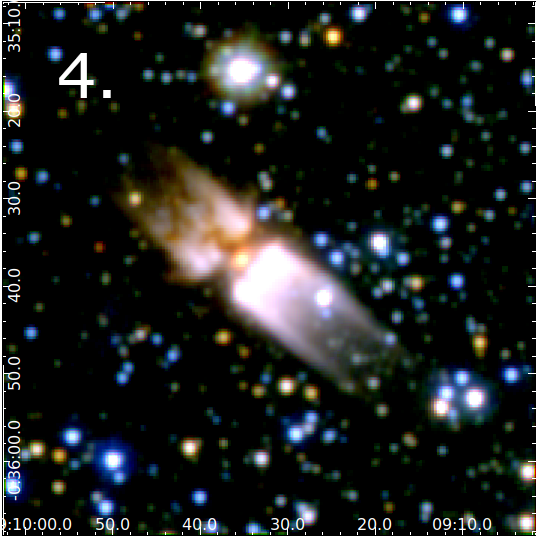}\\
\includegraphics[width=0.33\textwidth,height=0.31\textwidth]{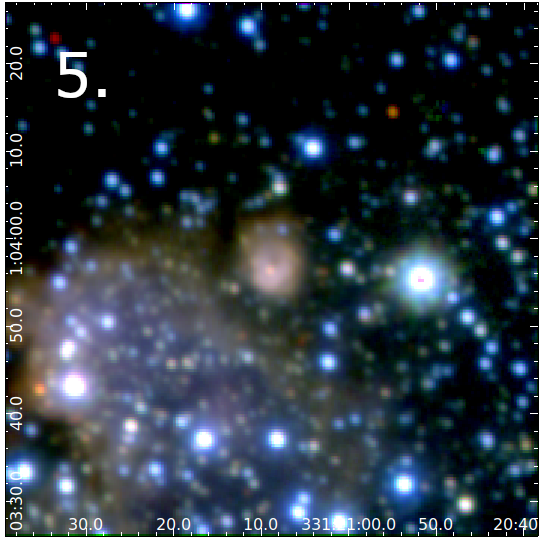}
\includegraphics[width=0.33\textwidth,height=0.31\textwidth]{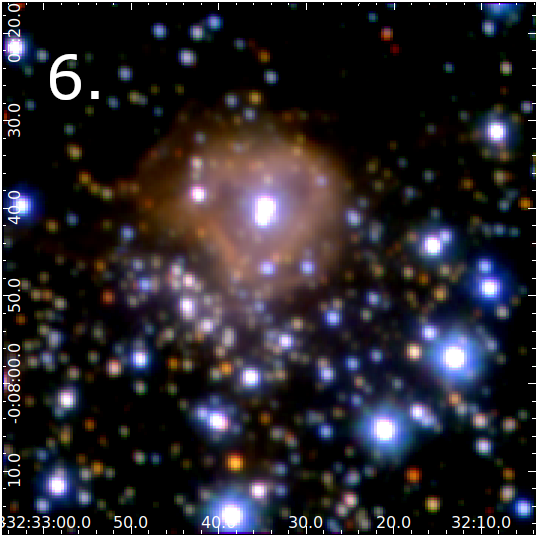}\\
\includegraphics[width=0.33\textwidth,height=0.31\textwidth]{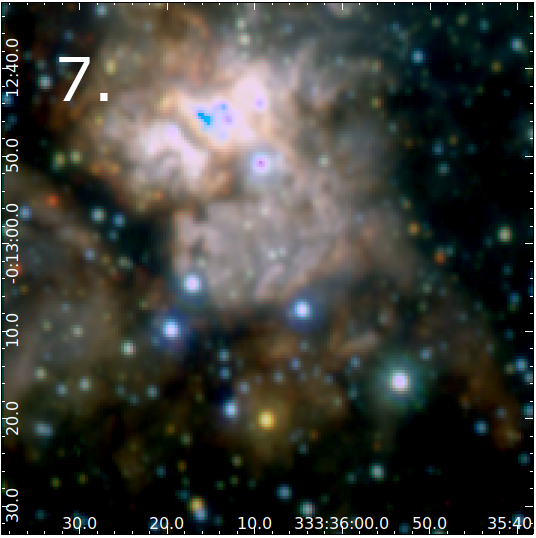}
\includegraphics[width=0.33\textwidth,height=0.31\textwidth]{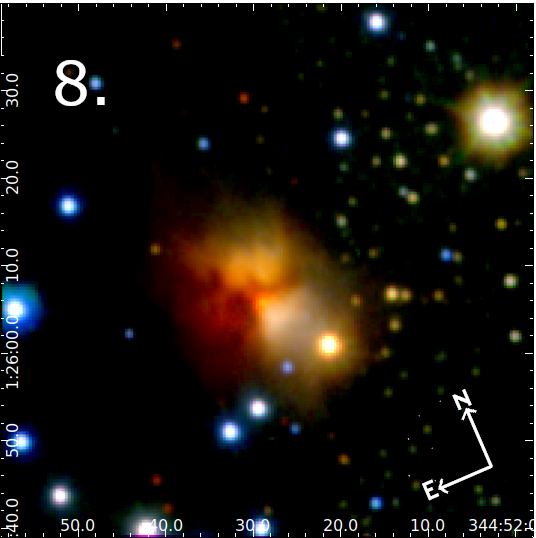}
\caption{The VVV colour composed images of some Galactic examples of the 2MASS extended sources. 
The panels show the sources: 1. J12100188-6250004, 2. J12411836-6144427, 3. J12530717-6350331, 
4. J13462058-6247597, 5. J16062307-5043275, 6. J16170265-5047083, 7. J16221035-5006168 and 8. J16574897-4034050. 
}\label{sources}
\end{figure*}

For the 164 2MASX extended objects that are not extragalactic, we searched for the closest object in a circle of 1 arcmin radius 
using the SIMBAD database\footnote{This research has made use of the SIMBAD database,
operated at CDS, Strasbourg, France.}. We also identified these sources with different surveys at
other frequencies.
Table~\ref{table2mass} shows interesting  information of the 164 2MASX extended objects, listing the 2MASX identification in column (1), the Galactic coordinates in columns (2) and (3) from the 2MASX  catalogue, the SIMBAD ID of the closest source in the database  and the corresponding literature references in columns (4) and (5), and the object description in column (6).  In general, we took the object description from the SIMBAD database, and for the cases without previous descriptions (about 16\% of the total) we added our comments after the visual inspection.   
The source 2MASX J13464910-6024299 is the known CenB, a well-studied radio source, and we included only 5 of the 59 available references in the table.   
The sources 2MASX J13462058-6247497, J16103869-4905591, J16152012-4913215 and J17134463-3705111 have the \cite{Schroder2019a} flags 8p?, 9, 9p and 8p?, respectively, in agreement with our  description. 
Figure~\ref{sources} shows the VVV colour composed images of some examples of these 2MASX objects, such as HII regions (1. 12100188-6250004 and 3. 12530717-6350331),  young stellar object (YSO) candidates (2. 12411836-6144427, 5. 16062307-5043275 and 8. 16574897-4034050),  IR sources (6. 16170265-5047083 and 7. 16221035-5006168), and emission line stars (4. 13462058-6247597).

\begin{table*}
\center
\caption{2MASX extragalactic sources}
\begin{tabular}{|ccccccc|}
\hline 
 2MASX ID           &   $l$    &  $b$     & 2MASX & WISE & Radial velocity  & REFERENCES \\
                    &   [deg]  & [deg]    & Type  & Type &   [km/sec]              & \\
 \hline 
  J11494881-6400073 & 296.2408 & -1.9336 &  IrS & G & 2119 $\pm$ 7 & 1, 2, 3, 4, 5, 6, 7, 14, 16 \\
  J11523245-5950248 & 295.5898 & 2.1891  &  G & G & 5400 $\pm$ 15 & 8, 14, 16\\
  J11565971-6019148 & 296.2382 & 1.8419  &  G & G & -- & 8\\
  J12092077-6229125 & 298.0924 & -0.0111 &  IrS &  IrS & 1449 $\pm$ 14 &  2, 4, 14, 16\\ 
  J12120649-6045031 & 298.1435 & 1.7548  &  IrS &  IrS & -- & 14 \\
  J12153835-6102448 & 298.6112 & 1.5253  &  G   &  G   & -- & 14 \\
  J12164837-6103579 & 298.7541 & 1.5244  &  G   &  G   & -- & 14 \\
  J12190068-6024475 & 298.9365 & 2.2067  &  G   &  G   & -- & -- \\
  J12280968-6054558 & 300.1055 & 1.8283  &  G   &  IrS & 5011 $\pm$  51 & 2, 14, 16\\
  J12294880-6107045 & 300.3227 & 1.6442  &  G   &  IrS & 5391 $\pm$ 319 & 2, 4, 14, 16\\
  J13281021-6022580 & 307.4590 & 2.1650  &  IrS &  IrS & -- & 9, 14 \\
  J13440358-6019350 & 309.4047 & 1.8822  &  IrS &  IrS & -- & 7, 9, 10, 14 \\
  J13464910-6024299 & 309.7214 & 1.7315  &  G  & G & 3872 $\pm$ 20 &  6, 9, 11, 14, 15\\
  J13471848-6034133 & 309.7457 & 1.5600  & IrS  & IrS & -- & 9, 12, 13, 14 \\
  J13502487-6005380 & 310.2259 & 1.9398  & G & IrS & -- & 14 \\
  J14062119-6025446 & 312.0511 & 1.1096  & G & IrS & -- & 2, 8, 13, 14\\
  J14155209-5855348 & 313.6519 & 2.1768  & G  & IrS & 5399  $\pm$ 4 & 2, 4, 8, 14 \\
  J14391130-5742293 & 316.9480 & 2.2014  & G  & IrS & -- & 6, 8, 14\\
  J15232096-5915266 & 321.2764 & -1.9076 & IrS & IrS & -- & 14 \\
  J15395406-5737113 & 323.9908 & -1.8069 & G   & G & -- & 8, 14\\
  J16174633-4751584 & 334.6618 & 1.8824 & G & G & -- &  8, 14\\
  \hline 
  \end{tabular}
  \begin{tablenotes}
      \small
      \item REFERENCES. 1: \citet{Kourkchi2017}, 2: \citet{Said2016}, 3: \citet{Courtois2015}, 4: \citet{Williams2014},
      \item  5: \citet{Courtois2009}, 6: \citet{Paturel2003}, 7: \citet{Vauglin2002},  8: \citet{Karachentseva2010},
      \item  9: \citet{Schroder2007}, \citet{Rousseau2000}, 11: \citet{Jarrett2004}, 12: \citet{Nagayama2004},
      \item 10:  \citet{Mitronova2004}, 14: \citet{Schroder2019a}, 15: \citet{West1989}, 16: \citet{Staveley2016}
    \end{tablenotes}
  \label{table2massgal}
  \end{table*}

\begin{table*}
\center
\caption{2MASX Galactic sources}
\begin{tabular}{|cccccc|}
\hline 
 2MASX ID   &    $l$    & $b$     & SIMBAD ID &  REFERENCES &  Description \\
            &   [deg]  & [deg]   &  & &        \\
 \hline 
  J11402953-6327520  & 295.1064 &  -1.6758 &  La Serena 003       & 1 & Possible (open) star cluster \\
  J11450474-6317459  & 295.5581 &  -1.3783 &  La Serena 009       & 1 & Possible (open) star cluster \\
  J11520923-6153348  & 296.0102 &  0.1794  &  IRAS 11496-6136     & 2 & Carbon Star \\
  J11535055-6420218  & 296.7457 &  -2.1616 &  VVV CL007           & 3 & Cluster of Stars \\
  J11585505-6337470  & 297.1394 &  -1.3502 &  G297.1390-01.3510   & 4 & IR source\\
  J12005699-6304097  & 297.2516 &  -0.7558 &  La Serena 017       & 1 & Possible (open) star cluster\\
  J12090125-6315596  & 298.1830 &  -0.7866 &  ESO 95-1            & 5 & HII (ionized) region \\
  J12095662-6249224  & 298.2143 &  -0.3319 &  Caswell CH3OH 298.22-00.33 & 6 & Maser \\
  J12100013-6250474  & 298.2247 &  -0.3541 &   --                 & --  & Star\\
  J12100188-6250004  & 298.2259 &  -0.3407 &  G298.2-00.3         & 7  & HII (ionized) region \\
  J12114684-6146291  & 298.2609 &  0.7368  &  VVV CL010           & 3  &  Cluster of stars \\
  J12152350-6301156  & 298.8570 &  -0.4348 &  [HLB98] SEST 54     & 8  &  Maser\\
  J12152500-6301207  & 298.8600 &  -0.4358  &  G298.8591-00.4372   & 14 &  YSO \\
  J12204537-6310426  & 299.4790 &  -0.5126 &   --                 & --   &  Star with a bright nearby star\\
  J12294170-6213085  & 300.4012 &  0.5458  &  IRAS 12268-6156     &  5 &  Dense core \\
  J12300399-6256387  & 300.5045 &  -0.1731 &  G300.5047-00.1745   &  4 &  HII (ionized) region \\
  J12324972-6135310  & 300.7200 &  1.1998  & [MHL2007] G300.7221+01.2007 & 14 &  YSO \\
  J12344923-6139290  & 300.9607 &  1.1501  &   --                 & --   &  Star\\
  J12345078-6139070  & 300.9633 &  1.1564  &  WRAY 16-118         &  9 &  PN \\
  J12345259-6140190  & 300.9682 &  1.1366  &  VVV CL015           &  3 &  Cluster of stars \\
  J12345752-6139401  & 300.9772 &  1.1480  &  G300.9785+01.1459   &  4 &  IR source \\
  J12350355-6141451  & 300.9913 &  1.1141   &  VVV CL016           &  3 &  Open (galactic) Cluster \\
  J12360541-6150347  & 301.1221 &  0.9748  &    --                & --   &  Star \\
  J12411836-6144427  & 301.7329 &  1.1030  &  G301.7319+01.1030   & 14 &  YSO candidate\\
  J12415682-6204187  & 301.8208 &  0.7794  &  G301821+007850      & 10 &  Bubble \\
  J12433200-6255130  & 302.0325 &  -0.0624 &  IRAS 12405-6238     & 5  &  HII (ionized) region \\
  J12501909-6134572  & 302.7987 &  1.2889  &  DBS2003 80          & 11 &  Open (galactic) Cluster\\
  J12502553-6135223  & 302.8115 &  1.2819  &  DBS2003 80          & 11 &  Open (galactic) Cluster\\
  J12530717-6350331  & 303.1173 &  -0.9714 &  IRAS 12500-6334     & 5  &  HII (ionized) region \\ 
  J12545111-6102526  & 303.3453 &  1.8211  &  IRAS 12518-6046     & 12 &  Far-IR source \\
  J12552193-6104510  & 303.4070 &  1.7873  &   --                 & --   &  Nebula\\
  J13083302-6214482  & 304.9227 &  0.5588 &  DBS2003 82          & 11 &  Cluster of stars \\
  J13084090-6215161  & 304.9375 &  0.5501 &  HTU2013 2           & 13 &  Dense core \\
  J13110865-6234415  & 305.1986 &  0.2068 &  G305.197+0.206      & 13 &  HII (ionized) region \\
  J13111431-6245005  & 305.1962 &  0.0345 &  DBS2003 83          & 11 &  Cluster of stars \\
  J13111621-6245505  & 305.1988 &  0.0204 &  DBS2003 84          & 11 &  Cluster of stars\\
  J13121738-6242198  & 305.3198 &  0.0695 &  DBS2003 132         & 11 &  Cluster of stars\\
  J13122717-6233047  & 305.3510 &  0.2217 &  G305.348+0.223      & 13 &  HII (ionized) region \\
  J13142096-6244230  & 305.5520 &  0.0154 &  IRAS 13111-6228     & 5  &  HII (ionized) region \\
  J13142635-6244309  & 305.5621 &  0.0123 &  G305.561+0.013      & 14 &  YSO candidate \\
  J13270813-6203201  & 307.1010 &  0.5269 &  DZOA 4638-16        & 15 &  IR source\\
  J13324267-6026541  & 308.0094 &  2.0147 &  G308.0023+02.0190   & 14 &  YSO Candidate\\
  J13405761-6145447  & 308.7543 &  0.5482 &  DZOA 4648-07        & 15 &  IR source \\
  J13455154-6009067  & 309.6594 &  2.0071 &   --                 & --  &  Star \\
  J13462058-6247597  & 309.1589 & -0.5939 &  IRAS 13428-6232     & 17 & Emission-line star\\
  J13463702-6239303  & 309.2196 &  -0.4622 &  AGAL G309.221-00.462 & 18 & Sub-millimetric source \\ 
  J13503488-6140199  & 309.8877 &  0.3982  &  DZOA 4655-12        & 15 & YSO \\
  J13510266-6130150  & 309.9796 &  0.5493  &  IRAS 13475-6115     & 12 & Star \\
  J13515956-6115394  & 310.1462 &  0.7597  &  Caswell OH 310.146+00.760   & 6 & Maser \\  
  J14020791-6148249  & 311.1783 &  -0.0723 &  [MHL2007] G311.1794-00.0720 & 14 & YSO candidate \\
  J14023620-6105450  & 311.4253 &  0.5969  &  DZOA 4664-05        & 15 & IR source \\
  J14023845-6118190  & 311.3727 &  0.3941 &    --                &  --  & Faint star with bright nearby star \\
  J14084243-6110419  & 312.1084 &  0.3092 &  AGAL G312.108+00.309 & 18 & Sub-millimetric source \\
  J14143959-6222433  & 312.4121 &  -1.0486 &  IRAS 14109-6208      & 12 & Far-IR source ($\lambda$ $>=$ 30 {$\mu$}m) \\
  J14242344-6205208  & 313.5772 &  -1.1540 &  AGAL G313.576-01.154 & 18   & YSO\\ 
  J14245658-6144520  & 313.7582 &  -0.8574 &  AKARI-IRC-V1 J1424563-614450 & 19 & Star \\
  J14275344-5844595  & 315.1742 &  1.8103  &    --                 & --    & Star with a bright nearby star\\
  J14420173-6030091  & 316.1402 &  -0.4984 &  BNM96 316.139-0.503 & 5 &  HII (ionized) region \\
  J14450462-5949094  & 316.7704 &  -0.0363 &  SMN83 G316.8-0.1 1   & 20 & Part of a cloud \\
  J14450514-5949050  & 316.7719 &  -0.0357 &  VGO2007 14416-5937 B & 21 & IR source  \\
  \hline 
  \end{tabular}
  \label{table2mass}
  \end{table*}
  
\begin{table*}
\contcaption{}
\label{table2mass:continued}
\center
\begin{tabular}{|cccccc|}
\hline 
 2MASX ID   &     $l$    &  $b$      & SIMBAD ID &  REFERENCES &  Description \\
            &   [deg]  & [deg]   &  &  &        \\
 \hline 
  J14452278-5949415 & 316.8011 & -0.0605 &  WBB2001 41 & 22 & IR source \\
  J14592915-5820102 & 319.0892 & 0.4604  &  IRAS 14556-5808 & 5 & HII (ionized) region \\
  J14593166-5749002 & 319.3392 & 0.9158 & IRAS 14557-5737 & 5 & HII (ionized) region \\  
  J15002412-5924522 & 318.6831 & -0.5455 &   --    &  -- & No object detected \\
  J15003503-5858101 & 318.9150 & -0.1650 & WBB2001 43 & 22 & IR source \\
  J15031365-5904294 & 319.1630 & -0.4206 & IRAS 14593-5852 & 5 & HII (ionized) region \\
  J15034121-5835075 & 319.4522 & -0.0213 & VVV CL044       & 3 & Cluster of stars \\
  J15051680-5731176 & 320.1555 & 0.8043 & WRAY 16-160     & 9 & PN \\ 
  J15160510-5811445 & 321.0508 & -0.5063 & AGAL G321.054-00.507  & 18  & Sub-millimetric source\\ 
  J15183778-5638505 & 322.1607 & 0.6262 & MSX5C G322.1587+00.6262 & 4 & IR source \\
  J15202255-5627119 & 322.4679 & 0.6611 & IRAS 15165-5616 & 23  & Star \\
  J15325265-5556106 & 324.1995 & 0.1206 & IRAS 15290-5546 & 5 & HII (ionized) region \\
  J15370333-5500156 & 325.2236 & 0.5332 &  --   & --  &    Star with a bright nearby star\\
  J15373472-5534467 & 324.9446 & 0.0246 &  --   & --  &    Star\\ 
  J15421778-5358308 & 326.4482 & 0.9056 & IRAS 15384-5348 & 5 & HII (ionized) region \\ 
  J15433657-5357522 & 326.6084 & 0.7972 & VVV CL055     & 3 & Cluster of stars\\ 
  J15444145-5405599 & 326.6518 & 0.5931 & [DBS2003] 95 & 5 & HII (ionized) region\\
  J15450053-5402059 & 326.7284 & 0.6160 & [RA2006b] IRS 2 & 24 & NIR source ($\lambda$ $<$ 10 {$\mu$}m)\\
  J15455314-5430042 & 326.5440 & 0.1691  & AGAL G326.544+00.169 & 18 & Sub-millimetric source \\
  J15462068-5410535 & 326.7932 & 0.3802  & IRAS 15425-5401 & 23 & Outflow candidate\\
  J15491623-5425461 & 326.9747 & -0.0788 & IRAS 15453-5416 & 23 & Star\\
  J15500315-5403503 & 327.2922 &  0.1353 &  --    & --   & Star \\
  J15530050-5434515 & 327.3015 & -0.5377 & MSX6C G327.3017-00.5382 & 4 & IR source\\
  J15530199-5435425 & 327.2953 & -0.5509 & AGAL G327.301-00.552 & 18 & YSO \\  
  J15530579-5435206 & 327.3063 & -0.5520 & GAL 327.3-00.5     & 25 & HII (ionized) region \\ 
  J15540626-5311391 & 328.3071 & 0.4310 & GAL 328.31+00.45   & 26 & HII (ionized) region \\ 
  J15540750-5323442 & 328.1812 & 0.2737 & SDC G328.178+0.285 & 27 & Dark cloud (nebula)\\ 
  J15560795-5419395 & 327.8128 & -0.6322 & MMB G327.808-00.634 & 28 & Maser \\
  J15575303-5402134 & 328.1960 & -0.5751 & IRAS 15539-5353 & 5 & HII (ionized) region  \\ 
  J15581658-5207370 & 329.4774 & 0.8426 &  [MHL2007] G329.4761+00.8414 2 & 14 & YSO candidate  \\
  J15593814-5345369 & 328.5721 & -0.5321 & GUM 50  & 29 &  HII (ionized) region \\ 
  J15594129-5344479 & 328.5868 & -0.5268 & TYC 8697-1749-1  & 30 & Star  \\ 
  J16005582-5236231 & 329.4725 & 0.2150 & IRAS 15570-5227  & 5 & HII (ionized) region\\ 
  J16011315-5234360 & 329.5252 & 0.2087 & MMB G329.526+00.216  & 28 & Maser\\
  J16015300-5356196 & 328.7059 & -0.8840 & IRAS 15579-5347  & 12 &  Far-IR source ($\lambda$ $>=$ 30 {$\mu$}m)\\
  J16055177-5048136 & 331.2379 & 1.0617 & AGAL G331.239+01.059  & 31 & Sub-millimetric source  \\
  J16062307-5043275 & 331.3524 & 1.0657 & 2MASS J16062311-5043276 & 14 & YSO candidate \\
  J16062662-5043302 & 331.3589 & 1.0589 & [CPA2006] S59 & 32 & Bubble\\
  J16075456-4918476 & 332.4815 & 1.9473 & -- & -- & No object detected \\ 
  J16103869-4905591 &  332.9552 & 1.8021 & IRAS 16069-4858 & 5 & HII (ionized) region \\  
  J16121310-5202389 & 331.1301 & -0.5220  & IRAS 16083-5154 & 5 & HII (ionized) region \\
  J16124234-5145015 & 331.3859 &  -0.3593 & VVV CL063 & 3 &  Cluster of stars\\
  J16150065-4950392 & 332.9614 &  0.7739 & [DBS2003] 102 & 33 &  Open (galactic) Cluster\\
  J16151362-4948505 & 333.0074 & 0.7717 & SSTGLMC G333.0075+00.7709 & 34 & AGB star candidate\\  
  J16151868-4948535 & 333.0167  & 0.7617 & [GGB2012] G333.0162+00.7615 & 35 & Outflow candidate\\ 
  J16152012-4913215 & 333.4291 & 1.1868 & PN Pe 1-5 & 36 & PN\\
  J16164238-5117132 & 332.1561 & -0.4532 & GAL 332.15-00.45 & 26 &  HII (ionized) region\\
  J16170265-5047083 & 332.5428  & -0.1297 & 2MASS J16170288-5047051 & 37 &  NIR source ($\lambda$ $<$ 10 {$\mu$}m) \\
  J16185735-5023591 & 333.0294 & -0.0650 & MSX6C G333.0299-00.0645 & 38 & HII (ionized) region \\
  J16190102-5020341 & 333.0763 & -0.0312 & -- & -- & Star with bright nearby stars \\ 
  J16192228-5009150 & 333.2489 & 0.0637 & IRAS 16156-5002 & 5 & HII (ionized) region\\
  J16195698-5045452 & 332.8874 & -0.4343 & -- & -- & Star with bright nearby star \\
  J16200971-4936012 & 333.7284 & 0.3687 & PN G333.7+00.3 & 4 & Possible PN\\
  J16201048-5053291 & 332.8221 & -0.5509 & 2MASS J16201045-5053286  & 14 & YSO candidate\\
  J16203332-5040423 & 333.0148  & -0.4419 & MSX5C G333.0149-00.4431  & 4 & IR source \\
  J16205968-5035141 & 333.1284 & -0.4263 & [MHL2007] G333.1306-00.4275 3 & 14 & YSO candidate\\
  J16210034-5036143 & 333.1179 & -0.4393 & [MKN2004] MMS 40 & 50 & Millimetric radio-source\\
  J16210050-5035089 & 333.1310 & -0.4268 & [MHL2007] G333.1306-00.4275 1 & 14 &  YSO candidate \\
  J16213188-5025042 & 333.3085 & -0.3664 & AGAL G333.308-00.366  & 18 & YSO\\
  J16220862-5005405 & 333.6062 & -0.2070 &  -- &  -- & IR source\\ 
\hline 
  \end{tabular}
  \end{table*}
  
\begin{table*}
\contcaption{}
\center
\begin{tabular}{|cccccc|}
\hline 
2MASX ID        &     $l$    &  $b$     & SIMBAD ID &  REFERENCES &   Description \\
            &   [deg]  & [deg]    &  &  &        \\  
 \hline 
  J16221035-5006168 & 333.6023 & -0.2174 & MSX5C G333.6044-00.2165 & 4 & IR source\\
  J16225317-5048489 & 333.1801 & -0.7981 & CD-50 10459 & 39 &  Emission-line star   \\
  J16265564-4909491 & 334.8132 & -0.1072 & MSX6C G334.8128-00.1059 & 40 & IR source \\
  J16395987-4851524 & 336.4891 & -1.4760 & -- & -- & YSO \\ 
  J16405900-4707080 & 337.9057 & -0.4423 & 2MASS J16405819-4706319 & 40 & AGB star candidate\\
  J16410569-4707383 & 337.9119 & -0.4621 & [PLW2012] G337.909-00.462-039.2 & 41 & Dense core \\
  J16410805-4706463 & 337.9272 & -0.4575 & [MHL2007] G337.9266-00.4588 1 & 14 & YSO candidate\\
  J16410990-4708074 & 337.9138 & -0.4764 & EGO G337.91-0.48 & 42 & Outflow candidate\\
  J16431592-4605409 & 338.9342 & -0.0620 & --  & -- &  IR source  \\
  J16521117-4703067 & 339.1934 & -1.8506 & 2MASS-GC03 & 43 & Globular Cluster\\
  J16534645-4316088 & 342.2984 & 0.3288  & VVV CL089  & 3  & Open (galactic) Cluster \\
  J16541460-4517282 & 340.7828 & -1.0124 & AGAL G340.784-01.016 & 31 & Sub-millimetric source \\ 
  J16541621-4519002 & 340.7659 & -1.0322 & -- & -- & Star    \\
  J16541764-4517122 & 340.7918 & -1.0165 & RCW 110B  & 51 &  HII (ionized) region \\
  J16560261-4304439 & 342.7066 &  0.1268 & [GBM2006] 16524-4300A         & 44 &  Radio-source \\ 
  J16563986-4013290 & 345.0065 &  1.8214 & [MHL2007] G345.0052+01.8209 1 & 14 & YSO candidate\\ 
  J16564527-4014280 & 345.0044 & 1.7977  & IRAS 16533-4009 & 4 &  HII (ionized) region \\
  J16574897-4034050 & 344.8746 & 1.4360  & [MHL2007] G344.8746+01.4347 1 & 14 & YSO candidate\\ 
  J16592078-4232376 & 343.5024 & -0.0142 & MSX6C G343.5024-00.0145 & 4 & IR source\\
  J16593051-4010419 & 345.3804 & 1.4247  & [CAB2011] G345.37+1.42  & 45 & Millimetric radio-source \\ 
  J16593668-4003251 & 345.4880 & 1.4842 & [GGR2014] 1 & 46 & Millimetric radio-source \\ 
  J16593754-4005271 & 345.4630 & 1.4611 & [MHL2007] G345.4640+01.4581 1 & 14 & YSO candidate\\
  J16594225-4003451 & 345.4946 & 1.4668 & IRAS 16562-3959 & 47 & Star forming region \\
  J16594625-4004141 & 345.4961 & 1.4518 & -- & -- & Nebula   \\ 
  J17003381-4034210 & 345.1940 & 1.0234 & -- & -- & Star     \\  
  J17003600-4033300 & 345.2095 & 1.0266 & [DBS2003] 113  & 33 & Cluster of stars \\ 
  J17003811-4032379 & 345.2249 & 1.0303 & MSX6C G345.2244+01.0304 & 4 & IR source \\ 
  J17041301-4220009 & 344.2194 & -0.5953 & 2MASX J17041301-4220009 & 43 & Cluster of stars \\ 
  J17042084-4044456 & 345.4955 & 0.3495 & [MCM2005b] 89 & 33 &  Cluster of stars \\
  J17042791-4046186 & 345.4885 & 0.3160 & MSX6C G345.4881+00.3148 & 4 & HII (ionized) region \\  
  J17043294-4039237 & 345.5898 & 0.3734 &  AGAL G345.589+00.372  & 48 & YSO \\ 
  J17051332-4101242 & 345.3751 & 0.0498 & -- & -- & Stars    \\  
  J17093532-4135589 & 345.4062 & -0.9519 & [DBS2003] 116 & 33 & Cluster of stars\\ 
  J17134463-3706111 & 349.5094 & 1.0554 & NGC 6302 & 36 & PN \\
  J17182142-3919134 & 348.2307 & -0.9700 & --  & -- & IR source \\
  J17182340-3918444 & 348.2410 & -0.9706 & [TPR2014] IRAS 17149 B3 & 49 & Star in Nebula\\ 
  J17182590-3919264 & 348.2361 & -0.9839 & MSX5C G348.2362-00.9804 & 52 &  IR source \\ 
  J17182961-3919024 & 348.2484 & -0.9899 & [BBC2006] IRAS 17149-3916 Clump 4 & 53 & Dense core \\ 
  J17191552-3904324 & 348.5314 & -0.9724 & [BNM96] 348.534-0.973 & 5 & HII (ionized) region \\
  J17192058-3903543 & 348.5494 &  -0.9797 & [GMB2007b] 17158-3901 C  & 50 & Millimetric radio-source\\ 
  J17195582-3900424  & 348.6585 & -1.0427 & [MHL2007] G348.6600-01.0446 1 & 14 & YSO candidate\\
  J17200331-3858039 & 348.7085 & -1.0374  & -- & -- & Star\\ 
  J17200631-3858379 & 348.7063 & -1.0507 & -- & -- & Star in Nebula \\  
  J17200691-3859539 & 348.6901 & -1.0644 & -- & -- & Stars  \\ 
\hline 
\end{tabular}
\begin{tablenotes}
      \small
      \item REFERENCES. 1: \cite{Barba2015}, 2: \cite{Kwok1997}, 3: \cite{Borissova2011}, 4: \cite{Egan2003}, 5: \cite{Bronfman1996},
      \item 6: \cite{Caswell1977}, 7: \cite{Goss1970}, 8: \cite{Harju1998}, 9: \cite{Wray1966}, 10: \cite{Simpson2012}, 11: \cite{Dutra2003}, 
      \item  12: \cite{Helou1988}, 13: \cite{Hindson2013}, 14: \cite{Mottram2007}, 15: \cite{Schroder2007},  16: \cite{Suarez2006},
      \item 17: \cite{Urquhart2014}, 18: \cite{Caratti2015}, 19: \cite{Shaver1983}, 20: \cite{Vig2007}, 21:\cite{Wals2001},
      \item 22: \cite{Ishihara2010}, 23: \cite{Roman2006}, 24: \cite{Giveon2002}, 25:  \cite{Jones2012},  26: \cite{Peretto2009},
      \item 27: \cite{Caswell2010}, 28: \cite{Gum1955}, 29: \cite{Pinheiro2010}, 30: \cite{Contreras2013}, 31: \cite{Churchwell2006}, 32: \cite{Morales2013}, 
       \item 33: \cite{Robitaille2008}, 34: \cite{Guzman2012}, 35: \cite{Frew2013}, 36: \cite{Cutri2003}, 37: \cite{Yu2016}, 38: \cite{Stephenson1971}, 
      \item  39: \cite{Robitaille2007}, 40: \cite{Purcell012}, 41: \cite{Cyganowski2008}, 
      42: \cite{Kharchenko2013}, 43: \cite{Garay2006},
      \item  44: \cite{Culverhouse2011}, 45: \cite{Guzman2014}, 46: \cite{Araya2005}, 47: \cite{Urquhart2018}, 48: \cite{Tapia2014}, 49: \cite{Garay2007},
      \item 50: \cite{Mookerjea2004}, 51: \cite{Kuchar1997}, 52: \cite{Egan2001}, 53: \cite{Beltran2006}
      
    \end{tablenotes}
\end{table*}

\bsp	
\label{lastpage}
\end{document}